\documentclass[aps,pre,reprint,amsmath,amssymb,superscriptaddress,showpacs,floatfix]{revtex4-2}
\usepackage{graphicx}
\usepackage{dcolumn}
\usepackage{bm}
\usepackage[colorlinks, allcolors=blue]{hyperref}
\usepackage[usenames, dvipsnames]{color}
\usepackage{float}

\usepackage{lipsum}
\usepackage{titlesec}
\titlespacing\section{0pt}{12pt plus 4pt minus 4pt}{1pt plus 20pt minus 2pt}
\usepackage{xcolor}
\usepackage{tabularx}
\usepackage{amsmath}
\usepackage{comment}
\usepackage{afterpage}
\usepackage{placeins}
\usepackage{booktabs}
\usepackage{multirow}
\usepackage{array}
\usepackage{setspace}
\graphicspath{{Figs/}}
\usepackage{siunitx}
\usepackage{hhline}
\usepackage{xfrac}
\usepackage{float,graphicx}
\usepackage{mathtools}
\usepackage{listings}
\usepackage{amssymb}
\usepackage[normalem]{ulem}
\usepackage{titlesec}
\usepackage{amsfonts}
\usepackage[version=4]{mhchem}

\usepackage{epstopdf}
\catcode`@11
\def\seceqaa{\@addtoreset{equation}{section}
principles\def\theequation{A\arabic{equation}}}
\def\seceqbb{\@addtoreset{equation}{section}
\def\theequation{B\arabic{equation}}}
\def\seceqcc{\@addtoreset{equation}{section}
\def\theequation{C\arabic{equation}}}
\def\seceqdd{\@addtoreset{equation}{section}
\def\theequation{D\arabic{equation}}}
\def\seceqee{\@addtoreset{equation}{section}
\def\theequation{E\arabic{equation}}}
\def\seceqff{\@addtoreset{equation}{section}
\def\theequation{F\arabic{equation}}}
\def\seceqgg{\@addtoreset{equation}{section}
\def\theequation{G\arabic{equation}}}
\def\seceqhh{\@addtoreset{equation}{section}
\def\theequation{H\arabic{equation}}}
\catcode`@11

\begin{document}

\title{Anomalous Hall effect induced by Berry curvature in topological nodal-line van der Waals ferromagnet Fe$_4$GeTe$_2$} 


\setcounter{footnote}{1}
\author{Satyabrata Bera}
\altaffiliation{These authors contributed equally to this work}
\affiliation{School of Physical Sciences, Indian Association for the Cultivation of Science, Jadavpur, Kolkata 700032, India}

\author{Sudipta Chatterjee}
\altaffiliation{These authors contributed equally to this work}
\affiliation{Department of Condensed Matter and Materials Physics, S. N. Bose National Centre for Basic Sciences, JD Block, Sector III, Salt Lake, Kolkata 700106, India}

\author{Subhadip Pradhan}
\altaffiliation{These authors contributed equally to this work}
\affiliation{School of Physical Sciences, National Institute of Science
Education and Research Bhubaneswar, An OCC
of Homi Bhabha National Institute, Khurda Road, Jatni,
Odisha 752050, India} 

\author{Suman Kalyan Pradhan}
\affiliation{School of Physical Sciences, Indian Association for the Cultivation of Science, Jadavpur, Kolkata 700032, India}

\author{Sk Kalimuddin}
\affiliation{School of Physical Sciences, Indian Association for the Cultivation of Science, Jadavpur, Kolkata 700032, India}

\author{Arnab Bera}
\affiliation{School of Physical Sciences, Indian Association for the Cultivation of Science, Jadavpur, Kolkata 700032, India}

\author{Ashis K. Nandy}
\email{aknandy@niser.ac.in}
\affiliation{School of Physical Sciences, National Institute of Science
Education and Research Bhubaneswar, An OCC
of Homi Bhabha National Institute, Khurda Road, Jatni,
Odisha 752050, India}

\author{Mintu Mondal}
\email{sspmm4@iacs.res.in}
\affiliation{School of Physical Sciences, Indian Association for the Cultivation of Science, Jadavpur, Kolkata 700032, India}

\date{\today}

\begin{abstract}
The exploration of nontrivial transport phenomena associated with the interplay between magnetic order and spin-orbit coupling (SOC), particularly in van der Waals (vdW) systems has gained a resurgence of interest due to their easy exfoliation, ideal for two-dimensional (2D) spintronics. We report the near room temperature quasi-2D ferromagnet, Fe$_4$GeTe$_2$ from the iron-based vdW family (Fe$_n$GeTe$_2$, $n$=3,4,5), exhibiting a large anomalous Hall conductivity (AHC), $\sigma^A_{xy}$ $\sim$ 490 $\Omega^{-1}\textrm{cm}^{-1}$ at 2 K. The near quadratic behavior of anomalous Hall resistivity ($\rho^{A}_{xy}$) with the longitudinal resistivity ($\rho_{xx}$) suggests that a dominant AHC contribution is coming from an intrinsic Berry curvature (BC) mechanism. Concomitantly, the electronic structure calculations reveal a large BC arising from SOC induced gaped nodal lines around the Fermi level, governing such large AHC property. Moreover, we also report an exceptionally large anomalous Hall angle ($\simeq$ 10.6\%) and Hall factor ($\simeq$ 0.22 V$^{-1}$) values which so far, are the largest in compared to those for other members in this vdW family.
\end{abstract}

\maketitle
Topological semimetals (TSMs) are characterized by gapless low-energy electronic states that can be protected by symmetries and simultaneously, offer an excellent platform for exploring unconventional transport phenomena \cite{wanPhysRevB.83.205101,XuPhysRevLett.107.186806,chiuRevModPhys.88.035005,ArmitageRevModPhys.90.015001,LvRevModPhys.93.025002}. By examining the electronic structures, the TSMs can be further classified into Dirac semimetals (DSMs), Weyl semimetals (WSMs), and nodal-line semimetals (NLSMs) depending on the degeneracy and dimensions of the band crossings in the Brillouin zone (BZ). 
The DSMs~\cite{wangPhysRevB.85.195320,hePhysRevLett.113.246402,neupane2014observation,liu2014discovery,Hosen_2018_dirac,PhysRevB.106.125124} with four-fold degenerate band crossing point may transform into two-fold degenerate WSMs~\cite{weng2015weyl,lv2015experimental,xu2015experimental,li2020giant} carrying chiral charges which are attributed to certain symmetry breaking. In the case of NLSMs, the band crossings either form a closed loop or a line instead of discrete points in the BZ \cite{burkov2011topological,fang2015topological,chan20163,liang2016node,kim2018large,singh2021anisotropic,guin20212d}.

Recently, various magnetic systems carrying topologically non-trivial quantum states protected by symmetries governed by their magnetic order, have attracted tremendous attention, owing to the unusual transport properties like anomalous Hall effect (AHE)~\cite{kim2018large,manna2018colossal,li2020giant,singh2021anisotropic,guin20212d, Liu2018a}, anomalous Nernst effect $etc$. \cite{sakai2018giant,xu2019large}. In particular, the AHE becomes one of the most intriguing transport phenomena in magnetic TSMs in which transverse charge current is generated without any external magnetic fields. Especially, the intrinsic contribution to AHE comes from the topologically non-trivial states via a geometrical quantity in the band structures\textemdash the Berry curvature (BC)~\cite{nagaosa2010anomalous}. A gapless nodal-line often does not exhibit a net BC and hence, the anomalous Hall conductivity (AHC) calculated for such nodal-line features near about the Fermi energy (E$_\textrm{F}$) turns out to be zero. These topological features are generally protected by mirror symmetry \cite{PhysRevLett.106.106802,Hsieh2012}, which may break in the presence of spin-orbit coupling (SOC) in magnetic systems~\cite{Wang2018,Fang2003}, then the nodal-line gets either gapped out fully or evolves into a pair of Weyl points~\cite{pradhan2022vector}. In both cases, the AHC can be observed and tuned by changing the electronic structures through suitable manipulation of the magnetic state~\cite{pradhan2022vector,li2020giant,shindou2001orbital,charanpreet}.

The class of quasi-two-dimensional (2D) van der Waals ferromagnet (vdW-FM) provides an excellent platform for investigating many of these novel topological transport phenomena in the 2D limit by probing AHE. However, in most cases, the 2D vdW-FMs such as Cr(Si, Ge)Te$_3$ \cite{PhysRevB.100.060402}, CrI$_3$ \cite{PhysRevB.97.014420} $etc$ show magnetic transition temperatures (T$_\textrm{c}$) well below the room temperature, making them unsuitable for 2D spintronic applications. Very recently, there has been renewed interest in Fe-based 2D vdW-FMs where the T$_\textrm{c}$ can be increased with the Fe concentration, by means of increasing $n$ from 3 to 5 in the Fe$_n$GeTe$_2$ (F$n$GT) vdW family~\cite{wang2017anisotropic,tan2018hard,may2019ferromagnetism,may2019physical,seo2020}. With increasing $n$, the Fe-rich slab making the bulk system exhibits increased strength in exchange interactions between Fe moments, and hence, the T$_\textrm{c}$ of the corresponding member increases~\cite{seo2020}. Additionally, their physical properties significantly differed based on the value of $n$ due to the changes in their electronic structures ~\cite{wang2017anisotropic,liuPhysRevB.97.165415,mondalPhysRevB.104.094405,yang2021strong,gaoPhysRevB.105.014426,kim2018large}. As an example, the $n=3$ member having the closest stoichiometric Fe concentration (Fe$_{2.91}$GeTe$_2$, T$_\textrm{c} \simeq$ 220$K$) shows moderately large anomalous Hall angle ($\Theta_\textrm{AH}$) and Hall factor ($S_\textrm{H}$), which are $\approx 9\%$ and 0.17~V$^{-1}$, respectively~\cite{kim2018large}. Note, in this vdW-FM family, the only nonzero AHC component is $\sigma^{A}_{xy}$~\cite{yang2021strong}. The experimentally reported $\sigma^{A}_{xy}$ for Fe$_3$GeTe$_2$ (F3GT) is $\approx$ 540 $\Omega^{-1}\textrm{cm}^{-1}$, much larger than the theoretically measured value, $\approx$ 180 $\Omega^{-1}\textrm{cm}^{-1}$~\cite{kim2018large}. The above room temperature candidate, Fe$_5$GeTe$_2$ (F5GT), however, shows the maximum AHC value about 120 $\Omega^{-1}\textrm{cm}^{-1}$ near temperatures of about 100 $K$~\cite{may2019physical}. In particular, the theoretically designed F4GT sample by Seo $et~al.$~\cite{seo2020} has gained tremendous attention when experimentally it was found as a good candidate showing large magnetization and high conductivity near room temperatures (T$_\textrm{c}\simeq$ 270~$K$) and these properties were well retained even when it is cleaved.

In this study, we investigate the topological aspects of AHE for the Fe$_4$GeTe$_2$ (F4GT) sample which is found to show a large AHC value, $\sim$~490~$\Omega^{-1}\textrm{cm}^{-1}$ at 2~$K$, comparable with the F3GT sample. Although, a few theoretical works have been primarily focused on comparing anomalous transport properties among F$n$GT members~\cite{yang2021strong,manoranjanpaper}, a well-connected experimental and theoretical study to understand transport properties in F4GT is still lacking. Experimentally, based on Kaplus–Luttinger (KL) mechanism we have identified that in F4GT, the most contribution to AHC is coming from intrinsic BCs, and the estimated value is $\approx$ 462 $\Omega^{-1}\textrm{cm}^{-1}$. This has been further confirmed by electronic structure calculations where the theoretically estimated value is about 485~$\Omega^{-1}\textrm{cm}^{-1}$ at E$_\textrm{F}$. We find that such a large AHC results from BC induced by the gaped nodal line near about $E_\textrm{F}$ in the presence of SOC. Furthermore, in our sample, exceptionally large $\Theta_\textrm{AH}$ ($\approx$~10.6~\%) and $S_\textrm{H}$ ($\approx$~0.22 V$^{-1}$) values are simultaneously observed. 

\begin{figure}
\centering
\includegraphics[width=.5\textwidth]{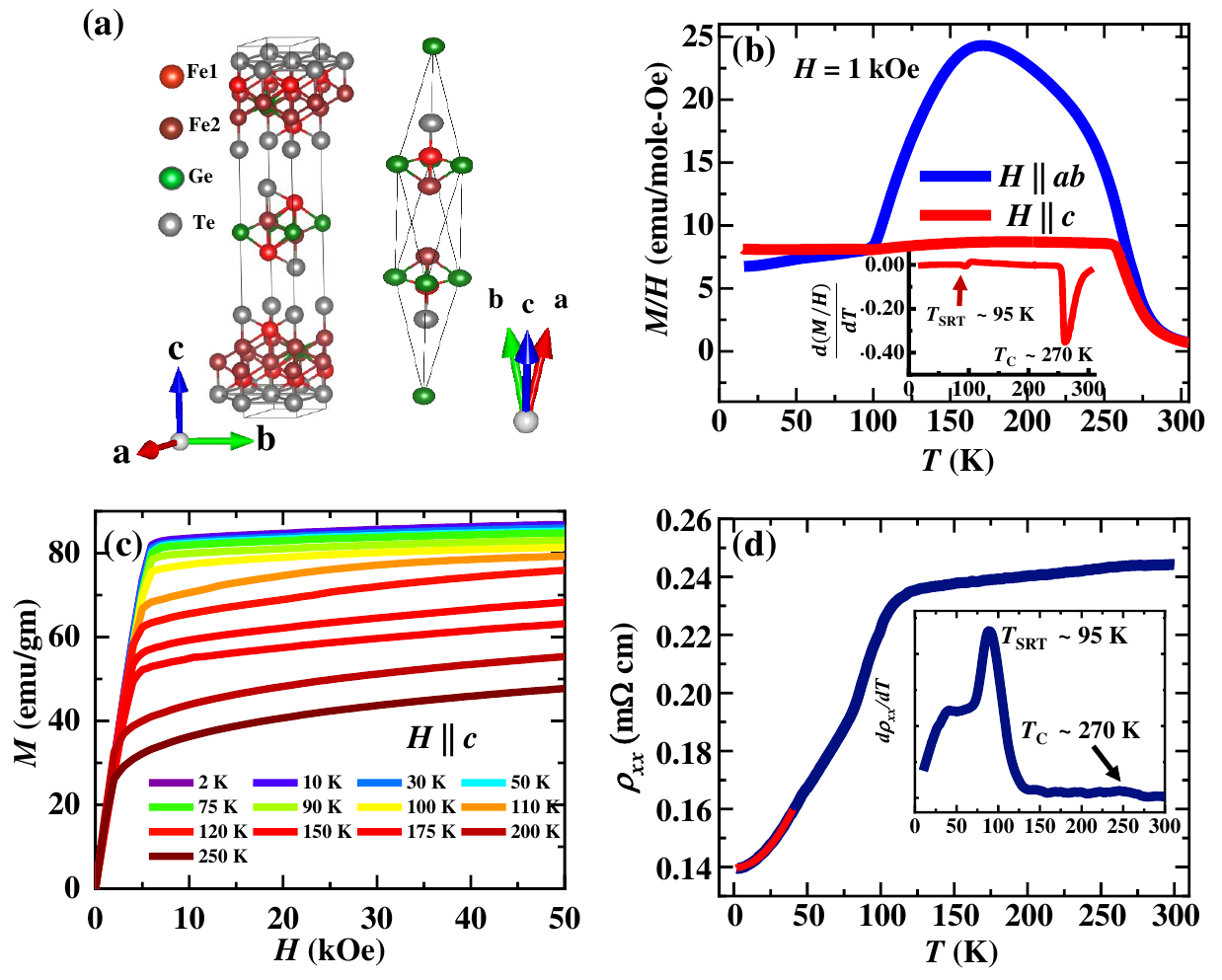}
\caption{(a) The crystal structures of F4GT are represented in the conventional (hexagonal) and primitive (rhombohedral) unit cells. 
(b) Temperature dependence of dc magnetization measured in the zero-field-cooled condition for applied magnetic field $H$ = 1 $kOe$ with $H $$\parallel$ $ab$ and $H$$\parallel$ $c$ respectively. Inset shows the derivative of $M$ $i.e.$ ($\frac{d(M/H)}{dT}$) vs. $T$ plot. (c) The $M$ variation as a function of an applied magnetic field along $c$-axis at different $T$s. (d) Temperature dependence zero field electrical resistivity curve with current through in-plane direction ($I$ $\parallel$ $ab$ plane). The solid red line shows the fit of $\rho_{xx}$$\propto$$T^2$. Inset shows the temperature derivative of $\rho_{xx}$.}
\label{fig1}
\vspace{-0.45cm}
\end{figure}

\begin{figure*}
\centering
\includegraphics[width=1.0\textwidth]{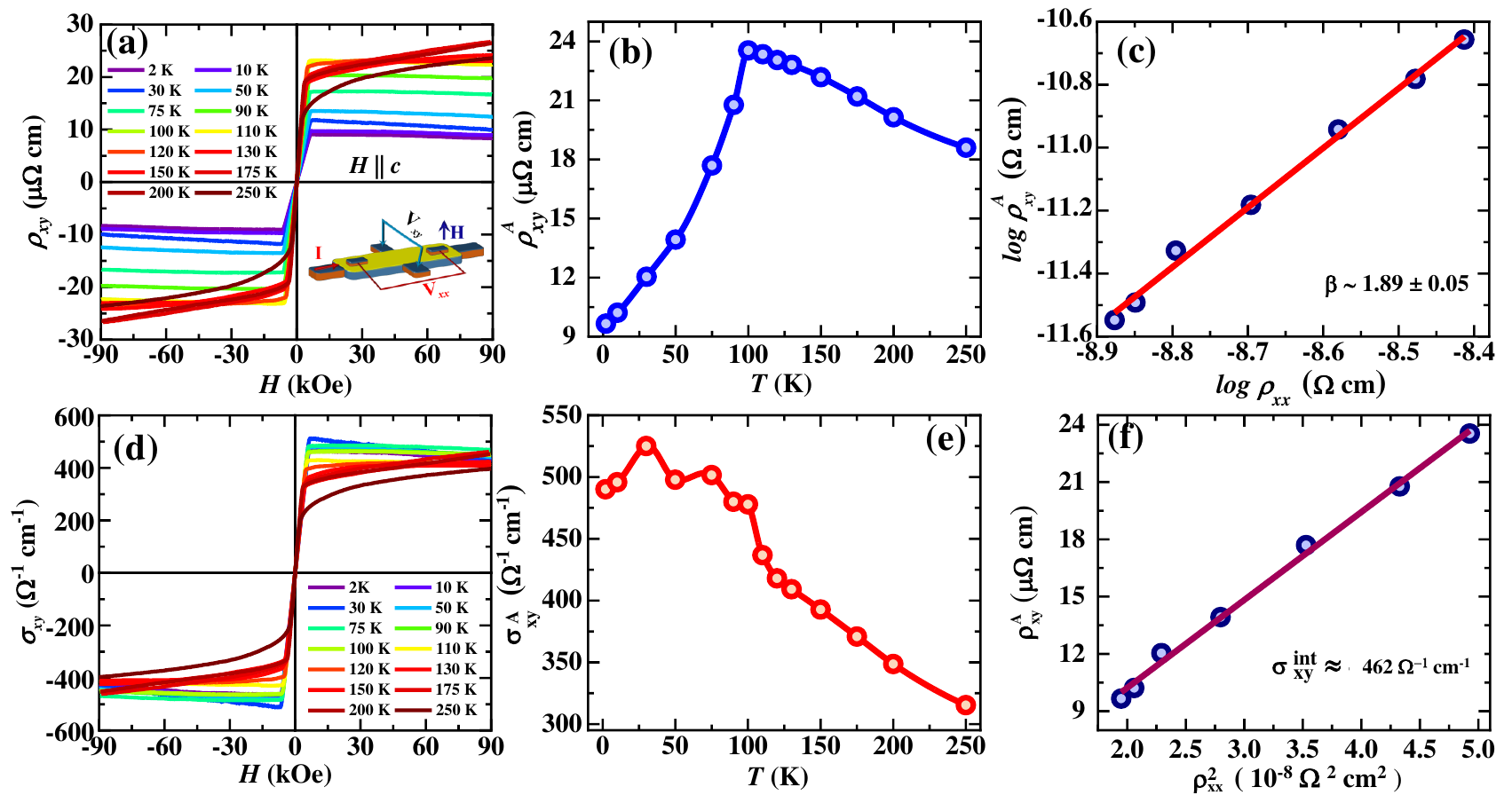}
\caption{(a) Magnetic field dependent Hall resistivity ($\rho_{xy}$) at various temperatures ranging from 2~$K$ to 250~$K$. Inset: The schematic diagram of the sample device used for longitudinal voltage $V_{xx}$ and the Hall voltage $V_{xy}$ measurements. (b) Temperature-dependence of anomalous Hall resistivity $\rho^{A}_{xy}$. (c) The plot of log$\rho^{A}_{xy}$(T) vs log$\rho_{xx}$(T), with the solid red line representing the fit using the relation $\rho^{A}_{xy}$ $\propto$ $\rho^{\beta}_{xx}$. (d) Field-dependent Hall conductivity ($\sigma_{xy}$) at various indicated temperatures. (e) Temperature-dependent anomalous Hall conductivity ($\sigma^{A}_{xy}$). (f) Plot between $\rho^{A}_{xy}$ and $\rho^{2}_{xx}$, and the fitting using Eq.~\ref{eq3} is shown in solid line.} 
\label{fig2}
\vspace{-0.5cm}
\end{figure*}

High-quality single crystals of F4GT were grown by the chemical vapor transport (CVT) method using I$_2$ as a transport agent (see Supplemental Material Sec:S1(A)\cite{supply}.)
The crystal structures of F4GT (both primitive and conventional unit cell) are depicted in Fig.~\ref{fig1}(a), and the X-ray diffraction (XRD) pattern of the F4GT single crystal at room temperature is given in Fig.~S2 in the Supplemental Material (SM)~\cite{supply}. The crystal morphology and chemical compositions of the samples were analyzed using a high-vacuum scanning electron microscope (SEM), specifically the JEOL LSM-6500 model. Results are discussed in  Supplemental Material Sec: S1(C)\cite{supply}.

The magnetization measurements of the sample were performed using a Magnetic Property Measurement System (MPMS-SQUID, Quantum Design, USA). The magnetization data up to an applied field of 5 T and temperature down to 2 $K$ were recorded. The sample used for the magnetic measurements was of approximate dimensions 1.4 $\times$ 1.4 $\times$ 0.06  mm$^3$. The electrical and magnetic transport measurements were carried out by using a 9~T Physical Property Measurement System (PPMS, Quantum Design, USA) using the electrical transport option (ETO). Measurements of both longitudinal ($\rho_{xx}$) and Hall ($\rho_{xy}$) resistivities were done in a standard four-probe method using conducting silver paint and gold wires. The current ($I$ = 100 $\mu$A) is applied along the $ab$-plane and the applied magnetic field along the $c$-axis of the sample. In order to effectively eliminate the $\rho_{xx}$ contribution coming from the voltage probe misalignment, the final $\rho_{xy}$ is obtained by the difference of transverse resistivity measured at the positive and negative magnetic fields.

Figure ~\ref{fig1}(b) shows the temperature-dependent dc magnetization ($M$) under 1~$kOe$ applied magnetic field in parallel to the $ab$ and $c$ directions of the F4GT single crystal. A sharp increase in $M$ is observed with decreasing temperature, which clearly signifies a paramagnetic (PM) to an FM phase transition. 
The observed magnetic behavior of F4GT is consistent with previous reports~\cite{may2019physical,BERA2023170257}. From the $\frac{d(M/H)}{dT}$ vs. $T$ curve shown in the inset of Fig.~\ref{fig1}(b), the estimated T$_\textrm{c}$ (Curie temperature) is about 270~$K$.
With further lowering the temperature the magnetic moment along the $ab$ plane continuously increases to maxima, then quickly decreases and becomes equivalent to the moment along the $c$ axis. In the temperature range 270 $K$–160 $K$, the majority of Fe atoms at sites other than Fe1 (i.e., Fe2) and minor proportions of Fe1 align along the direction of the field. Therefore, the resultant moment of the system is ferromagnetically ordered. Below 160 $K$ down to 100 $K$, a major proportion of the Fe atoms at Fe1 sites, initially, in a disordered state, reorients and interact with the other Fe atoms. This reduces the magnetic moment with decreasing temperature \cite{BERA2023170257}. By further decreasing temperature, another magnetic transition is observed at $T~\approx$ 95~$K$, matching well with the recently reported spin-reorientation transition temperatures (T$_\textrm{SRT}$)~\cite{may2019physical,seo2020}. 
 Below T$_\textrm{SRT}$, the moment value along $c$-axis is little larger than $ab$-plane which signifies the switching of the easy axis of the magnetization.  The above results reveal the reorientation of the magnetic easy axis towards $c$-direction from $ab$ plane below T$_\textrm{SRT}$\cite{BERA2023170257}.

In Fig.~\ref{fig1}(c), the isothermal magnetization as a function of an applied magnetic field for different temperatures is shown up to 5~T, along the $c$ direction. The $M-H$ data along the $ab$ direction is incorporated in the Supplemental Material (see Fig. S4)~\cite{supply}. At temperatures well below T$_\textrm{c}$, each $M(H)$ isotherm shows a sharp increase in the low field region and a saturation-like behavior in the high field region. The saturation magnetization decreases continuously as the temperature increases, and the overall nature of the $M(H)$ isotherms changes significantly as we approach T$_\textrm{c}$. The temperature-dependent $\rho_{xx}$ is shown in Fig.~\ref{fig1}(d) where the residual resistivity value is $\sim$ 0.1396 $m\Omega~cm$ at 2 $K$, yielding a residual resistivity ratio [RRR = $\rho_{xx}$(300 K)/$\rho_{xx}$(2 K)] of $\sim$ 1.75. The $d\rho_{xx}/dT$ vs. $T$ plot in the inset of Fig.~\ref{fig1}(d) further reveals clear slope changes near T$_\textrm{SRT} \sim$ 95 $K$ and T$_\textrm{c} \sim$ 270 $K$ which strongly supports the magnetic origin of phase transitions around the T$_\textrm{SRT}$ and T$_\textrm{c}$. A clear anomaly in the resistivity data is observed around 40 $K$. This anomaly is related to the Fermi-liquid behavior of the present compound where the $\rho$$_{xx}$ shows a quadratic temperature dependence~\cite{pal2023unusual}. To establish the Fermi liquid behavior for the present compound, the resistivity data is fitted from 2 $K$ to 40 $K$ using $\rho_{xx}$$\propto$$T^2$. The observed temperature behavior of transport properties is consistent with the previous report~\cite{seo2020}. 



We have now carried out detailed magneto-transport measurements over a wide range of temperatures to study the AHE in this compound. Figure \ref{fig2}(a) depicts the Hall resistivity ($\rho_{xy}$) as a function of the applied magnetic field  for temperatures ranging from 2 $K$ to 250 $K$. The $\rho_{xy} (H)$ grows sharply for a small applied magnetic field and anomalous behavior is observed up to $\sim$ 0.6 T. In the high field region, $\rho_{xy} (H)$ shows a weak linear field dependence up to a magnetic field of 9 T. Similarities in the forms of the $M(H)$ and $\rho_{xy} (H)$ curves clearly indicate the presence of AHE in the investigated studied compound. As we approach the T$_\textrm{c}$, the overall character of the $\rho_{xy} (H)$ curves changes substantially in both the low field and the high field regions. 
 Generally, in addition to the ordinary Hall effect, $\rho_{xy}$ in a FM material has an additional contribution from $M$ and expressed as,

\vspace{-0.5cm}
\begin{equation}
 \si{\rho}\textsubscript{xy}(H) = \si\rho^0_{xy}+\si\rho^A_{xy} = R_0\si{H} + R_s\mu_{0}M
\end{equation}
where, $\rho^{0}_{xy}$  and $\rho^{A}_{xy}$  are the ordinary and anomalous contributions to the Hall resistivity, with $R_0$ and $R_s$ being the normal and anomalous Hall coefficients, respectively. The values of $R_0$ and $\rho^{A}_{xy}$ are determined from the linear fit to the $\rho_{xy} (H)$ curve in the high field region, where the slope and the y-axis intercept of the linear fit correspond to $R_0$ and $\rho^{A}_{xy}$, respectively. Figure~\ref{fig2}(b) depicts the variation of the $\rho^{A}_{xy}$ as a function of temperature, demonstrating that the $\rho^{A}_{xy}$ increases with rising temperature from 2 $K$ to approximately 100 $K$, and then starts decreasing above 100 $K$.

To elucidate the mechanism responsible for the observed AHE in F4GT, we have plotted $\rho^{A}_{xy}$ vs $\rho_{xx}$ on a double logarithm scale, and a fitting was employed to evaluate the exponent $\beta$ using the relation $\rho^{A}_{xy}$ $\propto$ $\rho^{\beta}_{xx}$ \cite{nagaosa2010anomalous} as shown in Fig.~\ref{fig2}(c). As above $\sim$100 $K$, the magnetic transition comes into play and $\rho^{A}_{xy}$ starts decreasing at higher temperatures. So, for a good comparison, the plot between $\rho^{A}_{xy}$ and $\rho_{xx}$ is restricted in the temperature range from 2 $K$ to 100 $K$, and this has been addressed previously for several metallic ferromagnets \cite{manna2018colossal,royPhysRevB.102.085147,Chatterjee_2023}. The near-quadratic relationship between $\rho^{A}_{xy}$ and $\rho_{xx}$ ($i.e.$, $\beta$ $\approx$ 2) strongly indicates that the intrinsic KL or the extrinsic side-jump mechanism dominates in the AHE rather than the extrinsic skew-scattering mechanism. Furthermore, we have also calculated the Hall conductivity ($\sigma_{xy}$) using the tensor conversion relation \cite{manna2018colossal} as,

\vspace{-0.55cm}
\begin{equation}
 \si{\sigma}\textsubscript{xy} = \frac{\si\rho_{xy}}{(\si\rho_{xx}^2 + \si\rho_{xy}^2)}~. 
 \label{eq2}
 \vspace{-0.2cm}
\end{equation}

Figure ~\ref{fig2}(d) demonstrates the field dependence of Hall conductivity at various temperatures. The AHC is obtained by extrapolating the high-field $\sigma_{xy}$ data to zero fields on the $y$-axis. The temperature dependence of AHC is shown in Fig.~\ref{fig2}(e). At 2~$K$, a large value of AHC is observed to be $\sim$ 490~$\Omega^{-1}\textrm{cm}^{-1}$. The temperature-dependent AHC and $\rho^{A}_{xy}$ are presented in Fig. S5 of the Supplemental Material \cite{supply}, and the change of AHC is almost temperature independent, showing that the origin of AHE in F4GT is intrinsic in nature \cite{Chatterjee_2023}. 
To extract the intrinsic AHC value, we have fitted $\rho^{A}_{xy}$ vs. $\rho^2_{xx}$ plot as illustrated in Fig.~\ref{fig2}(f), using the following equation~\cite{manna2018colossal,Chatterjee_2023}, 

\vspace{-0.45cm}
\begin{equation}
\rho^A_{xy} = f(\rho_{xx0})+\sigma^{int}_{xy}\rho^2_{xx}~, 
\label{eq3}
\end{equation}

\begin{figure}
\centering
\includegraphics[width=0.5\textwidth]{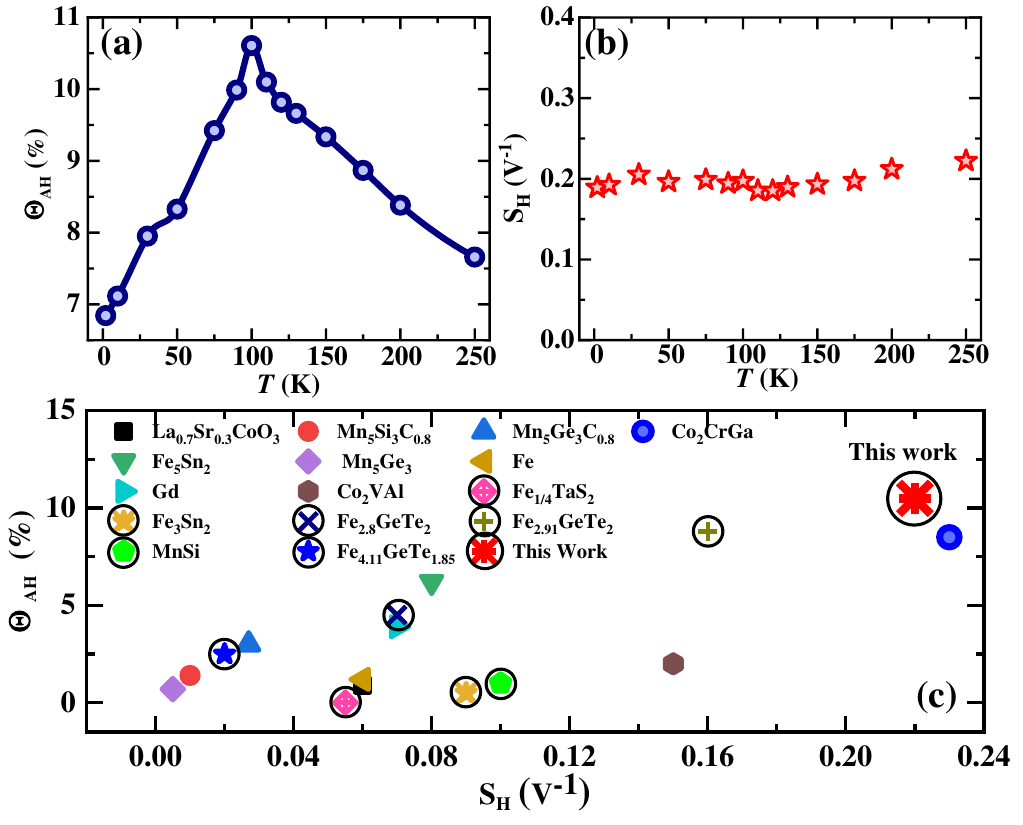}
\caption{(a) and (b) represent the $\Theta_\textrm{AH}$ and the $S_\textrm{H}$ as a function of temperature, respectively. (c) The $\Theta_\textrm{AH}$ and S$_\textrm{H}$ of the F4GT compound are plotted along with the other previously reported metallic FMs. Particularly, the symbols enclosed by the black empty circle represent vdW materials.}
\label{fig3}
\vspace{-0.2cm}
\end{figure}

\begin{figure*}
\centering
\includegraphics[width=0.95\textwidth]{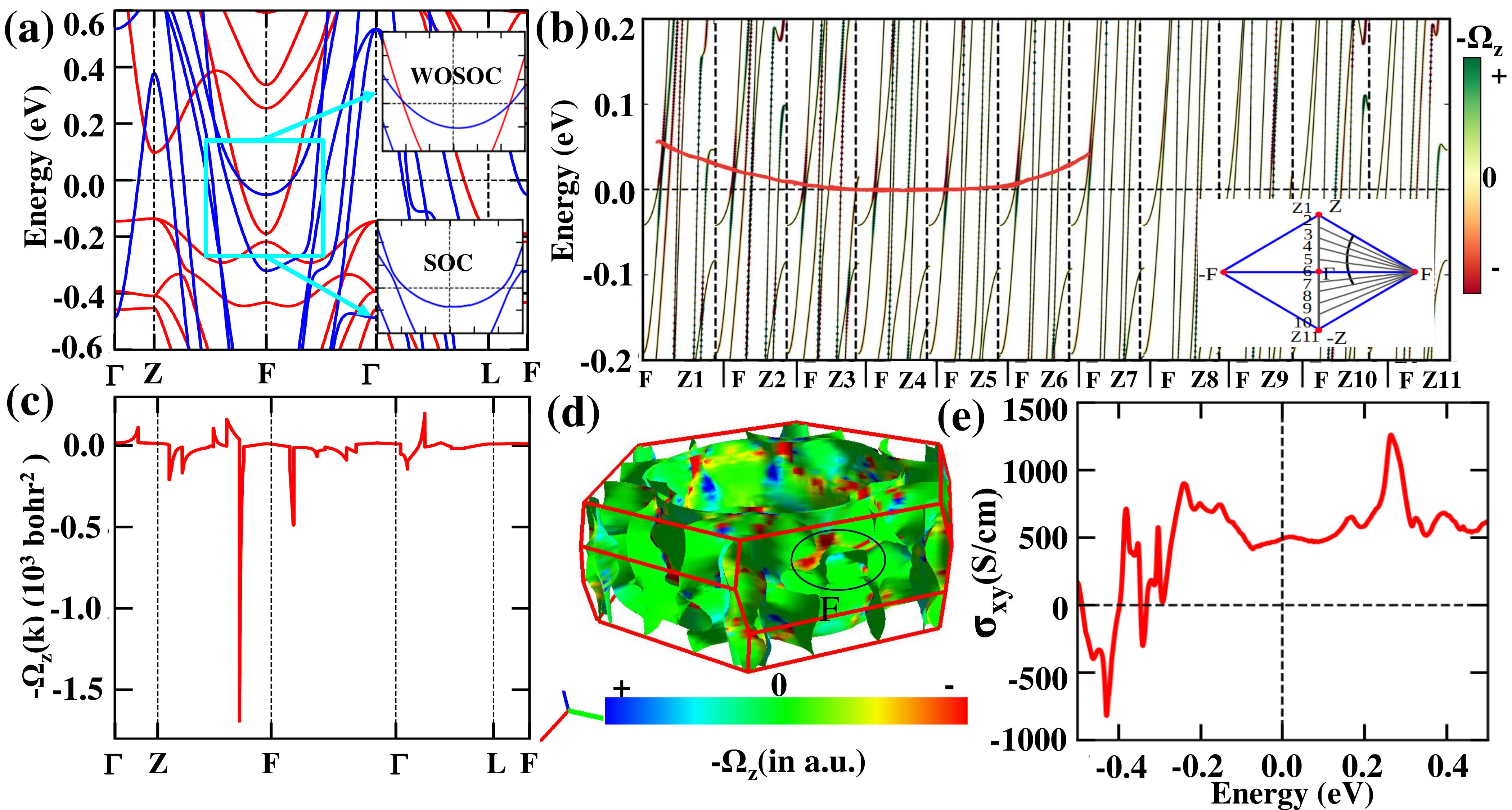}
\caption{(a) F4GT band structure is calculated for the primitive rhmbohedral unit-cell (Fig.~\ref{fig1}(a)) without SOC. In the inset, the zoomed band structures around the high-symmetric $F$ point in the BZ show the gap opening with SOC. (b) The BC induced by SOC for various $k$-path segments (the inset) is shown along the gaped nodal-line where the red curved line passing through all blue boxes is the guide to the eye. The color bar represents the amplitude of the $\Omega_{Z}$. (c) The BC plot along the high-symmetry path in the BZ for $E$=$E_\textrm{F}$. (d) The 3D map of $\Omega_{Z}$ in the full BZ shows the hot spots ($e.g.$, like the closed black circle) around $F$ point with a large BC value. (e) $\sigma^A_{xy}$ is calculated from BCs over full BZ within the energy range $-0.5~eV$ to $0.5~eV$ around $E_\textrm{F}$.}
\label{fig4}
\vspace{-0.45cm}
\end{figure*}

Where, the first term i.e. $f(\rho_{xx0})$ is a function of residual resistivity $\rho_{xx0}$, which includes the contributions due to skew scattering as well as side-jump mechanisms, and $\sigma^{int}_{xy}$ is the AHC purely due to the intrinsic Berry curvature mechanism \cite{Chatterjee_2023,manna2018colossal}. The estimated $\sigma^{int}_{xy}$ is found to be about~462~$\Omega^{-1}\textrm{cm}^{-1}$. As a result, the intrinsic Berry phase-driven mechanism is responsible for more than 90\% of the total AHC. The extrinsic side-jump contribution of AHC $|\sigma^A_{xy,sj}|$ has been shown to be in the order of ($e^2/(ha)$($\epsilon_\textrm{SOC}$/$E_F$)), where $\epsilon_\textrm{SOC}$ is the SOC strength~\cite{Nozieres1973simple,OnodaPhysRevLett.97.126602}.
For metallic FMs, ($\epsilon_\textrm{SOC}/E_\textrm{F}$) is usually in the order of 10$^{-2}$ \cite{royPhysRevB.102.085147,Chatterjee_2023}. Consequently, the extrinsic side-jump contribution to AHC should be very small or negligible in comparison to the intrinsic AHC. Therefore, the AHE in F4GT is primarily dominated by the intrinsic Berry phase-driven KL mechanism. Theoretically, the intrinsic AHC is of the order of $e^2/(ha)$ in the resonance condition, where $e $ is the electronic charge, $h$ is Planck’s constant, and $a$ is the lattice constant \cite{OnodaPhysRevLett.97.126602,PhysRevLett.99.086602}. Taking $a$ =V$^{1/3}$ $\sim$ 7.32 $\AA$  for the studied compound \cite{PhysRevB.94.075135}, we estimate the intrinsic AHC to be around $\sim$ 525~$\Omega^{-1}\textrm{cm}^{-1}$ for F4GT, which is closer to our experimentally observed value of 462~$\Omega^{-1}\textrm{cm}^{-1}$\cite{PhysRevB.107.125138}.

To further assess the strength of AHE in such vdW FM, we have estimated the two characteristic quantities, namely the anomalous Hall angle, $\Theta_\textrm{AH}$ and the anomalous Hall factor, $S_\textrm{H}$. The former can be determined as, $\Theta_\textrm{AH}$ = $\sigma^{A}_{xy}/\sigma_{xx}$, measuring the strength of the transverse current generated with respect to the applied normal current~\cite{nagaosa2010anomalous}. The latter one is defined as, $S_\textrm{H}$ = $\sigma^A_{xy}/{M}$, the measure of sensitivity of AHC with respect to $M$~\cite{PhysRevB.75.172403}. Figure~\ref{fig3}(a) and (b) show the plot of $\Theta_\textrm{AH}$ and $S_\textrm{H}$ as function of temperature, respectively. The $\Theta_\textrm{AH}$ value reaches maximum which is about 10.6\% at T$_\textrm{SRT}$ and it is also significantly large ($\sim 7.6\%$) even at $T=250~K$. On the other hand, the $S_\textrm{H}$ remains almost invariant with $T$, and a maximum value of about 0.22 $V^{-1}$ is obtained at 250 $K$. Such surprising coexistence of large $\Theta_\textrm{AH}$ and $S_\textrm{H}$ within a wide temperature window has led us to compare those quantities with other known metallic FMs, as shown in Fig.~\ref{fig3}(c). It is extremely rare to find a ferromagnetic metallic system with both large $\Theta_{AH}$ and S$_H$ \cite{kim2018large}. From Fig.~\ref{fig3}(c), we infer that the F4GT shows significantly larger $\Theta_{AH}$ and S$_H$ values compared to all previously reported metallic ferromagnets~\cite{PhysRevB.75.172403,PhysRevB.79.014431,PhysRevB.90.104421,PhysRevB.71.104407,PhysRevB.73.174421,PhysRevB.94.075135,Chatterjee_2023}.


In order to investigate the intrinsic origin of AHE for F4GT, we performed first-principles electronic structure calculations within density functional theory (DFT) formalism where the band structure is further parameterized using maximally-localized Wannier functions (ML-WFs) for AHC calculations. Details of the computational methodology and numerical AHC simulations are provided in the SM~\cite{supply}. All calculations are performed using the primitive rhombohedral unit cell and the band strictures are plotted along  the lines connecting high symmetry points in the BZ, as shown in  Fig.~S6. Within magnetic calculations, the spin-resolved atom projected density of states in Fig.~S6(c) from SM~\cite{supply} clearly indicate two types of Fe atoms having the magnetic moments, 1.94 and 2.64 $\mu_B$ for Fe1 and Fe2, respectively. The average magnetic moment per Fe atom is slightly higher than the experimentally reported value ($\sim$ 2.12 $\mu_\textrm{B}$/Fe)~\cite{BERA2023170257}. 

By analyzing the band structure without SOC, we identify two band crossing points along $FZ$ and $F\Gamma$ directions in the close vicinity of $E_\textrm{F}$, see the boxed region in Fig~\ref{fig4}(a). Two inset figures describe that by introducing SOC in our calculations, the crossing points are gaped out. In order to identify the nodal line, we have calculated band structures without SOC along several line segments connecting $F$ point and a few points on the $\bar{Z}\Gamma Z$ line, see Fig.~S8 of the SM \cite{supply}. Near $E_\textrm{F}$, the nodal line is depicted by joining the crossing points of bands (up and down) by a curved black line $i.e.$ a dispersive gapless nodal line. The projected orbital analysis, see Fig.~S7 in the SM \cite{supply}, suggests that those key bands have strong hybridization primarily between Fe-$d$ ($d_{yz}$, $d_{xz}$ and $d_{z^2}$) and Te-$p$ ($p_y$) orbitals. The twofold degeneracy of the nodal-line is lifted with the inclusion of SOC. Therefore, the gaped nodal line in the presence of SOC is found to generate non-zero BC, see Fig.~\ref{fig4}(b) as well as Fig.~S9 in the SM~\cite{supply}. It is evident from Fig.~S9(b) of the SM \cite{supply} that a strong BC is generated by the SOC-induced gap between bands at $E_\textrm{F}$.  Moreover, the color map denoting the BC contributions on band structures for various line segments, see the inset of Fig.~\ref{fig4}(b), infers that the gaped nodal line in the $FZ\Gamma$ plane plays the key role in generating the BC in F4GT. In Fig.~\ref{fig4}(c), the calculated BC at slightly above $E_\textrm{F}$ as a function of momentum $k$ along the high symmetric directions also reveals a large value around the $F$ point, and the largest value is found along the $FZ$ direction. The three-dimensional Fermi surface plot in conjunction with the BC color map in Fig.~\ref{fig4}(d) further strengthens these observations by identifying hot spots ($e.g.$, like black, circled one) near the $F$ point region point. This also yields that the major contributor to the BC in the BZ comes from those hot spots. After identifying the BC contribution from band structure, we have performed the numerical calculations for AHC (see Sec.~S4 of the SM for details~\cite{supply}). Note, the calculated AHC shown in Fig.~\ref{fig4}(e) will be completely intrinsic. We observe a nearly constant value within a range of 100 meV around $E_\textrm{F}$ and the calculated value is about 485~$\Omega^{-1}\textrm{cm}^{-1}$ at $E_\textrm{F}$. This is now attributed to the BC arising from the gaped nodal line and the value is in close agreement with the experimentally estimated intrinsic AHC value ($\sim$ 462 ~$\Omega^{-1}\textrm{cm}^{-1}$). It is also noteworthy that we observe a significant change in transport properties above the T$_\textrm{SRT}$, as demonstrated in Fig. 1(d) and Fig. 2(b). Additionally, we have explored from the first-principle electronic structure calculations that the AHC is changing gradually with a change in magnetization orientation (see Supplemental Material Sec. S5(E)\cite{supply}). This is due to the change in BC obtained from the calculated band structure.


In summary, we have systematically investigated the AHE in a quasi-2D vdW FM, Fe$_4$GeTe$_2$, a member in the Fe$_n$GeTe$_2$ family. The detailed analysis of our results reveals that F4GT exhibits a significantly large AHE along with a rare coexistence of a large anomalous Hall angle, $\Theta_{AH}$ $\simeq$ 10.6\%, and a large anomalous Hall factor, S$_H$ $\simeq$ 0.22 $V^{-1}$. The $\rho^{A}_{xy}$ scales near quadratically with $\rho_{xx}$, and our comprehensive experimental analysis reveals that the observed large AHE in the F4GT compound is attributed to an intrinsic BC-driven KL mechanism. Experimentally, a large AHC $\sigma_{xy}$ $\sim$ 500 $\Omega^{-1}\textrm{cm}^{-1}$ with an intrinsic contribution of $\sim$ 462 ~$\Omega^{-1}\textrm{cm}^{-1}$ is seen at~ 2 K. Theoretically, our band structure calculations establish that such a large BC is owing to the gaped nodal line in the presence of SOC. Notably, our first-principles calculation also predicts an intrinsic AHC value of 485~$\Omega^{-1}\textrm{cm}^{-1}$, which is in good agreement with the aforementioned experimentally observed value. Therefore, the coexistence of exceptionally large AHC,  $\Theta_\textrm{AH}$ and $S_\textrm{H}$ values along with a better T$_\textrm{c}$ make F4GT a suitable candidate for 2D spintronic devices in comparison to other vdW FMs.

Note that we are also aware of the experimental work on the very recent arXiv paper \cite{pal2023unusual} dealing with, primarily, magnetoresistance and briefly on AHE in the same system but on the F4GT flake instead of a bulk single crystal presented here. 

This work was supported by the (i) `Department of Science and Technology (DST)', Government of India (Grant No. SRG/2019/000674 and EMR/2016/005437),  S.Bera $\&$ A.Bera thanks CSIR Govt. of India for Research Fellowship with Grant No. 09/080(1110)/2019-EMR-I  $\&$ 09/080(1109)/2019-EMR-I, respectively. S. C. thanks Dr. B Ghosh and A. Panda for their help. S.P. and A.K.N. acknowledge the computational resources, Kalinga cluster, at National Institute of Science Education and Research, Bhubaneswar, India. A.K.N. and S.P. thanks Prof. P. M. Oppeneer for the Swedish National Infrastructure for Computing (SNIC) facility. 
The authors also thank the DST, India for the financial support during the experiments at the Indian Beamline, PF-18B, KEK, Japan, and Proposal No.2021-IB-28

%


\clearpage
\newpage

\vspace{2cm}

\begin{center}

\onecolumngrid
\textbf{\underline{\large{Supplemental Material}}} \\
\vspace{0.5cm}
\textbf{Anomalous Hall effect induced by Berry curvature in topological nodal-line van der Waals ferromagnet Fe$_4$GeTe$_2$} \\ 
\vspace{0.5cm}

\end{center}

\tableofcontents

\renewcommand{\thesection}{S\arabic{section}}
\renewcommand{\thefigure}{S\arabic{figure}}
\setcounter{section}{0}
\setcounter{figure}{0}

\section{Sample synthesis and characterization}
\subsection{Sample synthesis}
\begin{figure}[hb!]
\centering
\includegraphics[width=0.5\textwidth]{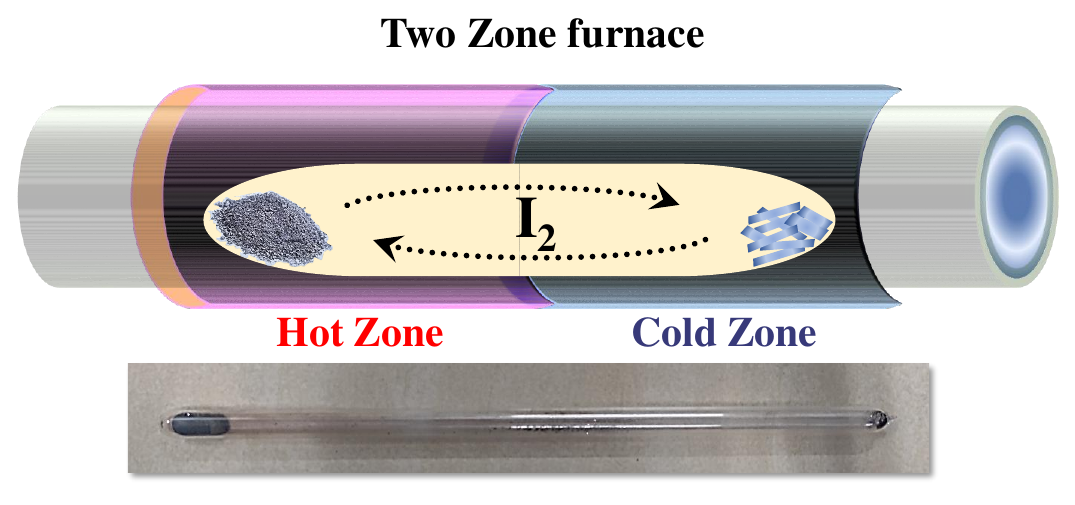}
\caption{The schematic diagram of the synthesis of Fe$_4$GeTe$_2$ single crystal by chemical vapor transport (CVT) method.}
\label{CVT}
\end{figure}

High-quality single crystals of Fe$_4$GeTe$_2$ (F4GT) were grown by the chemical vapor transport (CVT) method using I$_2$ as a transport agent. The schismatic representation of CVT method is presented in Figure~\ref{CVT}. The mixture of [Fe (99.99$\%$ pure), Ge (99.99$\%$ pure), and Te (99.99$\%$ pure)] powders in a molar ratio Fe:Ge: Te = 4.5:1:2 along with the transport agent, was vacuum-sealed in a quartz tube. Excess Fe was used to maintain the stoichiometric ratio, as there is a natural tendency for having vacancy at the Fe site in FGT compounds. Then the vacuum-sealed quartz tube was placed inside a two-zone horizontal tube furnace with source and growth temperatures $\sim$ 750 $^o$C and 700$^o$C, respectively \cite{SupBERA2023170257,SupmondalPhysRevB.104.094405}. The temperatures were kept fixed with minimal temperature fluctuations ($\pm$ 1 $^o$C), maintaining a constant temperature gradient for seven days to complete the growth process. After the growth process, the furnace temperature was reduced to room temperature (RT). After that, the quartz tube was broken under ambient conditions, and a few large, thin, shine-surface, good-quality single crystals with lateral dimensions up to six millimeters were obtained\cite{SupBERA2023170257,SupmondalPhysRevB.104.094405}. 

\subsection{X-ray diffraction }

\begin{figure*}
\centering
\includegraphics[width=0.9\textwidth]{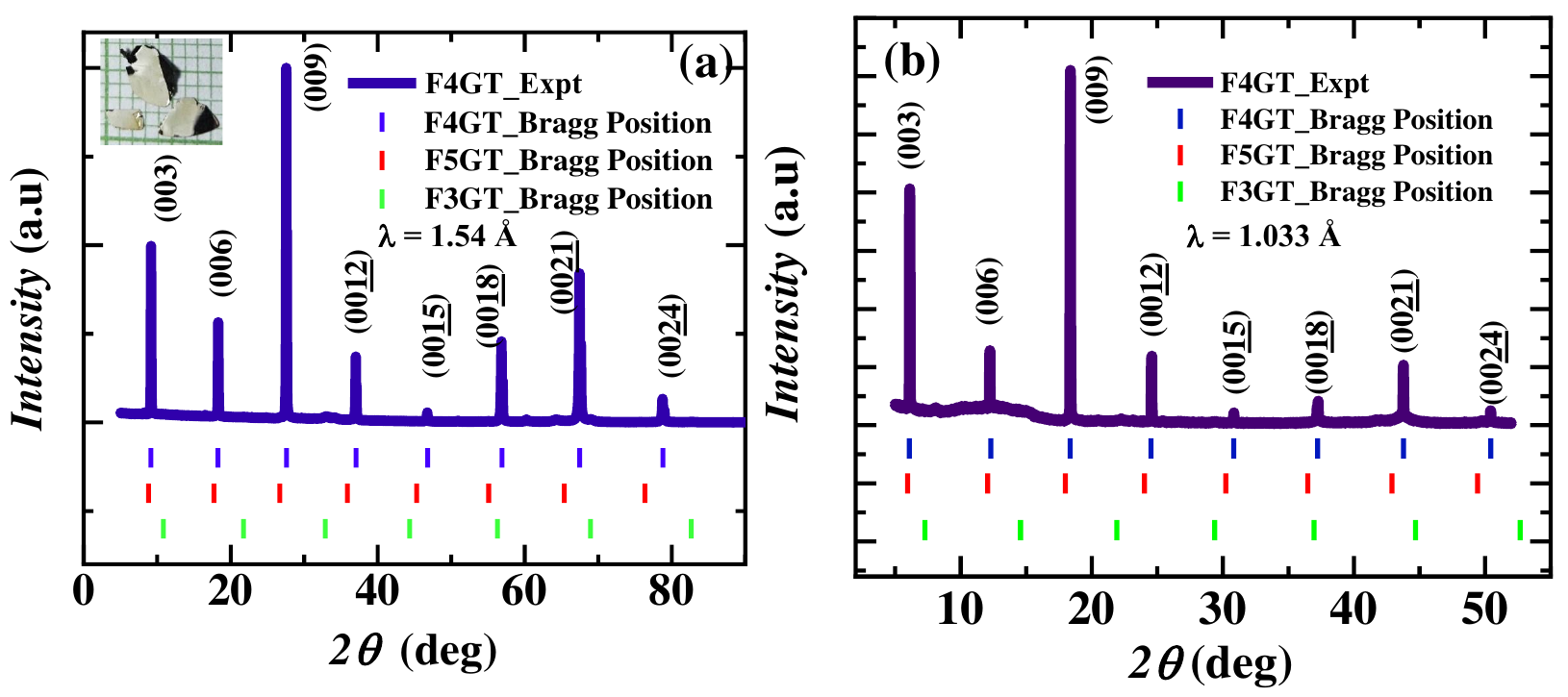}
\caption{X-ray diffraction pattern of Fe$_4$GeTe$_2$ for two different wavelength (a)~$\lambda$ = 1.54 $\AA$ (b)~$\lambda$ = 1.033 $\AA$ at room temperature and the Bragg position of F3GT, F4GT and F5GT. The shiny single crystals are shown inside.}
\label{S1}
\end{figure*}

Room temperature X-ray diffraction pattern for F4GT single crystal  was carried out by using the X-ray wavelength of 1.54 $\AA$ using our in-house X-ray diffractometer. We also performed XRD using the synchrotron X-ray radiation facility of Beamline PF-18B at Photon Factory, KEK, Japan  (X-ray wavelength of 1.033 $\AA$). The obtained results are illustrated in Figure.~\ref{S1} (a-b). All the Bragg reflection peaks come from the $(00l)$ plane which indicates that the crystal surface is normal to the $c$ axis with the plate-shaped surface parallel to the $ab$ plane. A real picture of a few single crystals with a moderate dimension of  4 × 3 × 0.3 mm$^{3}$ is shown in the inset of Figure.~ \ref{S1}. 

Bragg position of all Fe$_4$GeTe$_2$, Fe$_5$GeTe$_2$\cite{May2019,PhysRevMaterials.4.074008} and Fe$_3$GeTe$_2$ \cite{Alghamdi2019,PhysRevB.96.134428} are shown in the Figure~\ref{S1}. All Bragg reflection peaks are shown in the Figure~\ref{S1} related to $(00l)$ plane. From the experimental data, we clearly observe that all peaks related to F4GT crystal, there are no other peaks come from F5GT and F3GT. The estimated lattice parameters are found to be of $a$ = $b$ = 4.04 $\AA$, and $c$ = 29.14 $\AA$ \cite{SupBERA2023170257,SupmondalPhysRevB.104.094405}.

\subsection{Chemical characterization: EDX}
\begin{figure*}
\centering
\includegraphics[width=1\textwidth]{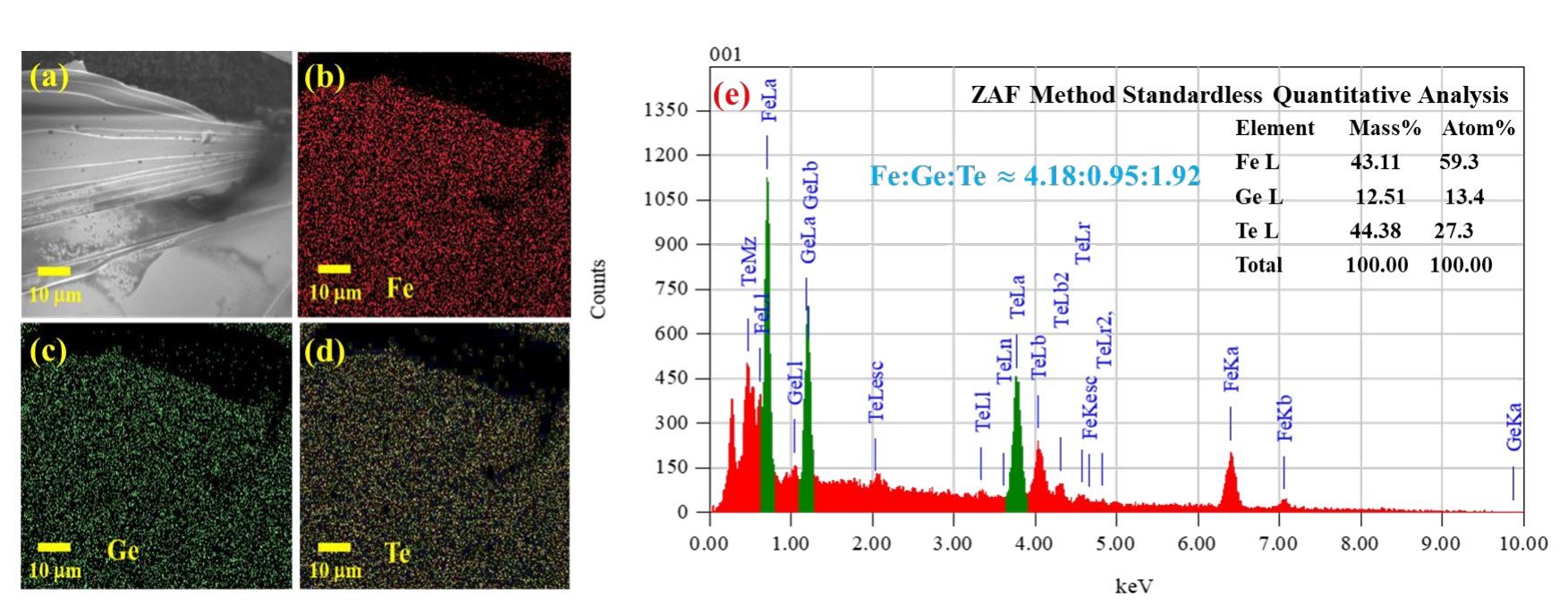}
\caption{Microstructural characterization of as-grown F4GT single crystal. (a) High-resolution SEM image of this crystal and typical energy-dispersive X-ray (EDX) spectrum of the F4GT with their elemental mapping of (b) Fe, (c) Ge, and (d) Te respectively. (e) A typical
EDX spectrum of the Fe$_4$GeTe$_2$ crystal and the inset table shows atom and mass percentages of all elements.}
\label{EDX}
\end{figure*}

Figure~\ref{EDX}(a) shows a low-magnification FESEM image of the F4GT single crystal. The layered structures are clearly visible in the image. To examine the elemental composition and chemical homogeneity, energy dispersive X-ray (EDX) spectroscopy on the crystal has been carried out. The elemental distributions of Iron (Fe), Germanium (Ge), and Tellurium (Te) obtained from EDX, are shown in Figure~\ref{EDX}(b-d), which confirms that the constituent elements are evenly distributed. The obtained atomic weight percentage of the Fe, Ge, and Te are 59.4, 13.5, and 27.3, (see Figure~\ref{EDX}(e)) confirming the presence of Fe:Ge:Te in a ratio of 4.18:0.95:1.92 without other impurities and matches the earlier report \cite{SupBERA2023170257,SupmondalPhysRevB.104.094405}.

\section{Field dependent magnetization}

\begin{figure}[h!]
\centering
\includegraphics[width=0.5\textwidth]{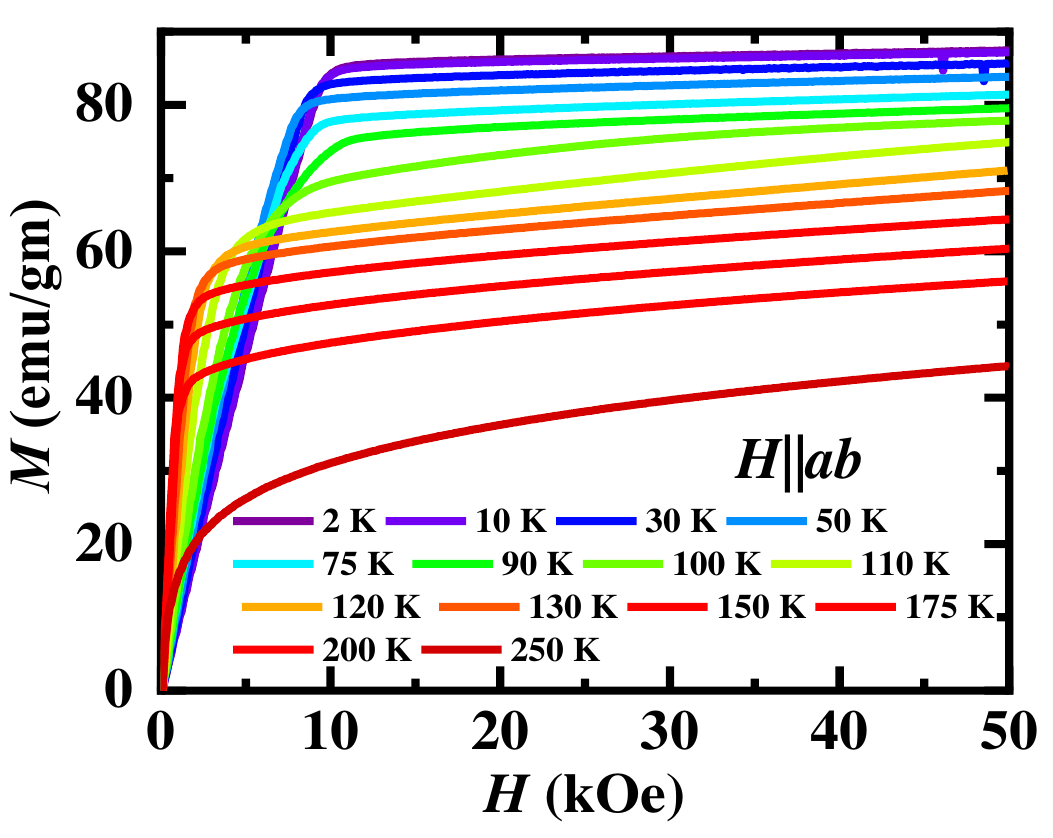}
\caption{The $M$ variation as a function of an applied magnetic field along $H$$\parallel$$ab$ at different Temperatures    .}
\label{MH_ab}
\end{figure}

As the Hall measurements were performed along the $c$ direction for the present compound, we have given the M-H data along the $c$ direction in the main text. However, the M-H data along the $ab$ direction upto 50 kOe applied magnetic field is incorporated in the Figure~\ref{MH_ab}. The saturation magnetic moment at 
$T$=2 K along the $ab$ plane is about 2.15$\mu$$_B$/Fe.

\section{Temperature-dependent anomalous Hall conductivity ($\sigma$$^A_{xy}$)}

The anomalous Hall effect (AHE) in a compound arises because of the intrinsic Berry phase mechanism then the anomalous Hall conductivity (AHC) must be temperature independent, and this has been highlighted in many of the previously reported metallic ferromagnets where the AHE is dominated by the intrinsic Berry phase mechanism\cite{Roy2020,SupPhysRevB.107.125138,SupChatterjee_2023}, although they have noticed a temperature dependence in the anomalous Hall resistivity.



\begin{figure}[h!]
\centering
\includegraphics[width=0.6\textwidth]{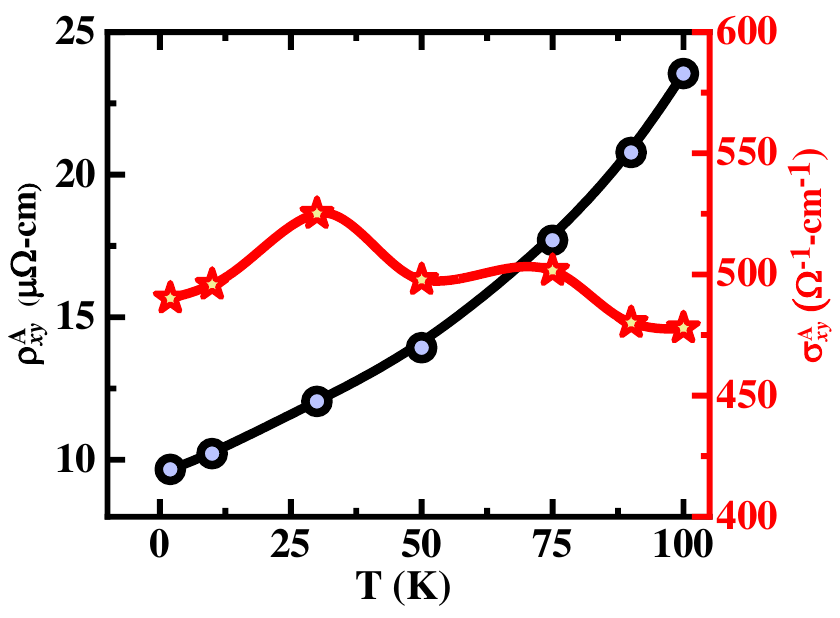}
\caption{Anomalous Hall resistivity $\rho$$^A_{xy}$ and anomalous Hall conductivity $\sigma$$^A_{xy}$ as a function of temperature.}
\label{AHC}
\end{figure}

Generally, the AHC ($\sigma$$^A_{xy}$) changes significantly close to any magnetic transition. For this system, the AHC changes above the spin reorientation transition i.e. T$_\textrm{SRT}$. This has also been observed in many other magnetic materials\cite{SupPhysRevB.94.075135,PhysRevB.102.085147}. The temperature-dependent $\sigma$$^A_{xy}$ is plotted in Fig.~\ref{AHC}, and the variation of $\sigma$$^A_{xy}$ up to spin reorientation transition is nearly temperature independent, clearly indicating that the origin of anomalous hall effect (AHE) in F4GT is intrinsic in nature\cite{SupWang2018,Liu2019}.


\section{Computational methodology}
Two approaches are utilized in conducting Density Functional Theory (DFT) calculations: the full-potential linearized augmented plane wave (FLAPW) method, which is incorporated in the FLEUR \cite{Supfleur} code, and the plane-wave projected augmented wave (PAW) method as implemented in the Vienna ab initio Simulation Package (VASP) \cite{Supvasp}. Our calculations for the density of states and band structure have been cross-checked with the two different basis sets. We have used the experimental crystal structure details throughout the calculation. For self-consistent calculations in VASP, we employed the Perdew–Burke–Ernzerhof (PBE)\cite{Supburke} functional within the generalized gradient approximation (GGA)\cite{Supperdew}. A $\Gamma$-centered k-mesh of $8\times8\times9$ and a plane-wave cutoff of 600 eV provide good convergence of total energy. Whereas the self-consistent calculations in FLEUR employed a plane-wave cutoff value of $k_{max}= 3.8$ a.u.$^{-1}$ for the linearized augmented plane wave (LAPW) basis functions.  To converge the charge densities, we used a Monkhorst-Pack \cite{Supmonk} $k$-mesh of $8 \times 8\times 8$ spanning the whole Brillouin zone (BZ). Our calculations include the effect of SOC self-consistently. The electronic band structures incorporating SOC are further parameterized using maximally-localized Wannier functions (MLWFs) in the FLEUR code. The tight-binding Hamiltonian is constructed using atomic orbital-like MLWFs of Fe-d and Te-p states which accurately reproduces the system's spectrum within a large energy window (approximately 6.5 eV) around the Fermi energy. The Berry curvature is calculated to a high precision using Wannier90 code\cite{Suppizzi2020wannier90,SupmarzariPhysRevB.56.12847}.
Finally, with the tight-binding model Hamiltonian, the linear response Kubo formula\cite{Supkubo} in the clean limit was used to calculate the intrinsic AHC as follows,
\begin{eqnarray}
\Omega_{n}(\mathbf{k}) = -\hbar^{2} \sum_{n \neq m} \frac{\operatorname{2 Im} \langle u_{n\mathbf{k}}|\hat{v}_{i}|u_{m\mathbf{k}}\rangle \langle u_{n\mathbf{k}}|\hat{v}_{j}|u_{m\mathbf{k}}\rangle}{(\epsilon_{n\mathbf{k}}-\epsilon_{m\mathbf{k}})^{2}}
\end{eqnarray}
where $\Omega_{n}(\mathbf{k})$ is the Berry curvature of band $n$, $\hat{v}_{i} =\frac{1}{\hbar}{\partial \hat{H}(\mathbf{k})}/{\partial k_{i}} $ is the velocity operator with $i\in \{x,y\}$, $u_{n\mathbf{k}}$ and $\epsilon_{n\mathbf{k}}$ are the eigenstates and eigenvalues of the Hamiltonian $\hat{H}(\mathbf{k})$, respectively.
We then calculate the anomalous Hall conductivity (AHC) by:
\begin{equation}
\begin{aligned}
\sigma^{A}_{H}=& -\hbar e^{2} \sum_{n} \int_{BZ} \frac{d{\bf k}}{\left(2 \pi\right)^3}f_{n}(\mathbf{k})\Omega_{n}(\mathbf{k})
\end{aligned}
\end{equation}
$\sigma_H^A$ is calculated in WANNIERTOOL code\cite{Wu_2018} with a highly dense $k$ mesh of $450 \times 450 \times 450$. This dense mesh is found to yield well-converged values for the AHC. 
\section{Electronic structure calculation results}
\begin{figure*}[h!]
\centering
\includegraphics[width=1\textwidth]{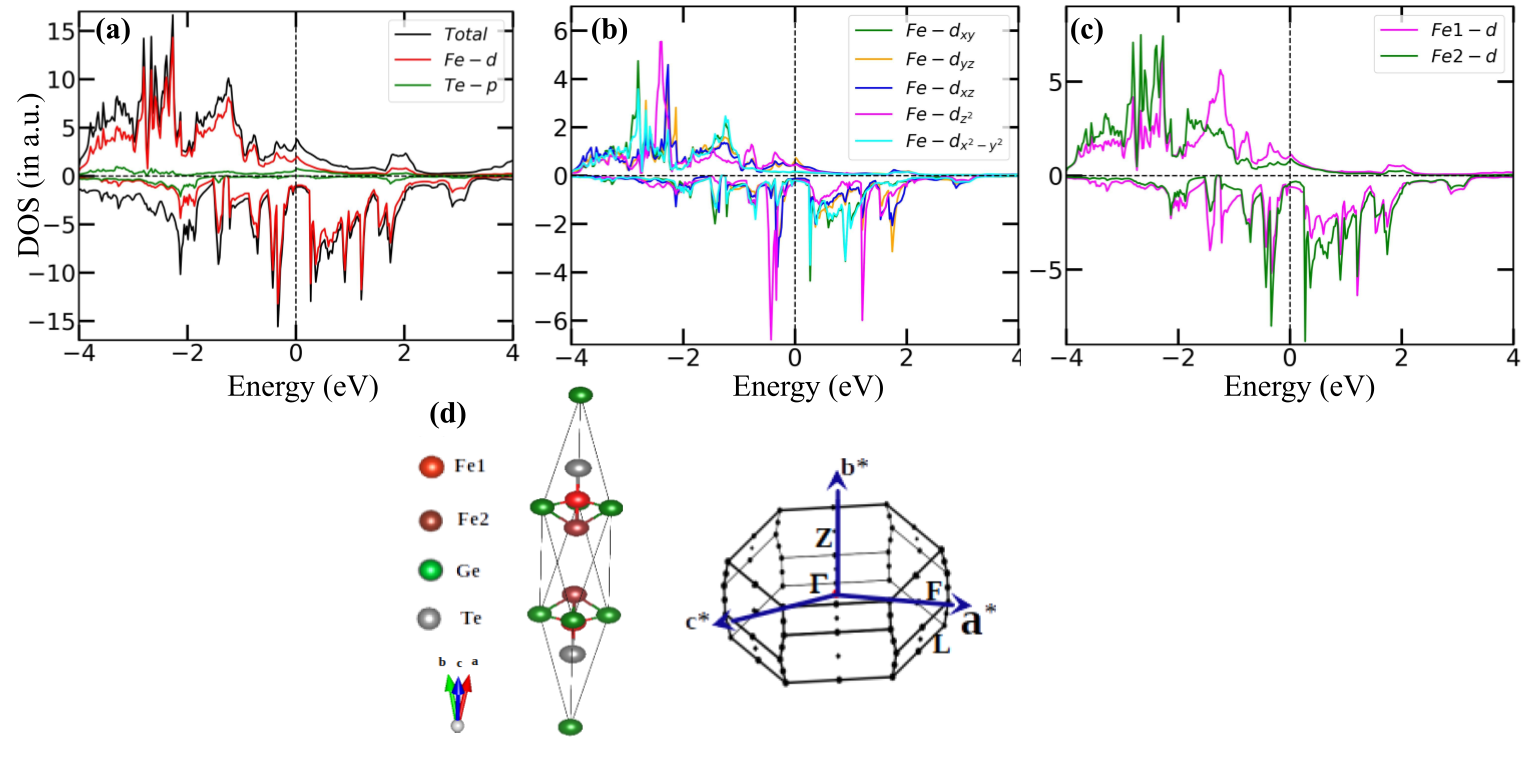}
\caption{\textbf{Theoretical calculation - DOS.}~{\bf(a)} Atom projected DOS. The up and down panel corresponds to the spin-up and spin-down DOS respectively. {\bf(b)} Different dominant Fe-d orbital characters near E$_F$.  {\bf(c)} DOS from two different types of Fe atoms. (d) Primitive unit cell and corresponding bulk Brillouin Zone (BZ) with all four high symmetry points: $\Gamma$, Z, F, and L of F4GT. }
\label{S2}
\end{figure*}

\subsection{Density of states (DOS)}

Using DFT calculations with spin polarization, we examined Fe$_4$GeTe$_2$'s electronic structure. The spin polarized DOS for the primitive structure of Fe$_4$GeTe$_2$(Figure~\ref{S2}(d), same in main text Figure.~(1)(a)) shown in Figure~\ref{S2}(a). The figure indicates the presence of a pseudogap or quasi-gap feature characterized by low DOS in the vicinity of the Fermi level, which suggests the system may exhibit a semimetallic behavior. The atom projected DOS clearly shows predominantly Fe-d character close to the Fermi level ($E_\textrm{F}$) originating from both Fe1 and Fe2 sites. There is also contain a significant contribution from Te-$p$ orbitals. However, there is no substantial contribution from Ge atom. 
Furthermore, the decomposed density of states for all Fe orbitals displayed in Figure~\ref{S2}(b) indicates a mixture of all Fe-3d orbitals around $E_\textrm{F}$. Within approximately 200 meV around the $E_\textrm{F}$, $d_{yz}$, $d_{z^2}$, and $d_{xz}$ orbitals are the most dominant. Figure.~\ref{S2}(d) (same in main text Figure.~(1)(a)) shows, there are two types of Fe atoms, each with a different coordination environment. As a result, both Fe atom types show significant differences in DOS as shown in Fig~\ref{S2}(c). It shows the d-bands of Fe2 are more extensively spread out in contrast to Fe1. 

\subsection{Orbital projected band structures and Wannier band structure}

\begin{figure*}[h!]
\centering
\includegraphics[width=1\textwidth]{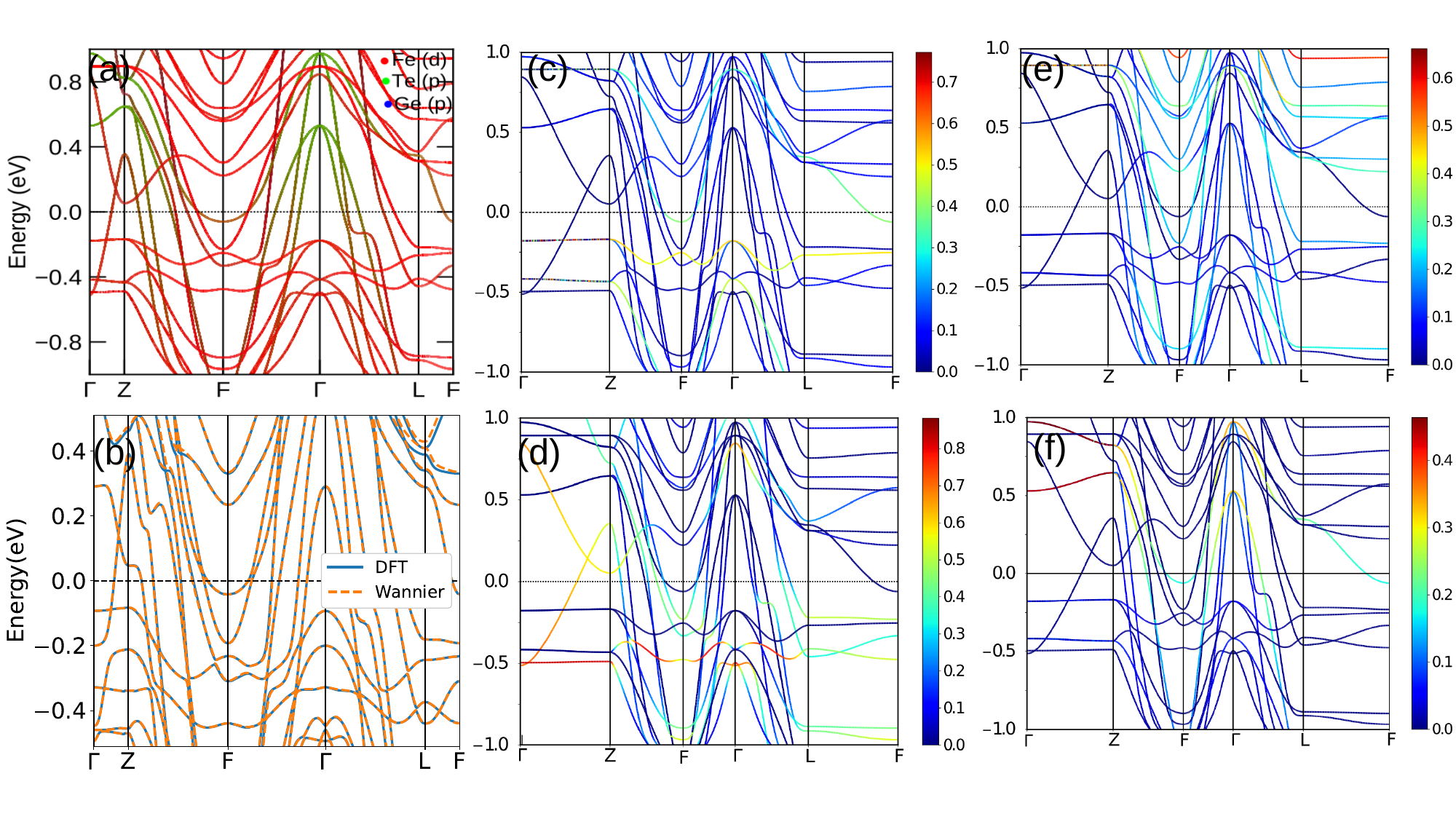}
\caption{\textbf{Projected Band without considering SOC:} {\bf (a)} Atom projected band structure {\bf (b)} DFT band and Wannier band comparison {\bf (c), (d) ,(e) } Fe-$d_{yz}$, Fe-$d_{xz}$ and Fe-$d_{z^2}$ orbital projected band structure. {\bf (f)} Te$-p_y$ orbital projected band structure.}
\label{S3}
\end{figure*}

Our detailed orbital projected electronic band structure calculations in Figure.~\ref{S3}(a) show that, the bands near to Fermi energy are the admixture of Fe-d and Te-p orbital. To deepen our comprehension of the electronic structure of the system, we have also generated projected band structure plot that features the l-m decomposed orbitals of the Fe-d and Te-p orbitals. Figure.~\ref{S3}(c), (d), (e), and (f) depict the respective contributions of Fe$-d_{yz}$, Fe$-d_{xz}$, Fe$-d_{z^2}$ and Te$-p_{y}$ orbitals to the  band structure. The other orbitals are ignored because their contribution near fermi label is very small. The two bands that cross each other near the Fermi energy and are responsible for the topological feature exhibit a significant hybridization of various Fe-3d and Te-p orbitals. This hybridization gives rise to a mixed orbital character, which is the primary characteristic of these crossing bands.

To construct the tight-binding Hamiltonian we considers only the Fe-d and Te-p orbitals because they are the primary contributors to the density of states near the Fermi level. In Figure.~\ref{S3}(b), we have presented a comparison of the Wannier band and the DFT band, obtained using carefully chosen parameters. Notably, we observe an exact overlap between these two bands in the vicinity of the Fermi level, indicating a high degree of agreement between the two models. This finding provides strong support for the validity of our theoretical approach.
\subsection{Nodal line in the FZ$\Gamma$ plan}
\begin{figure*}
\centering
\includegraphics[width=\textwidth]{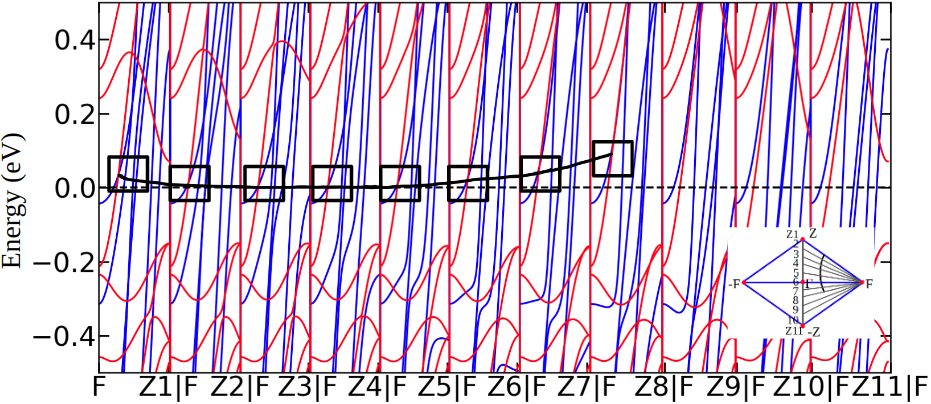}
\caption{\textbf{Band structure without considering SOC:} Band structures without considering SOC for various k-path segments (the inset)}
\label{S4}
\end{figure*}
To verify the existence of nodal lines of band touching in the FZ$\Gamma$ plane, we have performed a band structure calculation without considering SOC along several line segments connecting $F$ point and few points on the $\bar{Z}\Gamma Z$ line. We have observed unavoidable band crossings around Fermi energy in all directions from $F-Z1$ to $F-Z7$, as indicated by black boxes in Figure.~\ref{S4}. The collective band crossing points have formed a nodal line type of band touching in the material, illustrated by a black line. As discussed in the main text, the nodal line topological band crossing is gapped out in the presence of SOC.

\subsection{Band structure with SOC and Berry curvature}
\begin{figure}[h]
\centering
\includegraphics[width=1\linewidth]{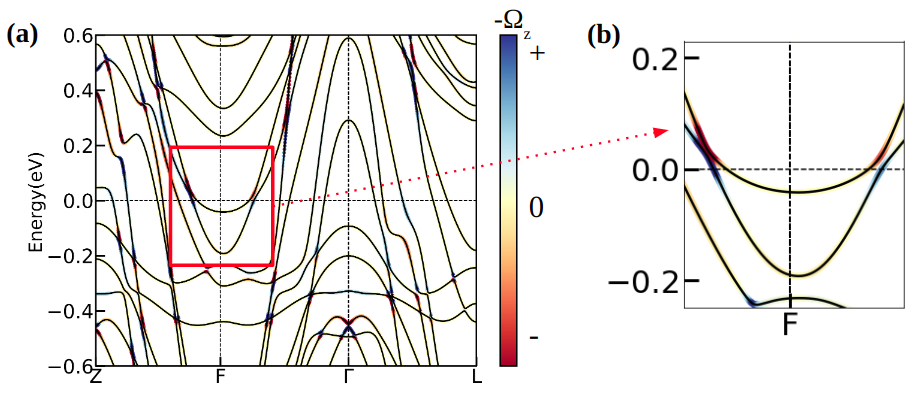}
\caption{\textbf{Band structure with Berry curvature.} (a) Band structure with SOC incorporating Berry curvature. (b) Zoomed-in view near soc-induced gapped bands.  }
\label{S5}
\end{figure}
Figure~\ref{S5}(a) illustrates the band structure of F4GT along the high-symmetry $k$-path direction, accompanied by the plotted projected Berry curvature component ($\Omega_z$). The plot indicates that the maximum value of the Berry curvature occurs around the Fermi energy in both the $ZF$ and $F\Gamma$ directions, precisely at the position where the band gap opens. Here, the magnetization is oriented along out-of-plane direction, $i.e.$ along $z$-axis.

\subsection{Exploring AHC as a function of magnetization direction}
\begin{figure}[h]
\centering
\includegraphics[width=1\linewidth]{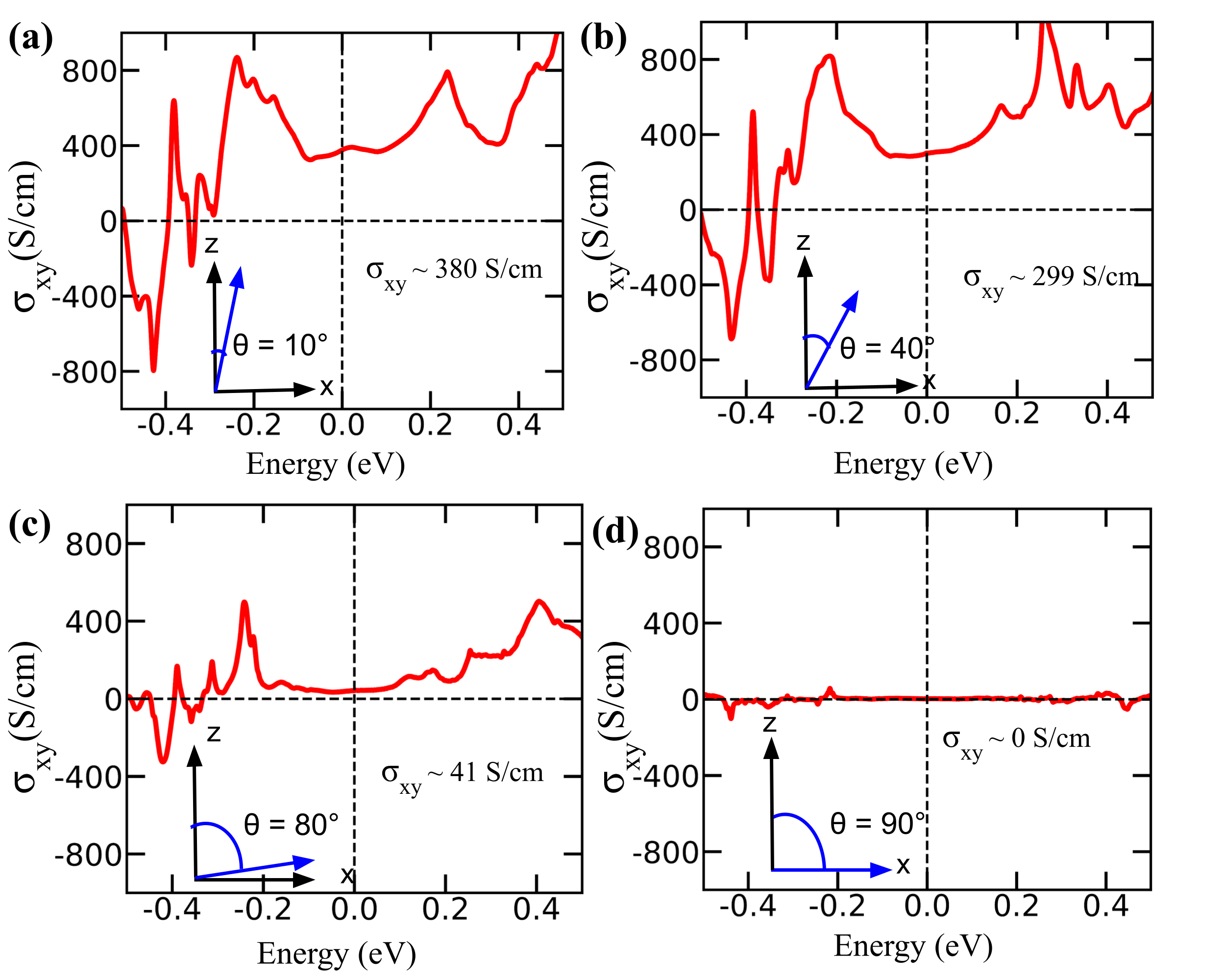}
\caption{\textbf{Magnetic orientation dependent AHC:~} $\sigma_{xy}$ as a function of energy around the Fermi energy for different magnetization orientations. The rotation of the magnetization from the $z$-axis towards the $x$-axis are $10^\circ$, $40^\circ$, $80^\circ$, and $90^\circ$ and (a), (b), (c), and (d) depict the corresponding $\sigma_{xy}$'s.}
\label{S6}
\end{figure}
\begin{figure}[h]
\centering
\includegraphics[width=1\linewidth]{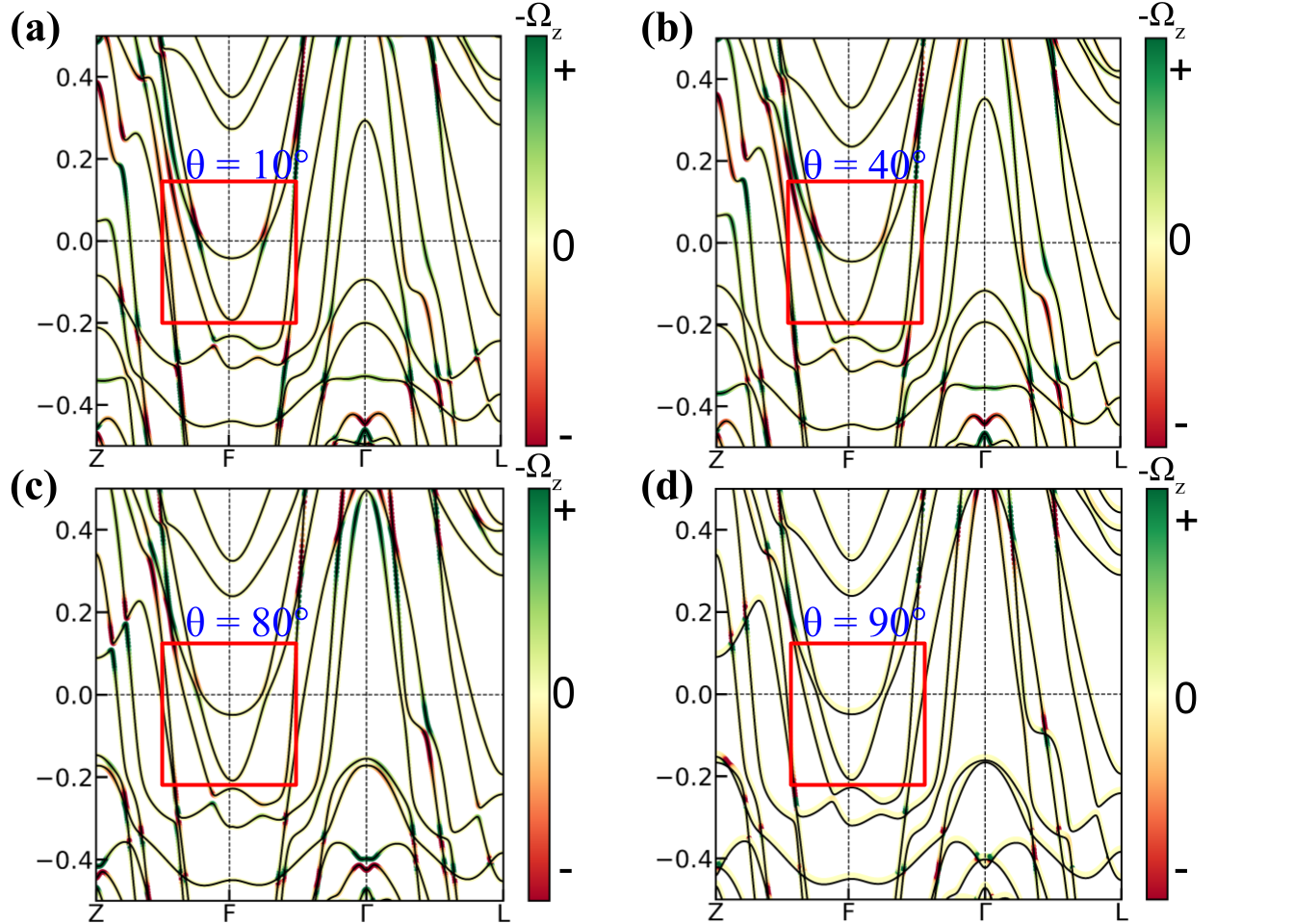}
\caption{\textbf{Band structure with Berry curvature for different orientations of magnetization:~}  Band Structure with SOC and incorporating Berry curvature for different magnetization orientations. Magnetization along $z$-axis is presented in Fig.~S9 and rotating the magnetization from the $z$-axis to the $x$-axis direction by $10^\circ$, $40^\circ$, $80^\circ$ and $90^\circ$ in (a), (b), (c) and (d), respectively.}
\label{S7}
\end{figure}

While the exact magnetic structure of F4GT near the spin reorientation transition (SRT) remains unclear, the AHE data provides clear evidence of a significant temperature effect at approximately 100~K ($\approx$ T$_\textrm{SRT}$). Our experimental findings reveal that there is no substantial change in the intrinsic AHC value up to the T$_\textrm{SRT}$, as illustrated in Fig.~\ref{AHC}. Identifying the magnetic structure at elevated temperatures presents challenges, as it necessitates the use of neutron diffraction techniques under higher temperature conditions.
However, we have performed theoretical investigations to examine how varying magnetization orientations can affect the material's anomalous transport properties. Below the T$_\textrm{SRT}$, the material exhibits an out-of-plane magnetocrystalline anisotropy where the Fe magnetic moments are oriented along the $z$-axis. Under the condition of preserving the crystal structure unchanged with temperature, we assume that the magnetic moment of Fe in F4GT undergoes canting from their low-temperature orientation, resulting in an inclination of magnetization with respect to the out-of-plane axis. Indeed, the AHC value, especially $\sigma_\textrm{xy}$, which is determined by the Berry curvatures derived from band structure calculations, is significantly impacted by the magnetization direction. We are now carrying out calculations for different magnetic orientation angles, measured relative to the $z$-axis, while maintaining the crystal structure unchanged. Fig.~\ref{S6} clearly illustrates that as the magnetization orientation rotates from the $z$-axis towards the in-plane $x$-axis direction, the AHC decreases, eventually reaching zero for magnetization along the $x$-axis. The AHC value exhibits minimal change around the Fermi energy, within a 150 meV energy range. In the figure, we have included the $\sigma_{xy}$ values at the Fermi energy for reference. So, in case of in-plane magnetization AHC vanishes. In Fig.~\ref{S7}, we have plotted the band structure along the high-symmetry $k$-path direction, showcasing the projected Berry curvature component $\Omega_z$ for different magnetization orientation directions. The plots clearly illustrate that the maximum value of the Berry curvature occurs at the position of the band gap opening, near the Fermi energy. When the magnetization direction is rotated by a mere $10^0$ relative to the $z$-axis, we observe a decrease in the Berry curvature around the nodal gap opening near the Fermi energy, as depicted in Fig.~\ref{S7}(a). Furthermore, as the rotation angle increases, the Berry curvature diminishes gradually, approaching almost zero for the in-plane magnetization, as depicted in Figs.~\ref{S7}(b)-(d). Indeed, these findings indicate that the orientation of the magnetization plays a crucial role in shaping the distribution of the Berry curvature and, consequently, has a significant impact on the value of $\sigma_\textrm{xy}$.
Theoretical exploration indicates a temperature-dependent gradual change in F4GT's magnetization orientation, necessitating more precise experiments in the future. 
\newpage
%


\begin{thebibliography}{67}%
\makeatletter

\providecommand \@ifxundefined [1]{%
 \@ifx{#1\undefined}
}%
\providecommand \@ifnum [1]{%
 \ifnum #1\expandafter \@firstoftwo
 \else \expandafter \@secondoftwo
 \fi
}%
\providecommand \@ifx [1]{%
 \ifx #1\expandafter \@firstoftwo
 \else \expandafter \@secondoftwo
 \fi
}%
\providecommand \natexlab [1]{#1}%
\providecommand \enquote  [1]{``#1''}%
\providecommand \bibnamefont  [1]{#1}%
\providecommand \bibfnamefont [1]{#1}%
\providecommand \citenamefont [1]{#1}%
\providecommand \href@noop [0]{\@secondoftwo}%
\providecommand \href [0]{\begingroup \@sanitize@url \@href}%
\providecommand \@href[1]{\@@startlink{#1}\@@href}%
\providecommand \@@href[1]{\endgroup#1\@@endlink}%
\providecommand \@sanitize@url [0]{\catcode `\\12\catcode `\$12\catcode
  `\&12\catcode `\#12\catcode `\^12\catcode `\_12\catcode `\%12\relax}%
\providecommand \@@startlink[1]{}%
\providecommand \@@endlink[0]{}%
\providecommand \url  [0]{\begingroup\@sanitize@url \@url }%
\providecommand \@url [1]{\endgroup\@href {#1}{\urlprefix }}%
\providecommand \urlprefix  [0]{URL }%
\providecommand \Eprint [0]{\href }%
\providecommand \doibase [0]{https://doi.org/}%
\providecommand \selectlanguage [0]{\@gobble}%
\providecommand \bibinfo  [0]{\@secondoftwo}%
\providecommand \bibfield  [0]{\@secondoftwo}%
\providecommand \translation [1]{[#1]}%
\providecommand \BibitemOpen [0]{}%
\providecommand \bibitemStop [0]{}%
\providecommand \bibitemNoStop [0]{.\EOS\space}%
\providecommand \EOS [0]{\spacefactor3000\relax}%
\providecommand \BibitemShut  [1]{\csname bibitem#1\endcsname}%
\let\auto@bib@innerbib\@empty
\bibitem [{\citenamefont {Wan}\ \emph {et~al.}(2011)\citenamefont {Wan},
  \citenamefont {Turner}, \citenamefont {Vishwanath},\ and\ \citenamefont
  {Savrasov}}]{wanPhysRevB.83.205101}%
  \BibitemOpen
  \bibfield  {author} {\bibinfo {author} {\bibfnamefont {X.}~\bibnamefont
  {Wan}}, \bibinfo {author} {\bibfnamefont {A.~M.}\ \bibnamefont {Turner}},
  \bibinfo {author} {\bibfnamefont {A.}~\bibnamefont {Vishwanath}},\ and\
  \bibinfo {author} {\bibfnamefont {S.~Y.}\ \bibnamefont {Savrasov}},\
  }
  \href
  {https://doi.org/10.1103/PhysRevB.83.205101} {\bibfield  {journal} {\bibinfo
  {journal} {Phys. Rev. B}\ }\textbf {\bibinfo {volume} {83}},\ \bibinfo
  {pages} {205101} (\bibinfo {year} {2011})}\BibitemShut {NoStop}%
\bibitem [{\citenamefont {Xu}\ \emph {et~al.}(2011)\citenamefont {Xu},
  \citenamefont {Weng}, \citenamefont {Wang}, \citenamefont {Dai},\ and\
  \citenamefont {Fang}}]{XuPhysRevLett.107.186806}%
  \BibitemOpen
  \bibfield  {author} {\bibinfo {author} {\bibfnamefont {G.}~\bibnamefont
  {Xu}}, \bibinfo {author} {\bibfnamefont {H.}~\bibnamefont {Weng}}, \bibinfo
  {author} {\bibfnamefont {Z.}~\bibnamefont {Wang}}, \bibinfo {author}
  {\bibfnamefont {X.}~\bibnamefont {Dai}},\ and\ \bibinfo {author}
  {\bibfnamefont {Z.}~\bibnamefont {Fang}},\ }
  \href
{https://doi.org/10.1103/PhysRevLett.107.186806} {\bibfield  {journal}
  {\bibinfo  {journal} {Phys. Rev. Lett.}\ }\textbf {\bibinfo {volume} {107}},\
  \bibinfo {pages} {186806} (\bibinfo {year} {2011})}\BibitemShut {NoStop}%
\bibitem [{\citenamefont {Chiu}\ \emph {et~al.}(2016)\citenamefont {Chiu},
  \citenamefont {Teo}, \citenamefont {Schnyder},\ and\ \citenamefont
  {Ryu}}]{chiuRevModPhys.88.035005}%
  \BibitemOpen
  \bibfield  {author} {\bibinfo {author} {\bibfnamefont {C.-K.}\ \bibnamefont
  {Chiu}}, \bibinfo {author} {\bibfnamefont {J.~C.~Y.}\ \bibnamefont {Teo}},
  \bibinfo {author} {\bibfnamefont {A.~P.}\ \bibnamefont {Schnyder}},\ and\
  \bibinfo {author} {\bibfnamefont {S.}~\bibnamefont {Ryu}},\ }
  \href {https://doi.org/10.1103/RevModPhys.88.035005}
  {\bibfield  {journal} {\bibinfo  {journal} {Rev. Mod. Phys.}\ }\textbf
  {\bibinfo {volume} {88}},\ \bibinfo {pages} {035005} (\bibinfo {year}
  {2016})}\BibitemShut {NoStop}%
\bibitem [{\citenamefont {Armitage}\ \emph {et~al.}(2018)\citenamefont
  {Armitage}, \citenamefont {Mele},\ and\ \citenamefont
  {Vishwanath}}]{ArmitageRevModPhys.90.015001}%
  \BibitemOpen
  \bibfield  {author} {\bibinfo {author} {\bibfnamefont {N.~P.}\ \bibnamefont
  {Armitage}}, \bibinfo {author} {\bibfnamefont {E.~J.}\ \bibnamefont {Mele}},\
  and\ \bibinfo {author} {\bibfnamefont {A.}~\bibnamefont {Vishwanath}},\
  } \href{https://doi.org/10.1103/RevModPhys.90.015001} {\bibfield  {journal}
  {\bibinfo  {journal} {Rev. Mod. Phys.}\ }\textbf {\bibinfo {volume} {90}},\
  \bibinfo {pages} {015001} (\bibinfo {year} {2018})}\BibitemShut {NoStop}%
\bibitem [{\citenamefont {Lv}\ \emph {et~al.}(2021)\citenamefont {Lv},
  \citenamefont {Qian},\ and\ \citenamefont {Ding}}]{LvRevModPhys.93.025002}%
  \BibitemOpen
  \bibfield  {author} {\bibinfo {author} {\bibfnamefont {B.~Q.}\ \bibnamefont
  {Lv}}, \bibinfo {author} {\bibfnamefont {T.}~\bibnamefont {Qian}},\ and\
  \bibinfo {author} {\bibfnamefont {H.}~\bibnamefont {Ding}},\ }
  \href  {https://doi.org/10.1103/RevModPhys.93.025002} {\bibfield  {journal}
  {\bibinfo  {journal} {Rev. Mod. Phys.}\ }\textbf {\bibinfo {volume} {93}},\
  \bibinfo {pages} {025002} (\bibinfo {year} {2021})}\BibitemShut {NoStop}%
\bibitem [{\citenamefont {Wang}\ \emph {et~al.}(2012)\citenamefont {Wang},
  \citenamefont {Sun}, \citenamefont {Chen}, \citenamefont {Franchini},
  \citenamefont {Xu}, \citenamefont {Weng}, \citenamefont {Dai},\ and\
  \citenamefont {Fang}}]{wangPhysRevB.85.195320}%
  \BibitemOpen
  \bibfield  {author} {\bibinfo {author} {\bibfnamefont {Z.}~\bibnamefont
  {Wang}}, \bibinfo {author} {\bibfnamefont {Y.}~\bibnamefont {Sun}}, \bibinfo
  {author} {\bibfnamefont {X.-Q.}\ \bibnamefont {Chen}}, \bibinfo {author}
  {\bibfnamefont {C.}~\bibnamefont {Franchini}}, \bibinfo {author}
  {\bibfnamefont {G.}~\bibnamefont {Xu}}, \bibinfo {author} {\bibfnamefont
  {H.}~\bibnamefont {Weng}}, \bibinfo {author} {\bibfnamefont {X.}~\bibnamefont
  {Dai}},\ and\ \bibinfo {author} {\bibfnamefont {Z.}~\bibnamefont {Fang}},\
  }  
  \href {https://doi.org/10.1103/PhysRevB.85.195320} {\bibfield  {journal} {\bibinfo
  {journal} {Phys. Rev. B}\ }\textbf {\bibinfo {volume} {85}},\ \bibinfo
  {pages} {195320} (\bibinfo {year} {2012})}\BibitemShut {NoStop}%
\bibitem [{\citenamefont {He}\ \emph {et~al.}(2014)\citenamefont {He},
  \citenamefont {Hong}, \citenamefont {Dong}, \citenamefont {Pan},
  \citenamefont {Zhang}, \citenamefont {Zhang},\ and\ \citenamefont
  {Li}}]{hePhysRevLett.113.246402}%
  \BibitemOpen
  \bibfield  {author} {\bibinfo {author} {\bibfnamefont {L.~P.}\ \bibnamefont
  {He}}, \bibinfo {author} {\bibfnamefont {X.~C.}\ \bibnamefont {Hong}},
  \bibinfo {author} {\bibfnamefont {J.~K.}\ \bibnamefont {Dong}}, \bibinfo
  {author} {\bibfnamefont {J.}~\bibnamefont {Pan}}, \bibinfo {author}
  {\bibfnamefont {Z.}~\bibnamefont {Zhang}}, \bibinfo {author} {\bibfnamefont
  {J.}~\bibnamefont {Zhang}},\ and\ \bibinfo {author} {\bibfnamefont {S.~Y.}\
  \bibnamefont {Li}},\ }
  \href {https://doi.org/10.1103/PhysRevLett.113.246402} {\bibfield  {journal}
  {\bibinfo  {journal} {Phys. Rev. Lett.}\ }\textbf {\bibinfo {volume} {113}},\
  \bibinfo {pages} {246402} (\bibinfo {year} {2014})}\BibitemShut {NoStop}%
\bibitem [{\citenamefont {Neupane}\ \emph {et~al.}(2014)\citenamefont
  {Neupane}, \citenamefont {Xu}, \citenamefont {Sankar}, \citenamefont
  {Alidoust}, \citenamefont {Bian}, \citenamefont {Liu}, \citenamefont
  {Belopolski}, \citenamefont {Chang}, \citenamefont {Jeng}, \citenamefont
  {Lin}, \citenamefont {Bansil}, \citenamefont {Chou},\ and\ \citenamefont
  {Hasan}}]{neupane2014observation}%
  \BibitemOpen
  \bibfield  {author} {\bibinfo {author} {\bibfnamefont {M.}~\bibnamefont
  {Neupane}}, \bibinfo {author} {\bibfnamefont {S.-Y.}\ \bibnamefont {Xu}},
  \bibinfo {author} {\bibfnamefont {R.}~\bibnamefont {Sankar}}, \bibinfo
  {author} {\bibfnamefont {N.}~\bibnamefont {Alidoust}}, \bibinfo {author}
  {\bibfnamefont {G.}~\bibnamefont {Bian}}, \bibinfo {author} {\bibfnamefont
  {C.}~\bibnamefont {Liu}}, \bibinfo {author} {\bibfnamefont {I.}~\bibnamefont
  {Belopolski}}, \bibinfo {author} {\bibfnamefont {T.-R.}\ \bibnamefont
  {Chang}}, \bibinfo {author} {\bibfnamefont {H.-T.}\ \bibnamefont {Jeng}},
  \bibinfo {author} {\bibfnamefont {H.}~\bibnamefont {Lin}}, \bibinfo {author}
  {\bibfnamefont {A.}~\bibnamefont {Bansil}}, \bibinfo {author} {\bibfnamefont
  {F.}~\bibnamefont {Chou}},\ and\ \bibinfo {author} {\bibfnamefont {M.~Z.}\
  \bibnamefont {Hasan}},\ }\href {https://doi.org/10.1038/ncomms4786} {\bibfield  {journal}
  {\bibinfo  {journal} {Nature Communications}\ }\textbf {\bibinfo {volume}
  {5}},\ \bibinfo {pages} {3786} (\bibinfo {year} {2014})}\BibitemShut
  {NoStop}%
\bibitem [{\citenamefont {Liu}\ \emph {et~al.}(2014)\citenamefont {Liu},
  \citenamefont {Zhou}, \citenamefont {Zhang}, \citenamefont {Wang},
  \citenamefont {Weng}, \citenamefont {Prabhakaran}, \citenamefont {Mo},
  \citenamefont {Shen}, \citenamefont {Fang}, \citenamefont {Dai},
  \citenamefont {Hussain},\ and\ \citenamefont {Chen}}]{liu2014discovery}%
  \BibitemOpen
  \bibfield  {author} {\bibinfo {author} {\bibfnamefont {Z.~K.}\ \bibnamefont
  {Liu}}, \bibinfo {author} {\bibfnamefont {B.}~\bibnamefont {Zhou}}, \bibinfo
  {author} {\bibfnamefont {Y.}~\bibnamefont {Zhang}}, \bibinfo {author}
  {\bibfnamefont {Z.~J.}\ \bibnamefont {Wang}}, \bibinfo {author}
  {\bibfnamefont {H.~M.}\ \bibnamefont {Weng}}, \bibinfo {author}
  {\bibfnamefont {D.}~\bibnamefont {Prabhakaran}}, \bibinfo {author}
  {\bibfnamefont {S.-K.}\ \bibnamefont {Mo}}, \bibinfo {author} {\bibfnamefont
  {Z.~X.}\ \bibnamefont {Shen}}, \bibinfo {author} {\bibfnamefont
  {Z.}~\bibnamefont {Fang}}, \bibinfo {author} {\bibfnamefont {X.}~\bibnamefont
  {Dai}}, \bibinfo {author} {\bibfnamefont {Z.}~\bibnamefont {Hussain}},\ and\
  \bibinfo {author} {\bibfnamefont {Y.~L.}\ \bibnamefont {Chen}},\ }\href
  {https://doi.org/10.1126/science.1245085} {\bibfield  {journal} {\bibinfo
  {journal} {Science}\ }\textbf {\bibinfo {volume} {343}},\ \bibinfo {pages}
  {864} (\bibinfo {year} {2014})}\BibitemShut {NoStop}%
\bibitem [{\citenamefont {Hosen}\ \emph {et~al.}(2018)\citenamefont {Hosen},
  \citenamefont {Dimitri}, \citenamefont {Nandy}, \citenamefont {Aperis},
  \citenamefont {Sankar}, \citenamefont {Dhakal}, \citenamefont {Maldonado},
  \citenamefont {Kabir}, \citenamefont {Sims}, \citenamefont {Chou},
  \citenamefont {Kaczorowski}, \citenamefont {Durakiewicz}, \citenamefont
  {Oppeneer},\ and\ \citenamefont {Neupane}}]{Hosen_2018_dirac}%
  \BibitemOpen
  \bibfield  {author} {\bibinfo {author} {\bibfnamefont {M.~M.}\ \bibnamefont
  {Hosen}}, \bibinfo {author} {\bibfnamefont {K.}~\bibnamefont {Dimitri}},
  \bibinfo {author} {\bibfnamefont {A.~K.}\ \bibnamefont {Nandy}}, \bibinfo
  {author} {\bibfnamefont {A.}~\bibnamefont {Aperis}}, \bibinfo {author}
  {\bibfnamefont {R.}~\bibnamefont {Sankar}}, \bibinfo {author} {\bibfnamefont
  {G.}~\bibnamefont {Dhakal}}, \bibinfo {author} {\bibfnamefont
  {P.}~\bibnamefont {Maldonado}}, \bibinfo {author} {\bibfnamefont
  {F.}~\bibnamefont {Kabir}}, \bibinfo {author} {\bibfnamefont
  {C.}~\bibnamefont {Sims}}, \bibinfo {author} {\bibfnamefont {F.}~\bibnamefont
  {Chou}}, \bibinfo {author} {\bibfnamefont {D.}~\bibnamefont {Kaczorowski}},
  \bibinfo {author} {\bibfnamefont {T.}~\bibnamefont {Durakiewicz}}, \bibinfo
  {author} {\bibfnamefont {P.~M.}\ \bibnamefont {Oppeneer}},\ and\ \bibinfo
  {author} {\bibfnamefont {M.}~\bibnamefont {Neupane}},\ }\href {https://doi.org/10.1038/s41467-018-05233-1} {\bibfield
  {journal} {\bibinfo  {journal} {Nature Communications}\ }\textbf {\bibinfo
  {volume} {9}},\ \bibinfo {pages} {3002} (\bibinfo {year} {2018})}\BibitemShut
  {NoStop}%
\bibitem [{\citenamefont {Dhakal}\ \emph {et~al.}(2022)\citenamefont {Dhakal},
  \citenamefont {Kabir}, \citenamefont {Nandy}, \citenamefont {Aperis},
  \citenamefont {Sakhya}, \citenamefont {Pradhan}, \citenamefont {Dimitri},
  \citenamefont {Sims}, \citenamefont {Regmi}, \citenamefont {Hosen},
  \citenamefont {Liu}, \citenamefont {Persaud}, \citenamefont {Kaczorowski},
  \citenamefont {Oppeneer},\ and\ \citenamefont
  {Neupane}}]{PhysRevB.106.125124}%
  \BibitemOpen
  \bibfield  {author} {\bibinfo {author} {\bibfnamefont {G.}~\bibnamefont
  {Dhakal}}, \bibinfo {author} {\bibfnamefont {F.}~\bibnamefont {Kabir}},
  \bibinfo {author} {\bibfnamefont {A.~K.}\ \bibnamefont {Nandy}}, \bibinfo
  {author} {\bibfnamefont {A.}~\bibnamefont {Aperis}}, \bibinfo {author}
  {\bibfnamefont {A.~P.}\ \bibnamefont {Sakhya}}, \bibinfo {author}
  {\bibfnamefont {S.}~\bibnamefont {Pradhan}}, \bibinfo {author} {\bibfnamefont
  {K.}~\bibnamefont {Dimitri}}, \bibinfo {author} {\bibfnamefont
  {C.}~\bibnamefont {Sims}}, \bibinfo {author} {\bibfnamefont {S.}~\bibnamefont
  {Regmi}}, \bibinfo {author} {\bibfnamefont {M.~M.}\ \bibnamefont {Hosen}},
  \bibinfo {author} {\bibfnamefont {Y.}~\bibnamefont {Liu}}, \bibinfo {author}
  {\bibfnamefont {L.}~\bibnamefont {Persaud}}, \bibinfo {author} {\bibfnamefont
  {D.}~\bibnamefont {Kaczorowski}}, \bibinfo {author} {\bibfnamefont {P.~M.}\
  \bibnamefont {Oppeneer}},\ and\ \bibinfo {author} {\bibfnamefont
  {M.}~\bibnamefont {Neupane}},\ }\href {https://doi.org/10.1103/PhysRevB.106.125124} {\bibfield  {journal} {\bibinfo
   {journal} {Phys. Rev. B}\ }\textbf {\bibinfo {volume} {106}},\ \bibinfo
  {pages} {125124} (\bibinfo {year} {2022})}\BibitemShut {NoStop}%
\bibitem [{\citenamefont {Weng}\ \emph {et~al.}(2015)\citenamefont {Weng},
  \citenamefont {Fang}, \citenamefont {Fang}, \citenamefont {Bernevig},\ and\
  \citenamefont {Dai}}]{weng2015weyl}%
  \BibitemOpen
  \bibfield  {author} {\bibinfo {author} {\bibfnamefont {H.}~\bibnamefont
  {Weng}}, \bibinfo {author} {\bibfnamefont {C.}~\bibnamefont {Fang}}, \bibinfo
  {author} {\bibfnamefont {Z.}~\bibnamefont {Fang}}, \bibinfo {author}
  {\bibfnamefont {B.~A.}\ \bibnamefont {Bernevig}},\ and\ \bibinfo {author}
  {\bibfnamefont {X.}~\bibnamefont {Dai}},\ }\href {https://doi.org/10.1103/PhysRevX.5.011029}
  {\bibfield  {journal} {\bibinfo  {journal} {Phys. Rev. X}\ }\textbf {\bibinfo
  {volume} {5}},\ \bibinfo {pages} {011029} (\bibinfo {year}
  {2015})}\BibitemShut {NoStop}%
\bibitem [{\citenamefont {Lv}\ \emph {et~al.}(2015)\citenamefont {Lv},
  \citenamefont {Weng}, \citenamefont {Fu}, \citenamefont {Wang}, \citenamefont
  {Miao}, \citenamefont {Ma}, \citenamefont {Richard}, \citenamefont {Huang},
  \citenamefont {Zhao}, \citenamefont {Chen}, \citenamefont {Fang},
  \citenamefont {Dai}, \citenamefont {Qian},\ and\ \citenamefont
  {Ding}}]{lv2015experimental}%
  \BibitemOpen
  \bibfield  {author} {\bibinfo {author} {\bibfnamefont {B.~Q.}\ \bibnamefont
  {Lv}}, \bibinfo {author} {\bibfnamefont {H.~M.}\ \bibnamefont {Weng}},
  \bibinfo {author} {\bibfnamefont {B.~B.}\ \bibnamefont {Fu}}, \bibinfo
  {author} {\bibfnamefont {X.~P.}\ \bibnamefont {Wang}}, \bibinfo {author}
  {\bibfnamefont {H.}~\bibnamefont {Miao}}, \bibinfo {author} {\bibfnamefont
  {J.}~\bibnamefont {Ma}}, \bibinfo {author} {\bibfnamefont {P.}~\bibnamefont
  {Richard}}, \bibinfo {author} {\bibfnamefont {X.~C.}\ \bibnamefont {Huang}},
  \bibinfo {author} {\bibfnamefont {L.~X.}\ \bibnamefont {Zhao}}, \bibinfo
  {author} {\bibfnamefont {G.~F.}\ \bibnamefont {Chen}}, \bibinfo {author}
  {\bibfnamefont {Z.}~\bibnamefont {Fang}}, \bibinfo {author} {\bibfnamefont
  {X.}~\bibnamefont {Dai}}, \bibinfo {author} {\bibfnamefont {T.}~\bibnamefont
  {Qian}},\ and\ \bibinfo {author} {\bibfnamefont {H.}~\bibnamefont {Ding}},\
  }\href {https://doi.org/10.1103/PhysRevX.5.031013}
  {\bibfield  {journal} {\bibinfo  {journal} {Phys. Rev. X}\ }\textbf {\bibinfo
  {volume} {5}},\ \bibinfo {pages} {031013} (\bibinfo {year}
  {2015})}\BibitemShut {NoStop}%
\bibitem [{\citenamefont {Xu}\ \emph {et~al.}(2015)\citenamefont {Xu},
  \citenamefont {Belopolski}, \citenamefont {Sanchez}, \citenamefont {Zhang},
  \citenamefont {Chang}, \citenamefont {Guo}, \citenamefont {Bian},
  \citenamefont {Yuan}, \citenamefont {Lu}, \citenamefont {Chang} \emph
  {et~al.}}]{xu2015experimental}%
  \BibitemOpen
  \bibfield  {author} {\bibinfo {author} {\bibfnamefont {S.-Y.}\ \bibnamefont
  {Xu}}, \bibinfo {author} {\bibfnamefont {I.}~\bibnamefont {Belopolski}},
  \bibinfo {author} {\bibfnamefont {D.~S.}\ \bibnamefont {Sanchez}}, \bibinfo
  {author} {\bibfnamefont {C.}~\bibnamefont {Zhang}}, \bibinfo {author}
  {\bibfnamefont {G.}~\bibnamefont {Chang}}, \bibinfo {author} {\bibfnamefont
  {C.}~\bibnamefont {Guo}}, \bibinfo {author} {\bibfnamefont {G.}~\bibnamefont
  {Bian}}, \bibinfo {author} {\bibfnamefont {Z.}~\bibnamefont {Yuan}}, \bibinfo
  {author} {\bibfnamefont {H.}~\bibnamefont {Lu}}, \bibinfo {author}
  {\bibfnamefont {T.-R.}\ \bibnamefont {Chang}}, \emph {et~al.},\ }\href {https://doi.org/10.1126/sciadv.1501092}
  {\bibfield  {journal} {\bibinfo  {journal} {Science advances}\ }\textbf
  {\bibinfo {volume} {1}},\ \bibinfo {pages} {e1501092} (\bibinfo {year}
  {2015})}\BibitemShut {NoStop}%
\bibitem [{\citenamefont {Li}\ \emph {et~al.}(2020)\citenamefont {Li},
  \citenamefont {Koo}, \citenamefont {Ning}, \citenamefont {Li}, \citenamefont
  {Miao}, \citenamefont {Min}, \citenamefont {Zhu}, \citenamefont {Wang},
  \citenamefont {Alem}, \citenamefont {Liu} \emph {et~al.}}]{li2020giant}%
  \BibitemOpen
  \bibfield  {author} {\bibinfo {author} {\bibfnamefont {P.}~\bibnamefont
  {Li}}, \bibinfo {author} {\bibfnamefont {J.}~\bibnamefont {Koo}}, \bibinfo
  {author} {\bibfnamefont {W.}~\bibnamefont {Ning}}, \bibinfo {author}
  {\bibfnamefont {J.}~\bibnamefont {Li}}, \bibinfo {author} {\bibfnamefont
  {L.}~\bibnamefont {Miao}}, \bibinfo {author} {\bibfnamefont {L.}~\bibnamefont
  {Min}}, \bibinfo {author} {\bibfnamefont {Y.}~\bibnamefont {Zhu}}, \bibinfo
  {author} {\bibfnamefont {Y.}~\bibnamefont {Wang}}, \bibinfo {author}
  {\bibfnamefont {N.}~\bibnamefont {Alem}}, \bibinfo {author} {\bibfnamefont
  {C.-X.}\ \bibnamefont {Liu}}, \emph {et~al.},\ }\href
  {https://doi.org/10.1038/s41467-020-17174-9} {\bibfield  {journal} {\bibinfo
  {journal} {Nature communications}\ }\textbf {\bibinfo {volume} {11}},\
  \bibinfo {pages} {1} (\bibinfo {year} {2020})}\BibitemShut {NoStop}%
\bibitem [{\citenamefont {Burkov}\ \emph {et~al.}(2011)\citenamefont {Burkov},
  \citenamefont {Hook},\ and\ \citenamefont {Balents}}]{burkov2011topological}%
  \BibitemOpen
  \bibfield  {author} {\bibinfo {author} {\bibfnamefont {A.~A.}\ \bibnamefont
  {Burkov}}, \bibinfo {author} {\bibfnamefont {M.~D.}\ \bibnamefont {Hook}},\
  and\ \bibinfo {author} {\bibfnamefont {L.}~\bibnamefont {Balents}},\
  }\href {https://doi.org/10.1103/PhysRevB.84.235126} {\bibfield  {journal} {\bibinfo
  {journal} {Phys. Rev. B}\ }\textbf {\bibinfo {volume} {84}},\ \bibinfo
  {pages} {235126} (\bibinfo {year} {2011})}\BibitemShut {NoStop}%
\bibitem [{\citenamefont {Fang}\ \emph {et~al.}(2015)\citenamefont {Fang},
  \citenamefont {Chen}, \citenamefont {Kee},\ and\ \citenamefont
  {Fu}}]{fang2015topological}%
  \BibitemOpen
  \bibfield  {author} {\bibinfo {author} {\bibfnamefont {C.}~\bibnamefont
  {Fang}}, \bibinfo {author} {\bibfnamefont {Y.}~\bibnamefont {Chen}}, \bibinfo
  {author} {\bibfnamefont {H.-Y.}\ \bibnamefont {Kee}},\ and\ \bibinfo {author}
  {\bibfnamefont {L.}~\bibnamefont {Fu}},\ }\href {https://doi.org/10.1103/PhysRevB.92.081201} {\bibfield
  {journal} {\bibinfo  {journal} {Phys. Rev. B}\ }\textbf {\bibinfo {volume}
  {92}},\ \bibinfo {pages} {081201} (\bibinfo {year} {2015})}\BibitemShut
  {NoStop}%
\bibitem [{\citenamefont {Chan}\ \emph {et~al.}(2016)\citenamefont {Chan},
  \citenamefont {Chiu}, \citenamefont {Chou},\ and\ \citenamefont
  {Schnyder}}]{chan20163}%
  \BibitemOpen
  \bibfield  {author} {\bibinfo {author} {\bibfnamefont {Y.-H.}\ \bibnamefont
  {Chan}}, \bibinfo {author} {\bibfnamefont {C.-K.}\ \bibnamefont {Chiu}},
  \bibinfo {author} {\bibfnamefont {M.~Y.}\ \bibnamefont {Chou}},\ and\
  \bibinfo {author} {\bibfnamefont {A.~P.}\ \bibnamefont {Schnyder}},\
  }\href {https://doi.org/10.1103/PhysRevB.93.205132} {\bibfield
  {journal} {\bibinfo  {journal} {Phys. Rev. B}\ }\textbf {\bibinfo {volume}
  {93}},\ \bibinfo {pages} {205132} (\bibinfo {year} {2016})}\BibitemShut
  {NoStop}%
\bibitem [{\citenamefont {Liang}\ \emph {et~al.}(2016)\citenamefont {Liang},
  \citenamefont {Zhou}, \citenamefont {Yu}, \citenamefont {Wang},\ and\
  \citenamefont {Weng}}]{liang2016node}%
  \BibitemOpen
  \bibfield  {author} {\bibinfo {author} {\bibfnamefont {Q.-F.}\ \bibnamefont
  {Liang}}, \bibinfo {author} {\bibfnamefont {J.}~\bibnamefont {Zhou}},
  \bibinfo {author} {\bibfnamefont {R.}~\bibnamefont {Yu}}, \bibinfo {author}
  {\bibfnamefont {Z.}~\bibnamefont {Wang}},\ and\ \bibinfo {author}
  {\bibfnamefont {H.}~\bibnamefont {Weng}},\ }\href {https://doi.org/10.1103/PhysRevB.93.085427} {\bibfield
  {journal} {\bibinfo  {journal} {Physical Review B}\ }\textbf {\bibinfo
  {volume} {93}},\ \bibinfo {pages} {085427} (\bibinfo {year}
  {2016})}\BibitemShut {NoStop}%
\bibitem [{\citenamefont {Kim}\ \emph {et~al.}(2018)\citenamefont {Kim},
  \citenamefont {Seo}, \citenamefont {Lee}, \citenamefont {Ko}, \citenamefont
  {Kim}, \citenamefont {Jang}, \citenamefont {Ok}, \citenamefont {Lee},
  \citenamefont {Jo}, \citenamefont {Kang} \emph {et~al.}}]{kim2018large}%
  \BibitemOpen
  \bibfield  {author} {\bibinfo {author} {\bibfnamefont {K.}~\bibnamefont
  {Kim}}, \bibinfo {author} {\bibfnamefont {J.}~\bibnamefont {Seo}}, \bibinfo
  {author} {\bibfnamefont {E.}~\bibnamefont {Lee}}, \bibinfo {author}
  {\bibfnamefont {K.-T.}\ \bibnamefont {Ko}}, \bibinfo {author} {\bibfnamefont
  {B.}~\bibnamefont {Kim}}, \bibinfo {author} {\bibfnamefont {B.~G.}\
  \bibnamefont {Jang}}, \bibinfo {author} {\bibfnamefont {J.~M.}\ \bibnamefont
  {Ok}}, \bibinfo {author} {\bibfnamefont {J.}~\bibnamefont {Lee}}, \bibinfo
  {author} {\bibfnamefont {Y.~J.}\ \bibnamefont {Jo}}, \bibinfo {author}
  {\bibfnamefont {W.}~\bibnamefont {Kang}}, \emph {et~al.},\ }\href
  {https://doi.org/10.1038/s41563-018-0132-3} {\bibfield  {journal} {\bibinfo
  {journal} {Nature materials}\ }\textbf {\bibinfo {volume} {17}},\ \bibinfo
  {pages} {794} (\bibinfo {year} {2018})}\BibitemShut {NoStop}%
\bibitem [{\citenamefont {Singh}\ \emph {et~al.}(2021)\citenamefont {Singh},
  \citenamefont {Noky}, \citenamefont {Bhattacharya}, \citenamefont {Vir},
  \citenamefont {Sun}, \citenamefont {Kumar}, \citenamefont {Felser},\ and\
  \citenamefont {Shekhar}}]{singh2021anisotropic}%
  \BibitemOpen
  \bibfield  {author} {\bibinfo {author} {\bibfnamefont {S.}~\bibnamefont
  {Singh}}, \bibinfo {author} {\bibfnamefont {J.}~\bibnamefont {Noky}},
  \bibinfo {author} {\bibfnamefont {S.}~\bibnamefont {Bhattacharya}}, \bibinfo
  {author} {\bibfnamefont {P.}~\bibnamefont {Vir}}, \bibinfo {author}
  {\bibfnamefont {Y.}~\bibnamefont {Sun}}, \bibinfo {author} {\bibfnamefont
  {N.}~\bibnamefont {Kumar}}, \bibinfo {author} {\bibfnamefont
  {C.}~\bibnamefont {Felser}},\ and\ \bibinfo {author} {\bibfnamefont
  {C.}~\bibnamefont {Shekhar}},\ }\href {https://doi.org/https://doi.org/10.1002/adma.202104126}
  {\bibfield  {journal} {\bibinfo  {journal} {Advanced Materials}\ }\textbf
  {\bibinfo {volume} {33}},\ \bibinfo {pages} {2104126} (\bibinfo {year}
  {2021})}\BibitemShut {NoStop}%
\bibitem [{\citenamefont {Guin}\ \emph {et~al.}(2021)\citenamefont {Guin},
  \citenamefont {Xu}, \citenamefont {Kumar}, \citenamefont {Kung},
  \citenamefont {Dufresne}, \citenamefont {Le}, \citenamefont {Vir},
  \citenamefont {Michiardi}, \citenamefont {Pedersen}, \citenamefont
  {Gorovikov} \emph {et~al.}}]{guin20212d}%
  \BibitemOpen
  \bibfield  {author} {\bibinfo {author} {\bibfnamefont {S.~N.}\ \bibnamefont
  {Guin}}, \bibinfo {author} {\bibfnamefont {Q.}~\bibnamefont {Xu}}, \bibinfo
  {author} {\bibfnamefont {N.}~\bibnamefont {Kumar}}, \bibinfo {author}
  {\bibfnamefont {H.-H.}\ \bibnamefont {Kung}}, \bibinfo {author}
  {\bibfnamefont {S.}~\bibnamefont {Dufresne}}, \bibinfo {author}
  {\bibfnamefont {C.}~\bibnamefont {Le}}, \bibinfo {author} {\bibfnamefont
  {P.}~\bibnamefont {Vir}}, \bibinfo {author} {\bibfnamefont {M.}~\bibnamefont
  {Michiardi}}, \bibinfo {author} {\bibfnamefont {T.}~\bibnamefont {Pedersen}},
  \bibinfo {author} {\bibfnamefont {S.}~\bibnamefont {Gorovikov}}, \emph
  {et~al.},\ }\href {https://doi.org/https://doi.org/10.1002/adma.202006301} {\bibfield
  {journal} {\bibinfo  {journal} {Advanced Materials}\ }\textbf {\bibinfo
  {volume} {33}},\ \bibinfo {pages} {2006301} (\bibinfo {year}
  {2021})}\BibitemShut {NoStop}%
\bibitem [{\citenamefont {Enke}\ \emph {et~al.}(2018)\citenamefont {Yan}, \citenamefont {Kumar}, \citenamefont {Muechler}, \citenamefont {Sun}, 
\ and\
\citenamefont {Felser}}]{Liu2018a}%
  \BibitemOpen
  \bibfield  {author} {\bibinfo {author} {\bibfnamefont {E.}~\bibnamefont
  {Liu}}, \bibinfo {author} {\bibfnamefont {Y.}~\bibnamefont {Sun}}, \bibinfo
  {author} {\bibfnamefont {K.}\ \bibnamefont {Kumar}}, \bibinfo {author}
  {\bibfnamefont {L.}~\bibnamefont {Muechler}}, \bibinfo {author}
  {\bibfnamefont {A.}~\bibnamefont {Sun}}, \bibinfo {author}
  {\bibfnamefont {L.}~\bibnamefont {Jiao}},  \bibinfo {author}
  {\bibfnamefont {S-Y.}~\bibnamefont {Yang}},  \bibinfo {author}
  {\bibfnamefont {D.}~\bibnamefont {Liu}}, \bibinfo {author}
  {\bibfnamefont {A.}~\bibnamefont {Liang}}, \bibinfo {author}
  {\bibfnamefont {Q.}~\bibnamefont {Xu}}, \bibinfo {author}
  {\bibfnamefont {J.}~\bibnamefont {Kroder}}, \bibinfo {author}
  {\bibfnamefont {V.}~\bibnamefont {Süß}}, \bibinfo {author}
  {\bibfnamefont {H.}~\bibnamefont {Borrmann}}, \bibinfo {author}
  {\bibfnamefont {C.}~\bibnamefont {Shekhar}}, \bibinfo {author}
  {\bibfnamefont {Z.}~\bibnamefont {Wang}}, \bibinfo {author}
  {\bibfnamefont {C.}~\bibnamefont {Xi}}, \bibinfo {author}
  {\bibfnamefont {W.}~\bibnamefont {Wang}}, \bibinfo {author}
  {\bibfnamefont {W.}~\bibnamefont {Schnelle}}, \bibinfo {author}
  {\bibfnamefont {S.}~\bibnamefont {Wirth}},
\bibinfo {author} {\bibfnamefont {Y.}~\bibnamefont {Chen}},
\bibinfo {author}{\bibfnamefont {S.T.B.}~\bibnamefont {Goennenwein}}, \ and\ \bibinfo {author}
  {\bibfnamefont {C.}\ \bibnamefont {Felser}},\ }\href {https://doi.org/10.1038/s41567-018-0234-5}
  {\bibfield  {journal} {\bibinfo  {journal} {Nature Physics}\
  }\textbf {\bibinfo {volume} {14}},\ \bibinfo {pages} {1125--1131} (\bibinfo {year}
  {2018})}\BibitemShut {NoStop}%
\bibitem [{\citenamefont {Manna}\ \emph {et~al.}(2018)\citenamefont {Manna},
  \citenamefont {Muechler}, \citenamefont {Kao}, \citenamefont {Stinshoff},
  \citenamefont {Zhang}, \citenamefont {Gooth}, \citenamefont {Kumar},
  \citenamefont {Kreiner}, \citenamefont {Koepernik}, \citenamefont {Car} \emph
  {et~al.}}]{manna2018colossal}%
  \BibitemOpen
  \bibfield  {author} {\bibinfo {author} {\bibfnamefont {K.}~\bibnamefont
  {Manna}}, \bibinfo {author} {\bibfnamefont {L.}~\bibnamefont {Muechler}},
  \bibinfo {author} {\bibfnamefont {T.-H.}\ \bibnamefont {Kao}}, \bibinfo
  {author} {\bibfnamefont {R.}~\bibnamefont {Stinshoff}}, \bibinfo {author}
  {\bibfnamefont {Y.}~\bibnamefont {Zhang}}, \bibinfo {author} {\bibfnamefont
  {J.}~\bibnamefont {Gooth}}, \bibinfo {author} {\bibfnamefont
  {N.}~\bibnamefont {Kumar}}, \bibinfo {author} {\bibfnamefont
  {G.}~\bibnamefont {Kreiner}}, \bibinfo {author} {\bibfnamefont
  {K.}~\bibnamefont {Koepernik}}, \bibinfo {author} {\bibfnamefont
  {R.}~\bibnamefont {Car}}, \emph {et~al.},\ }\href {https://doi.org/10.1103/PhysRevX.8.041045} {\bibfield  {journal} {\bibinfo
  {journal} {Physical Review X}\ }\textbf {\bibinfo {volume} {8}},\ \bibinfo
  {pages} {041045} (\bibinfo {year} {2018})}\BibitemShut {NoStop}%
\bibitem [{\citenamefont {Sakai}\ \emph {et~al.}(2018)\citenamefont {Sakai},
  \citenamefont {Mizuta}, \citenamefont {Nugroho}, \citenamefont {Sihombing},
  \citenamefont {Koretsune}, \citenamefont {Suzuki}, \citenamefont {Takemori},
  \citenamefont {Ishii}, \citenamefont {Nishio-Hamane}, \citenamefont {Arita}
  \emph {et~al.}}]{sakai2018giant}%
  \BibitemOpen
  \bibfield  {author} {\bibinfo {author} {\bibfnamefont {A.}~\bibnamefont
  {Sakai}}, \bibinfo {author} {\bibfnamefont {Y.~P.}\ \bibnamefont {Mizuta}},
  \bibinfo {author} {\bibfnamefont {A.~A.}\ \bibnamefont {Nugroho}}, \bibinfo
  {author} {\bibfnamefont {R.}~\bibnamefont {Sihombing}}, \bibinfo {author}
  {\bibfnamefont {T.}~\bibnamefont {Koretsune}}, \bibinfo {author}
  {\bibfnamefont {M.-T.}\ \bibnamefont {Suzuki}}, \bibinfo {author}
  {\bibfnamefont {N.}~\bibnamefont {Takemori}}, \bibinfo {author}
  {\bibfnamefont {R.}~\bibnamefont {Ishii}}, \bibinfo {author} {\bibfnamefont
  {D.}~\bibnamefont {Nishio-Hamane}}, \bibinfo {author} {\bibfnamefont
  {R.}~\bibnamefont {Arita}}, \emph {et~al.},\ }\href {https://doi.org/10.1038/s41567-018-0225-6}
  {\bibfield  {journal} {\bibinfo  {journal} {Nature Physics}\ }\textbf
  {\bibinfo {volume} {14}},\ \bibinfo {pages} {1119} (\bibinfo {year}
  {2018})}\BibitemShut {NoStop}%
\bibitem [{\citenamefont {Xu}\ \emph {et~al.}(2019)\citenamefont {Xu},
  \citenamefont {Phelan},\ and\ \citenamefont {Chien}}]{xu2019large}%
  \BibitemOpen
  \bibfield  {author} {\bibinfo {author} {\bibfnamefont {J.}~\bibnamefont
  {Xu}}, \bibinfo {author} {\bibfnamefont {W.~A.}\ \bibnamefont {Phelan}},\
  and\ \bibinfo {author} {\bibfnamefont {C.-L.}\ \bibnamefont {Chien}},\
  }\href {https://doi.org/10.1021/acs.nanolett.9b03739} {\bibfield  {journal}
  {\bibinfo  {journal} {Nano letters}\ }\textbf {\bibinfo {volume} {19}},\
  \bibinfo {pages} {8250} (\bibinfo {year} {2019})}\BibitemShut {NoStop}%
\bibitem [{\citenamefont {Nagaosa}\ \emph {et~al.}(2010)\citenamefont
  {Nagaosa}, \citenamefont {Sinova}, \citenamefont {Onoda}, \citenamefont
  {MacDonald},\ and\ \citenamefont {Ong}}]{nagaosa2010anomalous}%
  \BibitemOpen
  \bibfield  {author} {\bibinfo {author} {\bibfnamefont {N.}~\bibnamefont
  {Nagaosa}}, \bibinfo {author} {\bibfnamefont {J.}~\bibnamefont {Sinova}},
  \bibinfo {author} {\bibfnamefont {S.}~\bibnamefont {Onoda}}, \bibinfo
  {author} {\bibfnamefont {A.~H.}\ \bibnamefont {MacDonald}},\ and\ \bibinfo
  {author} {\bibfnamefont {N.~P.}\ \bibnamefont {Ong}},\ }\href {https://doi.org/10.1103/RevModPhys.82.1539} {\bibfield  {journal} {\bibinfo
  {journal} {Reviews of modern physics}\ }\textbf {\bibinfo {volume} {82}},\
  \bibinfo {pages} {1539} (\bibinfo {year} {2010})}\BibitemShut {NoStop}%
\bibitem [{\citenamefont {Fu}(2011)}]{PhysRevLett.106.106802}%
  \BibitemOpen
  \bibfield  {author} {\bibinfo {author} {\bibfnamefont {L.}~\bibnamefont
  {Fu}},\ }\href {https://doi.org/10.1103/PhysRevLett.106.106802}
  {\bibfield  {journal} {\bibinfo  {journal} {Phys. Rev. Lett.}\ }\textbf
  {\bibinfo {volume} {106}},\ \bibinfo {pages} {106802} (\bibinfo {year}
  {2011})}\BibitemShut {NoStop}%
\bibitem [{\citenamefont {Hsieh}\ \emph {et~al.}(2012)\citenamefont {Hsieh},
  \citenamefont {Lin}, \citenamefont {Liu}, \citenamefont {Duan}, \citenamefont
  {Bansil},\ and\ \citenamefont {Fu}}]{Hsieh2012}%
  \BibitemOpen
  \bibfield  {author} {\bibinfo {author} {\bibfnamefont {T.~H.}\ \bibnamefont
  {Hsieh}}, \bibinfo {author} {\bibfnamefont {H.}~\bibnamefont {Lin}}, \bibinfo
  {author} {\bibfnamefont {J.}~\bibnamefont {Liu}}, \bibinfo {author}
  {\bibfnamefont {W.}~\bibnamefont {Duan}}, \bibinfo {author} {\bibfnamefont
  {A.}~\bibnamefont {Bansil}},\ and\ \bibinfo {author} {\bibfnamefont
  {L.}~\bibnamefont {Fu}},\ }\href
  {https://doi.org/10.1038/ncomms1969} {\bibfield  {journal} {\bibinfo
  {journal} {Nature Communications}\ }\textbf {\bibinfo {volume} {3}},\
  \bibinfo {pages} {982} (\bibinfo {year} {2012})}\BibitemShut {NoStop}%
\bibitem [{\citenamefont {Wang}\ \emph {et~al.}(2018)\citenamefont {Wang},
  \citenamefont {Xu}, \citenamefont {Lou}, \citenamefont {Liu}, \citenamefont
  {Li}, \citenamefont {Huang}, \citenamefont {Shen}, \citenamefont {Weng},
  \citenamefont {Wang},\ and\ \citenamefont {Lei}}]{Wang2018}%
  \BibitemOpen
  \bibfield  {author} {\bibinfo {author} {\bibfnamefont {Q.}~\bibnamefont
  {Wang}}, \bibinfo {author} {\bibfnamefont {Y.}~\bibnamefont {Xu}}, \bibinfo
  {author} {\bibfnamefont {R.}~\bibnamefont {Lou}}, \bibinfo {author}
  {\bibfnamefont {Z.}~\bibnamefont {Liu}}, \bibinfo {author} {\bibfnamefont
  {M.}~\bibnamefont {Li}}, \bibinfo {author} {\bibfnamefont {Y.}~\bibnamefont
  {Huang}}, \bibinfo {author} {\bibfnamefont {D.}~\bibnamefont {Shen}},
  \bibinfo {author} {\bibfnamefont {H.}~\bibnamefont {Weng}}, \bibinfo {author}
  {\bibfnamefont {S.}~\bibnamefont {Wang}},\ and\ \bibinfo {author}
  {\bibfnamefont {H.}~\bibnamefont {Lei}},\ }\href
  {https://doi.org/10.1038/s41467-018-06088-2} {\bibfield  {journal} {\bibinfo
  {journal} {Nature Communications}\ }\textbf {\bibinfo {volume} {9}},\
  \bibinfo {pages} {3681} (\bibinfo {year} {2018})}\BibitemShut {NoStop}%
\bibitem [{\citenamefont {Fang}\ \emph {et~al.}(2003)\citenamefont {Fang},
  \citenamefont {Nagaosa}, \citenamefont {Takahashi}, \citenamefont {Asamitsu},
  \citenamefont {Mathieu}, \citenamefont {Ogasawara}, \citenamefont {Yamada},
  \citenamefont {Kawasaki}, \citenamefont {Tokura},\ and\ \citenamefont
  {Terakura}}]{Fang2003}%
  \BibitemOpen
  \bibfield  {author} {\bibinfo {author} {\bibfnamefont {Z.}~\bibnamefont
  {Fang}}, \bibinfo {author} {\bibfnamefont {N.}~\bibnamefont {Nagaosa}},
  \bibinfo {author} {\bibfnamefont {K.~S.}\ \bibnamefont {Takahashi}}, \bibinfo
  {author} {\bibfnamefont {A.}~\bibnamefont {Asamitsu}}, \bibinfo {author}
  {\bibfnamefont {R.}~\bibnamefont {Mathieu}}, \bibinfo {author} {\bibfnamefont
  {T.}~\bibnamefont {Ogasawara}}, \bibinfo {author} {\bibfnamefont
  {H.}~\bibnamefont {Yamada}}, \bibinfo {author} {\bibfnamefont
  {M.}~\bibnamefont {Kawasaki}}, \bibinfo {author} {\bibfnamefont
  {Y.}~\bibnamefont {Tokura}},\ and\ \bibinfo {author} {\bibfnamefont
  {K.}~\bibnamefont {Terakura}},\ }\href
  {https://doi.org/10.1126/science.1089408} {\bibfield  {journal} {\bibinfo
  {journal} {Science}\ }\textbf {\bibinfo {volume} {302}},\ \bibinfo {pages}
  {92} (\bibinfo {year} {2003})}\BibitemShut {NoStop}%
\bibitem [{\citenamefont {Pradhan}\ \emph {et~al.}(2022)\citenamefont
  {Pradhan}, \citenamefont {Samanta}, \citenamefont {Saha},\ and\ \citenamefont
  {Nandy}}]{pradhan2022vector}%
  \BibitemOpen
  \bibfield  {author} {\bibinfo {author} {\bibfnamefont {S.}~\bibnamefont
  {Pradhan}}, \bibinfo {author} {\bibfnamefont {K.}~\bibnamefont {Samanta}},
  \bibinfo {author} {\bibfnamefont {K.}~\bibnamefont {Saha}},\ and\ \bibinfo
  {author} {\bibfnamefont {A.~K.}\ \bibnamefont {Nandy}},\ }\href {https://arxiv.org/abs/2206.15044} {\bibfield
  {journal} {\bibinfo  {journal} {arXiv:2206.15044}\ } (\bibinfo {year}
  {2022})}\BibitemShut {NoStop}%
\bibitem [{\citenamefont {Shindou}\ and\ \citenamefont
  {Nagaosa}(2001)}]{shindou2001orbital}%
  \BibitemOpen
  \bibfield  {author} {\bibinfo {author} {\bibfnamefont {R.}~\bibnamefont
  {Shindou}}\ and\ \bibinfo {author} {\bibfnamefont {N.}~\bibnamefont
  {Nagaosa}},\ }\href {https://doi.org/10.1103/PhysRevLett.87.116801} {\bibfield  {journal}
  {\bibinfo  {journal} {Physical Review Letters}\ }\textbf {\bibinfo {volume}
  {87}},\ \bibinfo {pages} {116801} (\bibinfo {year} {2001})}\BibitemShut
  {NoStop}%
\bibitem [{\citenamefont {Singh}\ \emph {et~al.}(2020)\citenamefont {Singh},
  \citenamefont {Singh}, \citenamefont {Pradhan}, \citenamefont {Srihari},
  \citenamefont {Poswal}, \citenamefont {Nath}, \citenamefont {Nandy},\ and\
  \citenamefont {Nayak}}]{charanpreet}%
  \BibitemOpen
  \bibfield  {author} {\bibinfo {author} {\bibfnamefont {C.}~\bibnamefont
  {Singh}}, \bibinfo {author} {\bibfnamefont {V.}~\bibnamefont {Singh}},
  \bibinfo {author} {\bibfnamefont {G.}~\bibnamefont {Pradhan}}, \bibinfo
  {author} {\bibfnamefont {V.}~\bibnamefont {Srihari}}, \bibinfo {author}
  {\bibfnamefont {H.~K.}\ \bibnamefont {Poswal}}, \bibinfo {author}
  {\bibfnamefont {R.}~\bibnamefont {Nath}}, \bibinfo {author} {\bibfnamefont
  {A.~K.}\ \bibnamefont {Nandy}},\ and\ \bibinfo {author} {\bibfnamefont
  {A.~K.}\ \bibnamefont {Nayak}},\ }\href
{https://doi.org/10.1103/PhysRevResearch.2.043366} {\bibfield  {journal}
  {\bibinfo  {journal} {Phys. Rev. Res.}\ }\textbf {\bibinfo {volume} {2}},\
  \bibinfo {pages} {043366} (\bibinfo {year} {2020})}\BibitemShut {NoStop}%
\bibitem [{\citenamefont {Ito}\ \emph {et~al.}(2019)\citenamefont {Ito},
  \citenamefont {Kikkawa}, \citenamefont {Barker}, \citenamefont {Hirobe},
  \citenamefont {Shiomi},\ and\ \citenamefont {Saitoh}}]{PhysRevB.100.060402}%
  \BibitemOpen
  \bibfield  {author} {\bibinfo {author} {\bibfnamefont {N.}~\bibnamefont
  {Ito}}, \bibinfo {author} {\bibfnamefont {T.}~\bibnamefont {Kikkawa}},
  \bibinfo {author} {\bibfnamefont {J.}~\bibnamefont {Barker}}, \bibinfo
  {author} {\bibfnamefont {D.}~\bibnamefont {Hirobe}}, \bibinfo {author}
  {\bibfnamefont {Y.}~\bibnamefont {Shiomi}},\ and\ \bibinfo {author}
  {\bibfnamefont {E.}~\bibnamefont {Saitoh}},\ }\href {https://doi.org/10.1103/PhysRevB.100.060402} {\bibfield  {journal} {\bibinfo
   {journal} {Phys. Rev. B}\ }\textbf {\bibinfo {volume} {100}},\ \bibinfo
  {pages} {060402} (\bibinfo {year} {2019})}\BibitemShut {NoStop}%
\bibitem [{\citenamefont {Liu}\ and\ \citenamefont
  {Petrovic}(2018)}]{PhysRevB.97.014420}%
  \BibitemOpen
  \bibfield  {author} {\bibinfo {author} {\bibfnamefont {Y.}~\bibnamefont
  {Liu}}\ and\ \bibinfo {author} {\bibfnamefont {C.}~\bibnamefont {Petrovic}},\
  }\href  {https://doi.org/10.1103/PhysRevB.97.014420} {\bibfield  {journal} {\bibinfo
  {journal} {Phys. Rev. B}\ }\textbf {\bibinfo {volume} {97}},\ \bibinfo
  {pages} {014420} (\bibinfo {year} {2018})}\BibitemShut {NoStop}%
\bibitem [{\citenamefont {Wang}\ \emph {et~al.}(2017)\citenamefont {Wang},
  \citenamefont {Xian}, \citenamefont {Wang}, \citenamefont {Liu},
  \citenamefont {Ling}, \citenamefont {Zhang}, \citenamefont {Cao},
  \citenamefont {Qu},\ and\ \citenamefont {Xiong}}]{wang2017anisotropic}%
  \BibitemOpen
  \bibfield  {author} {\bibinfo {author} {\bibfnamefont {Y.}~\bibnamefont
  {Wang}}, \bibinfo {author} {\bibfnamefont {C.}~\bibnamefont {Xian}}, \bibinfo
  {author} {\bibfnamefont {J.}~\bibnamefont {Wang}}, \bibinfo {author}
  {\bibfnamefont {B.}~\bibnamefont {Liu}}, \bibinfo {author} {\bibfnamefont
  {L.}~\bibnamefont {Ling}}, \bibinfo {author} {\bibfnamefont {L.}~\bibnamefont
  {Zhang}}, \bibinfo {author} {\bibfnamefont {L.}~\bibnamefont {Cao}}, \bibinfo
  {author} {\bibfnamefont {Z.}~\bibnamefont {Qu}},\ and\ \bibinfo {author}
  {\bibfnamefont {Y.}~\bibnamefont {Xiong}},\ }\href {https://doi.org/10.1103/PhysRevB.96.134428} {\bibfield  {journal} {\bibinfo
  {journal} {Physical Review B}\ }\textbf {\bibinfo {volume} {96}},\ \bibinfo
  {pages} {134428} (\bibinfo {year} {2017})}\BibitemShut {NoStop}%
\bibitem [{\citenamefont {Tan}\ \emph {et~al.}(2018)\citenamefont {Tan},
  \citenamefont {Lee}, \citenamefont {Jung}, \citenamefont {Park},
  \citenamefont {Albarakati}, \citenamefont {Partridge}, \citenamefont {Field},
  \citenamefont {McCulloch}, \citenamefont {Wang},\ and\ \citenamefont
  {Lee}}]{tan2018hard}%
  \BibitemOpen
  \bibfield  {author} {\bibinfo {author} {\bibfnamefont {C.}~\bibnamefont
  {Tan}}, \bibinfo {author} {\bibfnamefont {J.}~\bibnamefont {Lee}}, \bibinfo
  {author} {\bibfnamefont {S.-G.}\ \bibnamefont {Jung}}, \bibinfo {author}
  {\bibfnamefont {T.}~\bibnamefont {Park}}, \bibinfo {author} {\bibfnamefont
  {S.}~\bibnamefont {Albarakati}}, \bibinfo {author} {\bibfnamefont
  {J.}~\bibnamefont {Partridge}}, \bibinfo {author} {\bibfnamefont {M.~R.}\
  \bibnamefont {Field}}, \bibinfo {author} {\bibfnamefont {D.~G.}\ \bibnamefont
  {McCulloch}}, \bibinfo {author} {\bibfnamefont {L.}~\bibnamefont {Wang}},\
  and\ \bibinfo {author} {\bibfnamefont {C.}~\bibnamefont {Lee}},\ }\href {https://doi.org/10.1038/s41467-018-04018-w}
  {\bibfield  {journal} {\bibinfo  {journal} {Nature communications}\ }\textbf
  {\bibinfo {volume} {9}},\ \bibinfo {pages} {1} (\bibinfo {year}
  {2018})}\BibitemShut {NoStop}%
\bibitem [{\citenamefont {May}\ \emph {et~al.}(2019{\natexlab{a}})\citenamefont
  {May}, \citenamefont {Ovchinnikov}, \citenamefont {Zheng}, \citenamefont
  {Hermann}, \citenamefont {Calder}, \citenamefont {Huang}, \citenamefont
  {Fei}, \citenamefont {Liu}, \citenamefont {Xu},\ and\ \citenamefont
  {McGuire}}]{may2019ferromagnetism}%
  \BibitemOpen
  \bibfield  {author} {\bibinfo {author} {\bibfnamefont {A.~F.}\ \bibnamefont
  {May}}, \bibinfo {author} {\bibfnamefont {D.}~\bibnamefont {Ovchinnikov}},
  \bibinfo {author} {\bibfnamefont {Q.}~\bibnamefont {Zheng}}, \bibinfo
  {author} {\bibfnamefont {R.}~\bibnamefont {Hermann}}, \bibinfo {author}
  {\bibfnamefont {S.}~\bibnamefont {Calder}}, \bibinfo {author} {\bibfnamefont
  {B.}~\bibnamefont {Huang}}, \bibinfo {author} {\bibfnamefont
  {Z.}~\bibnamefont {Fei}}, \bibinfo {author} {\bibfnamefont {Y.}~\bibnamefont
  {Liu}}, \bibinfo {author} {\bibfnamefont {X.}~\bibnamefont {Xu}},\ and\
  \bibinfo {author} {\bibfnamefont {M.~A.}\ \bibnamefont {McGuire}},\
  }\href
  {https://doi.org/10.1021/acsnano.8b09660} {\bibfield  {journal} {\bibinfo
  {journal} {ACS nano}\ }\textbf {\bibinfo {volume} {13}},\ \bibinfo {pages}
  {4436} (\bibinfo {year} {2019}{\natexlab{a}})}\BibitemShut {NoStop}%
\bibitem [{\citenamefont {May}\ \emph {et~al.}(2019{\natexlab{b}})\citenamefont
  {May}, \citenamefont {Bridges},\ and\ \citenamefont
  {McGuire}}]{may2019physical}%
  \BibitemOpen
  \bibfield  {author} {\bibinfo {author} {\bibfnamefont {A.~F.}\ \bibnamefont
  {May}}, \bibinfo {author} {\bibfnamefont {C.~A.}\ \bibnamefont {Bridges}},\
  and\ \bibinfo {author} {\bibfnamefont {M.~A.}\ \bibnamefont {McGuire}},\
  }\href {https://doi.org/10.1103/PhysRevMaterials.3.104401} {\bibfield  {journal}
  {\bibinfo  {journal} {Physical Review Materials}\ }\textbf {\bibinfo {volume}
  {3}},\ \bibinfo {pages} {104401} (\bibinfo {year}
  {2019}{\natexlab{b}})}\BibitemShut {NoStop}%
\bibitem [{\citenamefont {Seo}\ \emph {et~al.}(2020)\citenamefont {Seo},
  \citenamefont {Kim}, \citenamefont {An}, \citenamefont {Kim}, \citenamefont
  {Kim}, \citenamefont {Hwang}, \citenamefont {Kim}, \citenamefont {Jang},
  \citenamefont {Kim}, \citenamefont {Eom}, \citenamefont {Seo}, \citenamefont
  {Stania}, \citenamefont {Muntwiler}, \citenamefont {Lee}, \citenamefont
  {Watanabe}, \citenamefont {Taniguchi}, \citenamefont {Jo}, \citenamefont
  {Lee}, \citenamefont {Min}, \citenamefont {Jo}, \citenamefont {Yeom},
  \citenamefont {Choi}, \citenamefont {Shim},\ and\ \citenamefont
  {Kim}}]{seo2020}%
  \BibitemOpen
  \bibfield  {author} {\bibinfo {author} {\bibfnamefont {J.}~\bibnamefont
  {Seo}}, \bibinfo {author} {\bibfnamefont {D.~Y.}\ \bibnamefont {Kim}},
  \bibinfo {author} {\bibfnamefont {E.~S.}\ \bibnamefont {An}}, \bibinfo
  {author} {\bibfnamefont {K.}~\bibnamefont {Kim}}, \bibinfo {author}
  {\bibfnamefont {G.-Y.}\ \bibnamefont {Kim}}, \bibinfo {author} {\bibfnamefont
  {S.-Y.}\ \bibnamefont {Hwang}}, \bibinfo {author} {\bibfnamefont {D.~W.}\
  \bibnamefont {Kim}}, \bibinfo {author} {\bibfnamefont {B.~G.}\ \bibnamefont
  {Jang}}, \bibinfo {author} {\bibfnamefont {H.}~\bibnamefont {Kim}}, \bibinfo
  {author} {\bibfnamefont {G.}~\bibnamefont {Eom}}, \bibinfo {author}
  {\bibfnamefont {S.~Y.}\ \bibnamefont {Seo}}, \bibinfo {author} {\bibfnamefont
  {R.}~\bibnamefont {Stania}}, \bibinfo {author} {\bibfnamefont
  {M.}~\bibnamefont {Muntwiler}}, \bibinfo {author} {\bibfnamefont
  {J.}~\bibnamefont {Lee}}, \bibinfo {author} {\bibfnamefont {K.}~\bibnamefont
  {Watanabe}}, \bibinfo {author} {\bibfnamefont {T.}~\bibnamefont {Taniguchi}},
  \bibinfo {author} {\bibfnamefont {Y.~J.}\ \bibnamefont {Jo}}, \bibinfo
  {author} {\bibfnamefont {J.}~\bibnamefont {Lee}}, \bibinfo {author}
  {\bibfnamefont {B.~I.}\ \bibnamefont {Min}}, \bibinfo {author} {\bibfnamefont
  {M.~H.}\ \bibnamefont {Jo}}, \bibinfo {author} {\bibfnamefont {H.~W.}\
  \bibnamefont {Yeom}}, \bibinfo {author} {\bibfnamefont {S.-Y.}\ \bibnamefont
  {Choi}}, \bibinfo {author} {\bibfnamefont {J.~H.}\ \bibnamefont {Shim}},\
  and\ \bibinfo {author} {\bibfnamefont {J.~S.}\ \bibnamefont {Kim}},\
  }\href
  {https://doi.org/10.1126/sciadv.aay8912} {\bibfield  {journal} {\bibinfo
  {journal} {Science Advances}\ }\textbf {\bibinfo {volume} {6}},\ \bibinfo
  {pages} {eaay8912} (\bibinfo {year} {2020})}\BibitemShut {NoStop}%
\bibitem [{\citenamefont {Liu}\ \emph {et~al.}(2018)\citenamefont {Liu},
  \citenamefont {Stavitski}, \citenamefont {Attenkofer},\ and\ \citenamefont
  {Petrovic}}]{liuPhysRevB.97.165415}%
  \BibitemOpen
  \bibfield  {author} {\bibinfo {author} {\bibfnamefont {Y.}~\bibnamefont
  {Liu}}, \bibinfo {author} {\bibfnamefont {E.}~\bibnamefont {Stavitski}},
  \bibinfo {author} {\bibfnamefont {K.}~\bibnamefont {Attenkofer}},\ and\
  \bibinfo {author} {\bibfnamefont {C.}~\bibnamefont {Petrovic}},\ }\href {https://doi.org/10.1103/PhysRevB.97.165415} {\bibfield  {journal} {\bibinfo
  {journal} {Phys. Rev. B}\ }\textbf {\bibinfo {volume} {97}},\ \bibinfo
  {pages} {165415} (\bibinfo {year} {2018})}\BibitemShut {NoStop}%
\bibitem [{\citenamefont {Mondal}\ \emph {et~al.}(2021)\citenamefont {Mondal},
  \citenamefont {Khan}, \citenamefont {Mishra}, \citenamefont {Satpati},\ and\
  \citenamefont {Mandal}}]{mondalPhysRevB.104.094405}%
  \BibitemOpen
  \bibfield  {author} {\bibinfo {author} {\bibfnamefont {S.}~\bibnamefont
  {Mondal}}, \bibinfo {author} {\bibfnamefont {N.}~\bibnamefont {Khan}},
  \bibinfo {author} {\bibfnamefont {S.~M.}\ \bibnamefont {Mishra}}, \bibinfo
  {author} {\bibfnamefont {B.}~\bibnamefont {Satpati}},\ and\ \bibinfo {author}
  {\bibfnamefont {P.}~\bibnamefont {Mandal}},\ }\href {https://doi.org/10.1103/PhysRevB.104.094405} {\bibfield  {journal} {\bibinfo
   {journal} {Phys. Rev. B}\ }\textbf {\bibinfo {volume} {104}},\ \bibinfo
  {pages} {094405} (\bibinfo {year} {2021})}\BibitemShut {NoStop}%
\bibitem [{\citenamefont {Yang}\ \emph {et~al.}(2021)\citenamefont {Yang},
  \citenamefont {Zhou}, \citenamefont {Feng},\ and\ \citenamefont
  {Yao}}]{yang2021strong}%
  \BibitemOpen
  \bibfield  {author} {\bibinfo {author} {\bibfnamefont {X.}~\bibnamefont
  {Yang}}, \bibinfo {author} {\bibfnamefont {X.}~\bibnamefont {Zhou}}, \bibinfo
  {author} {\bibfnamefont {W.}~\bibnamefont {Feng}},\ and\ \bibinfo {author}
  {\bibfnamefont {Y.}~\bibnamefont {Yao}},\ }\href
{https://doi.org/10.1103/PhysRevB.104.104427} {\bibfield  {journal} {\bibinfo
   {journal} {Physical Review B}\ }\textbf {\bibinfo {volume} {104}},\ \bibinfo
  {pages} {104427} (\bibinfo {year} {2021})}\BibitemShut {NoStop}%
\bibitem [{\citenamefont {Gao}\ \emph {et~al.}(2022)\citenamefont {Gao},
  \citenamefont {Yan}, \citenamefont {Yin}, \citenamefont {Huang},
  \citenamefont {Li}, \citenamefont {Zhu}, \citenamefont {Cai}, \citenamefont
  {Shen}, \citenamefont {Lei}, \citenamefont {Zhang},\ and\ \citenamefont
  {Wang}}]{gaoPhysRevB.105.014426}%
  \BibitemOpen
  \bibfield  {author} {\bibinfo {author} {\bibfnamefont {Y.}~\bibnamefont
  {Gao}}, \bibinfo {author} {\bibfnamefont {S.}~\bibnamefont {Yan}}, \bibinfo
  {author} {\bibfnamefont {Q.}~\bibnamefont {Yin}}, \bibinfo {author}
  {\bibfnamefont {H.}~\bibnamefont {Huang}}, \bibinfo {author} {\bibfnamefont
  {Z.}~\bibnamefont {Li}}, \bibinfo {author} {\bibfnamefont {Z.}~\bibnamefont
  {Zhu}}, \bibinfo {author} {\bibfnamefont {J.}~\bibnamefont {Cai}}, \bibinfo
  {author} {\bibfnamefont {B.}~\bibnamefont {Shen}}, \bibinfo {author}
  {\bibfnamefont {H.}~\bibnamefont {Lei}}, \bibinfo {author} {\bibfnamefont
  {Y.}~\bibnamefont {Zhang}},\ and\ \bibinfo {author} {\bibfnamefont
  {S.}~\bibnamefont {Wang}},\ }\href {https://doi.org/10.1103/PhysRevB.105.014426} {\bibfield  {journal}
  {\bibinfo  {journal} {Phys. Rev. B}\ }\textbf {\bibinfo {volume} {105}},\
  \bibinfo {pages} {014426} (\bibinfo {year} {2022})}\BibitemShut {NoStop}%
\bibitem [{\citenamefont {Sau}\ \emph {et~al.}(2022)\citenamefont {Sau},
  \citenamefont {Hassan}, \citenamefont {Kumar},\ and\ \citenamefont
  {Kumar}}]{manoranjanpaper}%
  \BibitemOpen
  \bibfield  {author} {\bibinfo {author} {\bibfnamefont {J.}~\bibnamefont
  {Sau}}, \bibinfo {author} {\bibfnamefont {S.~R.}\ \bibnamefont {Hassan}},
  \bibinfo {author} {\bibfnamefont {N.}~\bibnamefont {Kumar}},\ and\ \bibinfo
  {author} {\bibfnamefont {M.}~\bibnamefont {Kumar}},\ }\href@noop {} (\bibinfo {year} {2022}),\ \Eprint
  {https://arxiv.org/abs/2207.03547} {arXiv:2207.03547 [cond-mat.str-el]}
  \BibitemShut {NoStop}%
\bibitem [{sup()}]{supply}%
  \BibitemOpen
  \href@noop {} {}\bibinfo {note} {See the Supplemental Material which includes
  Refs  \cite{fleur,vasp,burke,perdew,monk,pizzi2020wannier90,marzariPhysRevB.56.12847,kubo,wu2018wanniertools},
  for experimental method, {XRD}, \textit{ab initio} electronic structure
  calculations and {A}nomalous {H}all {C}onductivity {(AHC)}
  simulations.}\BibitemShut {Stop}%
\bibitem [{\citenamefont {Bera}\ \emph {et~al.}(2023)\citenamefont {Bera},
  \citenamefont {Pradhan}, \citenamefont {Khan}, \citenamefont {Pal},
  \citenamefont {Pal}, \citenamefont {Kalimuddin}, \citenamefont {Bera},
  \citenamefont {Das}, \citenamefont {Pal},\ and\ \citenamefont
  {Mondal}}]{BERA2023170257}%
  \BibitemOpen
  \bibfield  {author} {\bibinfo {author} {\bibfnamefont {S.}~\bibnamefont
  {Bera}}, \bibinfo {author} {\bibfnamefont {S.~K.}\ \bibnamefont {Pradhan}},
  \bibinfo {author} {\bibfnamefont {M.~S.}\ \bibnamefont {Khan}}, \bibinfo
  {author} {\bibfnamefont {R.}~\bibnamefont {Pal}}, \bibinfo {author}
  {\bibfnamefont {B.}~\bibnamefont {Pal}}, \bibinfo {author} {\bibfnamefont
  {S.}~\bibnamefont {Kalimuddin}}, \bibinfo {author} {\bibfnamefont
  {A.}~\bibnamefont {Bera}}, \bibinfo {author} {\bibfnamefont {B.}~\bibnamefont
  {Das}}, \bibinfo {author} {\bibfnamefont {A.~N.}\ \bibnamefont {Pal}},\ and\
  \bibinfo {author} {\bibfnamefont {M.}~\bibnamefont {Mondal}},\ }\href  {https://doi.org/https://doi.org/10.1016/j.jmmm.2022.170257} {\bibfield
  {journal} {\bibinfo  {journal} {Journal of Magnetism and Magnetic Materials}\
  }\textbf {\bibinfo {volume} {565}},\ \bibinfo {pages} {170257} (\bibinfo
  {year} {2023})}\BibitemShut {NoStop}%
\bibitem [{\citenamefont {Pal}\ \emph {et~al.}(2023)\citenamefont {Pal},
  \citenamefont {Pal}, \citenamefont {Mondal}, \citenamefont {Mandal},\ and\
  \citenamefont {Pal}}]{pal2023unusual}%
  \BibitemOpen
  \bibfield  {author} {\bibinfo {author} {\bibfnamefont {R.}~\bibnamefont
  {Pal}}, \bibinfo {author} {\bibfnamefont {B.}~\bibnamefont {Pal}}, \bibinfo
  {author} {\bibfnamefont {S.}~\bibnamefont {Mondal}}, \bibinfo {author}
  {\bibfnamefont {P.}~\bibnamefont {Mandal}},\ and\ \bibinfo {author}
  {\bibfnamefont {A.~N.}\ \bibnamefont {Pal}},\ }\href@noop {} {\bibfield  {journal} {\bibinfo  {journal} {arXiv preprint
  arXiv:2303.07440}\ } (\bibinfo {year} {2023})}\BibitemShut {NoStop}%
\bibitem [{\citenamefont {Roy}\ \emph {et~al.}(2020)\citenamefont {Roy},
  \citenamefont {Singha}, \citenamefont {Ghosh}, \citenamefont {Pariari},\ and\
  \citenamefont {Mandal}}]{royPhysRevB.102.085147}%
  \BibitemOpen
  \bibfield  {author} {\bibinfo {author} {\bibfnamefont {S.}~\bibnamefont
  {Roy}}, \bibinfo {author} {\bibfnamefont {R.}~\bibnamefont {Singha}},
  \bibinfo {author} {\bibfnamefont {A.}~\bibnamefont {Ghosh}}, \bibinfo
  {author} {\bibfnamefont {A.}~\bibnamefont {Pariari}},\ and\ \bibinfo {author}
  {\bibfnamefont {P.}~\bibnamefont {Mandal}},\ }\href {https://doi.org/10.1103/PhysRevB.102.085147} {\bibfield  {journal} {\bibinfo
   {journal} {Phys. Rev. B}\ }\textbf {\bibinfo {volume} {102}},\ \bibinfo
  {pages} {085147} (\bibinfo {year} {2020})}\BibitemShut {NoStop}%
\bibitem [{\citenamefont {Chatterjee}\ \emph {et~al.}(2022)\citenamefont
  {Chatterjee}, \citenamefont {Sau}, \citenamefont {Ghosh}, \citenamefont
  {Samanta}, \citenamefont {Ghosh}, \citenamefont {Kumar},\ and\ \citenamefont
  {Mandal}}]{Chatterjee_2023}%
  \BibitemOpen
  \bibfield  {author} {\bibinfo {author} {\bibfnamefont {S.}~\bibnamefont
  {Chatterjee}}, \bibinfo {author} {\bibfnamefont {J.}~\bibnamefont {Sau}},
  \bibinfo {author} {\bibfnamefont {S.}~\bibnamefont {Ghosh}}, \bibinfo
  {author} {\bibfnamefont {S.}~\bibnamefont {Samanta}}, \bibinfo {author}
  {\bibfnamefont {B.}~\bibnamefont {Ghosh}}, \bibinfo {author} {\bibfnamefont
  {M.}~\bibnamefont {Kumar}},\ and\ \bibinfo {author} {\bibfnamefont
  {K.}~\bibnamefont {Mandal}},\ }\href {https://doi.org/10.1088/1361-648X/aca0d7}
  {\bibfield  {journal} {\bibinfo  {journal} {Journal of Physics: Condensed
  Matter}\ }\textbf {\bibinfo {volume} {35}},\ \bibinfo {pages} {035601}
  (\bibinfo {year} {2022})}\BibitemShut {NoStop}%
\bibitem [{\citenamefont {Nozieres}\ and\ \citenamefont
  {Lewiner}(1973)}]{Nozieres1973simple}%
  \BibitemOpen
  \bibfield  {author} {\bibinfo {author} {\bibfnamefont {P.}~\bibnamefont
  {Nozieres}}\ and\ \bibinfo {author} {\bibfnamefont {C.}~\bibnamefont
  {Lewiner}},\ }\href  {https://doi.org/10.1051/jphys:019730034010090100} {\bibfield  {journal}
  {\bibinfo  {journal} {Journal de Physique}\ }\textbf {\bibinfo {volume}
  {34}},\ \bibinfo {pages} {901} (\bibinfo {year} {1973})}\BibitemShut
  {NoStop}%
\bibitem [{\citenamefont {Onoda}\ \emph {et~al.}(2006)\citenamefont {Onoda},
  \citenamefont {Sugimoto},\ and\ \citenamefont
  {Nagaosa}}]{OnodaPhysRevLett.97.126602}%
  \BibitemOpen
  \bibfield  {author} {\bibinfo {author} {\bibfnamefont {S.}~\bibnamefont
  {Onoda}}, \bibinfo {author} {\bibfnamefont {N.}~\bibnamefont {Sugimoto}},\
  and\ \bibinfo {author} {\bibfnamefont {N.}~\bibnamefont {Nagaosa}},\
  }\href
{https://doi.org/10.1103/PhysRevLett.97.126602} {\bibfield  {journal}
  {\bibinfo  {journal} {Phys. Rev. Lett.}\ }\textbf {\bibinfo {volume} {97}},\
  \bibinfo {pages} {126602} (\bibinfo {year} {2006})}\BibitemShut {NoStop}%
\bibitem [{\citenamefont {Miyasato}\ \emph {et~al.}(2007)\citenamefont {Abe},
  \citenamefont {Fujii}, \citenamefont {Asamitsu}, \citenamefont {Onoda}, \citenamefont {Onoda},\citenamefont {Nagaosa}, \ and\
  \citenamefont {Tokura}}]{PhysRevLett.99.086602}%
  \BibitemOpen
  \bibfield  {author} {\bibinfo {author} {\bibfnamefont {T.}~\bibnamefont
  {Miyasato}}, \bibinfo {author} {\bibfnamefont {N.}~\bibnamefont {Abe}}, \bibinfo
  {author} {\bibfnamefont {T.}\ \bibnamefont {Fujii}}, \bibinfo {author}
  {\bibfnamefont {A.}~\bibnamefont {Asamitsu}}, \bibinfo {author}
  {\bibfnamefont {S.}~\bibnamefont {Onoda}}, \bibinfo {author}
  {\bibfnamefont {Y.}~\bibnamefont {Onoda}}, \bibinfo {author}
  {\bibfnamefont {N.}~\bibnamefont {Nagaosa}},\ and\ \bibinfo {author}
  {\bibfnamefont {Y.}\ \bibnamefont {Tokura}},\ }\href {https://link.aps.org/doi/10.1103/PhysRevLett.99.086602}
  {\bibfield  {journal} {\bibinfo  {journal} {Phys. Rev. Lett.}\
  }\textbf {\bibinfo {volume} {99}},\ \bibinfo {pages} {086602} (\bibinfo {year}
  {2007})}\BibitemShut {NoStop}%
\bibitem [{\citenamefont {Wang}\ \emph {et~al.}(2016)\citenamefont {Wang},
  \citenamefont {Sun}, \citenamefont {Zhang}, \citenamefont {Pang},\ and\
  \citenamefont {Lei}}]{PhysRevB.94.075135}%
  \BibitemOpen
  \bibfield  {author} {\bibinfo {author} {\bibfnamefont {Q.}~\bibnamefont
  {Wang}}, \bibinfo {author} {\bibfnamefont {S.}~\bibnamefont {Sun}}, \bibinfo
  {author} {\bibfnamefont {X.}~\bibnamefont {Zhang}}, \bibinfo {author}
  {\bibfnamefont {F.}~\bibnamefont {Pang}},\ and\ \bibinfo {author}
  {\bibfnamefont {H.}~\bibnamefont {Lei}},\ }\href  {https://doi.org/10.1103/PhysRevB.94.075135} {\bibfield  {journal} {\bibinfo
  {journal} {Phys. Rev. B}\ }\textbf {\bibinfo {volume} {94}},\ \bibinfo
  {pages} {075135} (\bibinfo {year} {2016})}\BibitemShut {NoStop}%
\bibitem [{\citenamefont {Chatterjee}\ \emph {et~al.}(2023)\citenamefont {Sau},
  \citenamefont {Samanta}, \citenamefont {Ghosh}, \citenamefont {Kumar}, \citenamefont {Kumar}, \ and\
  \citenamefont {Mandal}}]{PhysRevB.107.125138}%
  \BibitemOpen
  \bibfield  {author} {\bibinfo {author} {\bibfnamefont {S.}~\bibnamefont
  {Chatterjee}}, \bibinfo {author} {\bibfnamefont {J.}~\bibnamefont {Sau}}, \bibinfo
  {author} {\bibfnamefont {S.}\ \bibnamefont {Samanta}}, \bibinfo {author}
  {\bibfnamefont {B.}~\bibnamefont {Ghosh}}, \bibinfo {author}
  {\bibfnamefont {N.}~\bibnamefont {Kumar}}, \bibinfo {author}
  {\bibfnamefont {M.}~\bibnamefont {Kumar}},\ and\ \bibinfo {author}
  {\bibfnamefont {K.}\ \bibnamefont {Mandal}},\ }\href {https://link.aps.org/doi/10.1103/PhysRevB.107.125138}
  {\bibfield  {journal} {\bibinfo  {journal} {Phys. Rev. B}\
  }\textbf {\bibinfo {volume} {107}},\ \bibinfo {pages} {125138} (\bibinfo {year}
  {2023})}\BibitemShut {NoStop}%
\bibitem [{\citenamefont {Lee}\ \emph {et~al.}(2007)\citenamefont {Lee},
  \citenamefont {Onose}, \citenamefont {Tokura},\ and\ \citenamefont
  {Ong}}]{PhysRevB.75.172403}%
  \BibitemOpen
  \bibfield  {author} {\bibinfo {author} {\bibfnamefont {M.}~\bibnamefont
  {Lee}}, \bibinfo {author} {\bibfnamefont {Y.}~\bibnamefont {Onose}}, \bibinfo
  {author} {\bibfnamefont {Y.}~\bibnamefont {Tokura}},\ and\ \bibinfo {author}
  {\bibfnamefont {N.~P.}\ \bibnamefont {Ong}},\ }\href {https://doi.org/10.1103/PhysRevB.75.172403} {\bibfield
  {journal} {\bibinfo  {journal} {Phys. Rev. B}\ }\textbf {\bibinfo {volume}
  {75}},\ \bibinfo {pages} {172403} (\bibinfo {year} {2007})}\BibitemShut
  {NoStop}%
\bibitem [{\citenamefont {Sangiao}\ \emph {et~al.}(2009)\citenamefont
  {Sangiao}, \citenamefont {Morellon}, \citenamefont {Simon}, \citenamefont
  {De~Teresa}, \citenamefont {Pardo}, \citenamefont {Arbiol},\ and\
  \citenamefont {Ibarra}}]{PhysRevB.79.014431}%
  \BibitemOpen
  \bibfield  {author} {\bibinfo {author} {\bibfnamefont {S.}~\bibnamefont
  {Sangiao}}, \bibinfo {author} {\bibfnamefont {L.}~\bibnamefont {Morellon}},
  \bibinfo {author} {\bibfnamefont {G.}~\bibnamefont {Simon}}, \bibinfo
  {author} {\bibfnamefont {J.~M.}\ \bibnamefont {De~Teresa}}, \bibinfo {author}
  {\bibfnamefont {J.~A.}\ \bibnamefont {Pardo}}, \bibinfo {author}
  {\bibfnamefont {J.}~\bibnamefont {Arbiol}},\ and\ \bibinfo {author}
  {\bibfnamefont {M.~R.}\ \bibnamefont {Ibarra}},\ }\href
{https://doi.org/10.1103/PhysRevB.79.014431} {\bibfield  {journal} {\bibinfo
  {journal} {Phys. Rev. B}\ }\textbf {\bibinfo {volume} {79}},\ \bibinfo
  {pages} {014431} (\bibinfo {year} {2009})}\BibitemShut {NoStop}%
\bibitem [{\citenamefont {S\"urgers}\ \emph {et~al.}(2014)\citenamefont
  {S\"urgers}, \citenamefont {Fischer}, \citenamefont {Winkel},\ and\
  \citenamefont {L\"ohneysen}}]{PhysRevB.90.104421}%
  \BibitemOpen
  \bibfield  {author} {\bibinfo {author} {\bibfnamefont {C.}~\bibnamefont
  {S\"urgers}}, \bibinfo {author} {\bibfnamefont {G.}~\bibnamefont {Fischer}},
  \bibinfo {author} {\bibfnamefont {P.}~\bibnamefont {Winkel}},\ and\ \bibinfo
  {author} {\bibfnamefont {H.~v.}\ \bibnamefont {L\"ohneysen}},\ }\href {https://doi.org/10.1103/PhysRevB.90.104421} {\bibfield  {journal}
  {\bibinfo  {journal} {Phys. Rev. B}\ }\textbf {\bibinfo {volume} {90}},\
  \bibinfo {pages} {104421} (\bibinfo {year} {2014})}\BibitemShut {NoStop}%
\bibitem [{\citenamefont {Baily}\ and\ \citenamefont
  {Salamon}(2005)}]{PhysRevB.71.104407}%
  \BibitemOpen
  \bibfield  {author} {\bibinfo {author} {\bibfnamefont {S.~A.}\ \bibnamefont
  {Baily}}\ and\ \bibinfo {author} {\bibfnamefont {M.~B.}\ \bibnamefont
  {Salamon}},\ }\href {https://doi.org/10.1103/PhysRevB.71.104407} {\bibfield  {journal} {\bibinfo
  {journal} {Phys. Rev. B}\ }\textbf {\bibinfo {volume} {71}},\ \bibinfo
  {pages} {104407} (\bibinfo {year} {2005})}\BibitemShut {NoStop}%
\bibitem [{\citenamefont {Onose}\ and\ \citenamefont
  {Tokura}(2006)}]{PhysRevB.73.174421}%
  \BibitemOpen
  \bibfield  {author} {\bibinfo {author} {\bibfnamefont {Y.}~\bibnamefont
  {Onose}}\ and\ \bibinfo {author} {\bibfnamefont {Y.}~\bibnamefont {Tokura}},\
  }\href {https://doi.org/10.1103/PhysRevB.73.174421} {\bibfield  {journal}
  {\bibinfo  {journal} {Phys. Rev. B}\ }\textbf {\bibinfo {volume} {73}},\
  \bibinfo {pages} {174421} (\bibinfo {year} {2006})}\BibitemShut {NoStop}%
\bibitem [{fle()}]{fleur}%
  \BibitemOpen
  \href {https://www.flapw.de/MaX-6.0/} {\bibinfo {title}
  {{www.flapw.de}}}\BibitemShut {NoStop}%
\bibitem [{\citenamefont {Kresse}\ and\ \citenamefont {Joubert}(1999)}]{vasp}%
  \BibitemOpen
  \bibfield  {author} {\bibinfo {author} {\bibfnamefont {G.}~\bibnamefont
  {Kresse}}\ and\ \bibinfo {author} {\bibfnamefont {D.}~\bibnamefont
  {Joubert}},\ }\href  {https://doi.org/10.1103/PhysRevB.59.1758} {\bibfield  {journal} {\bibinfo
  {journal} {Phys. Rev. B}\ }\textbf {\bibinfo {volume} {59}},\ \bibinfo
  {pages} {1758} (\bibinfo {year} {1999})}\BibitemShut {NoStop}%
\bibitem [{\citenamefont {Perdew}\ \emph {et~al.}(1996)\citenamefont {Perdew},
  \citenamefont {Burke},\ and\ \citenamefont {Ernzerhof}}]{burke}%
  \BibitemOpen
  \bibfield  {author} {\bibinfo {author} {\bibfnamefont {J.~P.}\ \bibnamefont
  {Perdew}}, \bibinfo {author} {\bibfnamefont {K.}~\bibnamefont {Burke}},\ and\
  \bibinfo {author} {\bibfnamefont {M.}~\bibnamefont {Ernzerhof}},\ }\href {https://doi.org/10.1103/PhysRevLett.77.3865} {\bibfield  {journal}
  {\bibinfo  {journal} {Phys. Rev. Lett.}\ }\textbf {\bibinfo {volume} {77}},\
  \bibinfo {pages} {3865} (\bibinfo {year} {1996})}\BibitemShut {NoStop}%
\bibitem [{\citenamefont {Perdew}\ and\ \citenamefont {Wang}(1992)}]{perdew}%
  \BibitemOpen
  \bibfield  {author} {\bibinfo {author} {\bibfnamefont {J.~P.}\ \bibnamefont
  {Perdew}}\ and\ \bibinfo {author} {\bibfnamefont {Y.}~\bibnamefont {Wang}},\
  }\href  {https://doi.org/10.1103/PhysRevB.45.13244} {\bibfield  {journal} {\bibinfo
  {journal} {Phys. Rev. B}\ }\textbf {\bibinfo {volume} {45}},\ \bibinfo
  {pages} {13244} (\bibinfo {year} {1992})}\BibitemShut {NoStop}%
\bibitem [{\citenamefont {Monkhorst}\ and\ \citenamefont {Pack}(1976)}]{monk}%
  \BibitemOpen
  \bibfield  {author} {\bibinfo {author} {\bibfnamefont {H.~J.}\ \bibnamefont
  {Monkhorst}}\ and\ \bibinfo {author} {\bibfnamefont {J.~D.}\ \bibnamefont
  {Pack}},\ }\href
 {https://doi.org/10.1103/PhysRevB.13.5188} {\bibfield  {journal} {\bibinfo
  {journal} {Phys. Rev. B}\ }\textbf {\bibinfo {volume} {13}},\ \bibinfo
  {pages} {5188} (\bibinfo {year} {1976})}\BibitemShut {NoStop}%
\bibitem [{\citenamefont {Pizzi}\ \emph {et~al.}(2020)\citenamefont {Pizzi},
  \citenamefont {Vitale}, \citenamefont {Arita}, \citenamefont {Bl{\"u}gel},
  \citenamefont {Freimuth}, \citenamefont {G{\'e}ranton}, \citenamefont
  {Gibertini}, \citenamefont {Gresch}, \citenamefont {Johnson}, \citenamefont
  {Koretsune} \emph {et~al.}}]{pizzi2020wannier90}%
  \BibitemOpen
  \bibfield  {author} {\bibinfo {author} {\bibfnamefont {G.}~\bibnamefont
  {Pizzi}}, \bibinfo {author} {\bibfnamefont {V.}~\bibnamefont {Vitale}},
  \bibinfo {author} {\bibfnamefont {R.}~\bibnamefont {Arita}}, \bibinfo
  {author} {\bibfnamefont {S.}~\bibnamefont {Bl{\"u}gel}}, \bibinfo {author}
  {\bibfnamefont {F.}~\bibnamefont {Freimuth}}, \bibinfo {author}
  {\bibfnamefont {G.}~\bibnamefont {G{\'e}ranton}}, \bibinfo {author}
  {\bibfnamefont {M.}~\bibnamefont {Gibertini}}, \bibinfo {author}
  {\bibfnamefont {D.}~\bibnamefont {Gresch}}, \bibinfo {author} {\bibfnamefont
  {C.}~\bibnamefont {Johnson}}, \bibinfo {author} {\bibfnamefont
  {T.}~\bibnamefont {Koretsune}}, \emph {et~al.},\ }\href {https://doi.org/10.1088/1361-648X/ab51ff} {\bibfield
  {journal} {\bibinfo  {journal} {Journal of Physics: Condensed Matter}\
  }\textbf {\bibinfo {volume} {32}},\ \bibinfo {pages} {165902} (\bibinfo
  {year} {2020})}\BibitemShut {NoStop}%
\bibitem [{\citenamefont {Marzari}\ and\ \citenamefont
  {Vanderbilt}(1997)}]{marzariPhysRevB.56.12847}%
  \BibitemOpen
  \bibfield  {author} {\bibinfo {author} {\bibfnamefont {N.}~\bibnamefont
  {Marzari}}\ and\ \bibinfo {author} {\bibfnamefont {D.}~\bibnamefont
  {Vanderbilt}},\ }\href
{https://doi.org/10.1103/PhysRevB.56.12847} {\bibfield  {journal} {\bibinfo
  {journal} {Phys. Rev. B}\ }\textbf {\bibinfo {volume} {56}},\ \bibinfo
  {pages} {12847} (\bibinfo {year} {1997})}\BibitemShut {NoStop}%
\bibitem [{\citenamefont {Yao}\ \emph {et~al.}(2004)\citenamefont {Yao},
  \citenamefont {Kleinman}, \citenamefont {MacDonald}, \citenamefont {Sinova},
  \citenamefont {Jungwirth}, \citenamefont {Wang}, \citenamefont {Wang},\ and\
  \citenamefont {Niu}}]{kubo}%
  \BibitemOpen
  \bibfield  {author} {\bibinfo {author} {\bibfnamefont {Y.}~\bibnamefont
  {Yao}}, \bibinfo {author} {\bibfnamefont {L.}~\bibnamefont {Kleinman}},
  \bibinfo {author} {\bibfnamefont {A.~H.}\ \bibnamefont {MacDonald}}, \bibinfo
  {author} {\bibfnamefont {J.}~\bibnamefont {Sinova}}, \bibinfo {author}
  {\bibfnamefont {T.}~\bibnamefont {Jungwirth}}, \bibinfo {author}
  {\bibfnamefont {D.-s.}\ \bibnamefont {Wang}}, \bibinfo {author}
  {\bibfnamefont {E.}~\bibnamefont {Wang}},\ and\ \bibinfo {author}
  {\bibfnamefont {Q.}~\bibnamefont {Niu}},\ }\href{https://doi.org/10.1103/PhysRevLett.92.037204} {\bibfield  {journal}
  {\bibinfo  {journal} {Phys. Rev. Lett.}\ }\textbf {\bibinfo {volume} {92}},\
  \bibinfo {pages} {037204} (\bibinfo {year} {2004})}\BibitemShut {NoStop}%
\bibitem [{\citenamefont {Wu}\ \emph {et~al.}(2018)\citenamefont {Wu},
  \citenamefont {Zhang}, \citenamefont {Song}, \citenamefont {Troyer},\ and\
  \citenamefont {Soluyanov}}]{wu2018wanniertools}%
  \BibitemOpen
  \bibfield  {author} {\bibinfo {author} {\bibfnamefont {Q.}~\bibnamefont
  {Wu}}, \bibinfo {author} {\bibfnamefont {S.}~\bibnamefont {Zhang}}, \bibinfo
  {author} {\bibfnamefont {H.-F.}\ \bibnamefont {Song}}, \bibinfo {author}
  {\bibfnamefont {M.}~\bibnamefont {Troyer}},\ and\ \bibinfo {author}
  {\bibfnamefont {A.~A.}\ \bibnamefont {Soluyanov}},\ }\href {https://doi.org/10.1016/j.cpc.2017.09.033}
  {\bibfield  {journal} {\bibinfo  {journal} {Computer Physics Communications}\
  }\textbf {\bibinfo {volume} {224}},\ \bibinfo {pages} {405} (\bibinfo {year}
  {2018})}\BibitemShut {NoStop}%
\end{thebibliography}

\begin{thebibliography}{22}%
\makeatletter
\providecommand \@ifxundefined [1]{%
 \@ifx{#1\undefined}
}%
\providecommand \@ifnum [1]{%
 \ifnum #1\expandafter \@firstoftwo
 \else \expandafter \@secondoftwo
 \fi
}%
\providecommand \@ifx [1]{%
 \ifx #1\expandafter \@firstoftwo
 \else \expandafter \@secondoftwo
 \fi
}%
\providecommand \natexlab [1]{#1}%
\providecommand \enquote  [1]{``#1''}%
\providecommand \bibnamefont  [1]{#1}%
\providecommand \bibfnamefont [1]{#1}%
\providecommand \citenamefont [1]{#1}%
\providecommand \href@noop [0]{\@secondoftwo}%
\providecommand \href [0]{\begingroup \@sanitize@url \@href}%
\providecommand \@href[1]{\@@startlink{#1}\@@href}%
\providecommand \@@href[1]{\endgroup#1\@@endlink}%
\providecommand \@sanitize@url [0]{\catcode `\\12\catcode `\$12\catcode
  `\&12\catcode `\#12\catcode `\^12\catcode `\_12\catcode `\%12\relax}%
\providecommand \@@startlink[1]{}%
\providecommand \@@endlink[0]{}%
\providecommand \url  [0]{\begingroup\@sanitize@url \@url }%
\providecommand \@url [1]{\endgroup\@href {#1}{\urlprefix }}%
\providecommand \urlprefix  [0]{URL }%
\providecommand \Eprint [0]{\href }%
\providecommand \doibase [0]{https://doi.org/}%
\providecommand \selectlanguage [0]{\@gobble}%
\providecommand \bibinfo  [0]{\@secondoftwo}%
\providecommand \bibfield  [0]{\@secondoftwo}%
\providecommand \translation [1]{[#1]}%
\providecommand \BibitemOpen [0]{}%
\providecommand \bibitemStop [0]{}%
\providecommand \bibitemNoStop [0]{.\EOS\space}%
\providecommand \EOS [0]{\spacefactor3000\relax}%
\providecommand \BibitemShut  [1]{\csname bibitem#1\endcsname}%
\let\auto@bib@innerbib\@empty
\bibitem [{\citenamefont {Bera}\ \emph {et~al.}(2023)\citenamefont {Bera},
  \citenamefont {Pradhan}, \citenamefont {Khan}, \citenamefont {Pal},
  \citenamefont {Pal}, \citenamefont {Kalimuddin}, \citenamefont {Bera},
  \citenamefont {Das}, \citenamefont {Pal},\ and\ \citenamefont
  {Mondal}}]{SupBERA2023170257}%
  \BibitemOpen
  \bibfield  {author} {\bibinfo {author} {\bibfnamefont {S.}~\bibnamefont
  {Bera}}, \bibinfo {author} {\bibfnamefont {S.~K.}\ \bibnamefont {Pradhan}},
  \bibinfo {author} {\bibfnamefont {M.~S.}\ \bibnamefont {Khan}}, \bibinfo
  {author} {\bibfnamefont {R.}~\bibnamefont {Pal}}, \bibinfo {author}
  {\bibfnamefont {B.}~\bibnamefont {Pal}}, \bibinfo {author} {\bibfnamefont
  {S.}~\bibnamefont {Kalimuddin}}, \bibinfo {author} {\bibfnamefont
  {A.}~\bibnamefont {Bera}}, \bibinfo {author} {\bibfnamefont {B.}~\bibnamefont
  {Das}}, \bibinfo {author} {\bibfnamefont {A.~N.}\ \bibnamefont {Pal}},\ and\
  \bibinfo {author} {\bibfnamefont {M.}~\bibnamefont {Mondal}},\ }\href
  {https://doi.org/https://doi.org/10.1016/j.jmmm.2022.170257} {\bibfield
  {journal} {\bibinfo  {journal} {Journal of Magnetism and Magnetic Materials}\
  }\textbf {\bibinfo {volume} {565}},\ \bibinfo {pages} {170257} (\bibinfo
  {year} {2023})}\BibitemShut {NoStop}%
\bibitem [{\citenamefont {Mondal}\ \emph {et~al.}(2021)\citenamefont {Mondal},
  \citenamefont {Khan}, \citenamefont {Mishra}, \citenamefont {Satpati},\ and\
  \citenamefont {Mandal}}]{SupmondalPhysRevB.104.094405}%
  \BibitemOpen
  \bibfield  {author} {\bibinfo {author} {\bibfnamefont {S.}~\bibnamefont
  {Mondal}}, \bibinfo {author} {\bibfnamefont {N.}~\bibnamefont {Khan}},
  \bibinfo {author} {\bibfnamefont {S.~M.}\ \bibnamefont {Mishra}}, \bibinfo
  {author} {\bibfnamefont {B.}~\bibnamefont {Satpati}},\ and\ \bibinfo {author}
  {\bibfnamefont {P.}~\bibnamefont {Mandal}},\ }\href
  {https://doi.org/10.1103/PhysRevB.104.094405} {\bibfield  {journal} {\bibinfo
   {journal} {Phys. Rev. B}\ }\textbf {\bibinfo {volume} {104}},\ \bibinfo
  {pages} {094405} (\bibinfo {year} {2021})}\BibitemShut {NoStop}%
\bibitem [{\citenamefont {May}\ \emph {et~al.}(2019)\citenamefont {May},
  \citenamefont {Ovchinnikov}, \citenamefont {Zheng}, \citenamefont {Hermann},
  \citenamefont {Calder}, \citenamefont {Huang}, \citenamefont {Fei},
  \citenamefont {Liu}, \citenamefont {Xu},\ and\ \citenamefont
  {McGuire}}]{May2019}%
  \BibitemOpen
  \bibfield  {author} {\bibinfo {author} {\bibfnamefont {A.~F.}\ \bibnamefont
  {May}}, \bibinfo {author} {\bibfnamefont {D.}~\bibnamefont {Ovchinnikov}},
  \bibinfo {author} {\bibfnamefont {Q.}~\bibnamefont {Zheng}}, \bibinfo
  {author} {\bibfnamefont {R.}~\bibnamefont {Hermann}}, \bibinfo {author}
  {\bibfnamefont {S.}~\bibnamefont {Calder}}, \bibinfo {author} {\bibfnamefont
  {B.}~\bibnamefont {Huang}}, \bibinfo {author} {\bibfnamefont
  {Z.}~\bibnamefont {Fei}}, \bibinfo {author} {\bibfnamefont {Y.}~\bibnamefont
  {Liu}}, \bibinfo {author} {\bibfnamefont {X.}~\bibnamefont {Xu}},\ and\
  \bibinfo {author} {\bibfnamefont {M.~A.}\ \bibnamefont {McGuire}},\ }\href
  {https://doi.org/10.1021/acsnano.8b09660} {\bibfield  {journal} {\bibinfo
  {journal} {ACS Nano}\ }\textbf {\bibinfo {volume} {13}},\ \bibinfo {pages}
  {4436} (\bibinfo {year} {2019})}\BibitemShut {NoStop}%
\bibitem [{\citenamefont {May}\ \emph {et~al.}(2020)\citenamefont {May},
  \citenamefont {Du}, \citenamefont {Cooper},\ and\ \citenamefont
  {McGuire}}]{PhysRevMaterials.4.074008}%
  \BibitemOpen
  \bibfield  {author} {\bibinfo {author} {\bibfnamefont {A.~F.}\ \bibnamefont
  {May}}, \bibinfo {author} {\bibfnamefont {M.-H.}\ \bibnamefont {Du}},
  \bibinfo {author} {\bibfnamefont {V.~R.}\ \bibnamefont {Cooper}},\ and\
  \bibinfo {author} {\bibfnamefont {M.~A.}\ \bibnamefont {McGuire}},\ }\href
  {https://doi.org/10.1103/PhysRevMaterials.4.074008} {\bibfield  {journal}
  {\bibinfo  {journal} {Phys. Rev. Mater.}\ }\textbf {\bibinfo {volume} {4}},\
  \bibinfo {pages} {074008} (\bibinfo {year} {2020})}\BibitemShut {NoStop}%
\bibitem [{\citenamefont {Alghamdi}\ \emph {et~al.}(2019)\citenamefont
  {Alghamdi}, \citenamefont {Lohmann}, \citenamefont {Li}, \citenamefont
  {Jothi}, \citenamefont {Shao}, \citenamefont {Aldosary}, \citenamefont {Su},
  \citenamefont {Fokwa},\ and\ \citenamefont {Shi}}]{Alghamdi2019}%
  \BibitemOpen
  \bibfield  {author} {\bibinfo {author} {\bibfnamefont {M.}~\bibnamefont
  {Alghamdi}}, \bibinfo {author} {\bibfnamefont {M.}~\bibnamefont {Lohmann}},
  \bibinfo {author} {\bibfnamefont {J.}~\bibnamefont {Li}}, \bibinfo {author}
  {\bibfnamefont {P.~R.}\ \bibnamefont {Jothi}}, \bibinfo {author}
  {\bibfnamefont {Q.}~\bibnamefont {Shao}}, \bibinfo {author} {\bibfnamefont
  {M.}~\bibnamefont {Aldosary}}, \bibinfo {author} {\bibfnamefont
  {T.}~\bibnamefont {Su}}, \bibinfo {author} {\bibfnamefont {B.~P.~T.}\
  \bibnamefont {Fokwa}},\ and\ \bibinfo {author} {\bibfnamefont
  {J.}~\bibnamefont {Shi}},\ }\href
  {https://doi.org/10.1021/acs.nanolett.9b01043} {\bibfield  {journal}
  {\bibinfo  {journal} {Nano Lett.}\ }\textbf {\bibinfo {volume} {19}},\
  \bibinfo {pages} {4400} (\bibinfo {year} {2019})}\BibitemShut {NoStop}%
\bibitem [{\citenamefont {Wang}\ \emph {et~al.}(2017)\citenamefont {Wang},
  \citenamefont {Xian}, \citenamefont {Wang}, \citenamefont {Liu},
  \citenamefont {Ling}, \citenamefont {Zhang}, \citenamefont {Cao},
  \citenamefont {Qu},\ and\ \citenamefont {Xiong}}]{PhysRevB.96.134428}%
  \BibitemOpen
  \bibfield  {author} {\bibinfo {author} {\bibfnamefont {Y.}~\bibnamefont
  {Wang}}, \bibinfo {author} {\bibfnamefont {C.}~\bibnamefont {Xian}}, \bibinfo
  {author} {\bibfnamefont {J.}~\bibnamefont {Wang}}, \bibinfo {author}
  {\bibfnamefont {B.}~\bibnamefont {Liu}}, \bibinfo {author} {\bibfnamefont
  {L.}~\bibnamefont {Ling}}, \bibinfo {author} {\bibfnamefont {L.}~\bibnamefont
  {Zhang}}, \bibinfo {author} {\bibfnamefont {L.}~\bibnamefont {Cao}}, \bibinfo
  {author} {\bibfnamefont {Z.}~\bibnamefont {Qu}},\ and\ \bibinfo {author}
  {\bibfnamefont {Y.}~\bibnamefont {Xiong}},\ }\href
  {https://doi.org/10.1103/PhysRevB.96.134428} {\bibfield  {journal} {\bibinfo
  {journal} {Phys. Rev. B}\ }\textbf {\bibinfo {volume} {96}},\ \bibinfo
  {pages} {134428} (\bibinfo {year} {2017})}\BibitemShut {NoStop}%
\bibitem [{\citenamefont {Roy}\ \emph {et~al.}(2020{\natexlab{a}})\citenamefont
  {Roy}, \citenamefont {Singha}, \citenamefont {Ghosh}, \citenamefont
  {Pariari},\ and\ \citenamefont {Mandal}}]{Roy2020}%
  \BibitemOpen
  \bibfield  {author} {\bibinfo {author} {\bibfnamefont {S.}~\bibnamefont
  {Roy}}, \bibinfo {author} {\bibfnamefont {R.}~\bibnamefont {Singha}},
  \bibinfo {author} {\bibfnamefont {A.}~\bibnamefont {Ghosh}}, \bibinfo
  {author} {\bibfnamefont {A.}~\bibnamefont {Pariari}},\ and\ \bibinfo {author}
  {\bibfnamefont {P.}~\bibnamefont {Mandal}},\ }\href
  {https://doi.org/10.1103/PhysRevB.102.085147} {\bibfield  {journal} {\bibinfo
   {journal} {Phys. Rev. B}\ }\textbf {\bibinfo {volume} {102}},\ \bibinfo
  {pages} {085147} (\bibinfo {year} {2020}{\natexlab{a}})}\BibitemShut
  {NoStop}%
\bibitem [{\citenamefont {Chatterjee}\ \emph {et~al.}(2023)\citenamefont
  {Chatterjee}, \citenamefont {Sau}, \citenamefont {Samanta}, \citenamefont
  {Ghosh}, \citenamefont {Kumar}, \citenamefont {Kumar},\ and\ \citenamefont
  {Mandal}}]{SupPhysRevB.107.125138}%
  \BibitemOpen
  \bibfield  {author} {\bibinfo {author} {\bibfnamefont {S.}~\bibnamefont
  {Chatterjee}}, \bibinfo {author} {\bibfnamefont {J.}~\bibnamefont {Sau}},
  \bibinfo {author} {\bibfnamefont {S.}~\bibnamefont {Samanta}}, \bibinfo
  {author} {\bibfnamefont {B.}~\bibnamefont {Ghosh}}, \bibinfo {author}
  {\bibfnamefont {N.}~\bibnamefont {Kumar}}, \bibinfo {author} {\bibfnamefont
  {M.}~\bibnamefont {Kumar}},\ and\ \bibinfo {author} {\bibfnamefont
  {K.}~\bibnamefont {Mandal}},\ }\href
  {https://doi.org/10.1103/PhysRevB.107.125138} {\bibfield  {journal} {\bibinfo
   {journal} {Phys. Rev. B}\ }\textbf {\bibinfo {volume} {107}},\ \bibinfo
  {pages} {125138} (\bibinfo {year} {2023})}\BibitemShut {NoStop}%
\bibitem [{\citenamefont {Chatterjee}\ \emph {et~al.}(2022)\citenamefont
  {Chatterjee}, \citenamefont {Sau}, \citenamefont {Ghosh}, \citenamefont
  {Samanta}, \citenamefont {Ghosh}, \citenamefont {Kumar},\ and\ \citenamefont
  {Mandal}}]{SupChatterjee_2023}%
  \BibitemOpen
  \bibfield  {author} {\bibinfo {author} {\bibfnamefont {S.}~\bibnamefont
  {Chatterjee}}, \bibinfo {author} {\bibfnamefont {J.}~\bibnamefont {Sau}},
  \bibinfo {author} {\bibfnamefont {S.}~\bibnamefont {Ghosh}}, \bibinfo
  {author} {\bibfnamefont {S.}~\bibnamefont {Samanta}}, \bibinfo {author}
  {\bibfnamefont {B.}~\bibnamefont {Ghosh}}, \bibinfo {author} {\bibfnamefont
  {M.}~\bibnamefont {Kumar}},\ and\ \bibinfo {author} {\bibfnamefont
  {K.}~\bibnamefont {Mandal}},\ }\href
  {https://doi.org/10.1088/1361-648X/aca0d7} {\bibfield  {journal} {\bibinfo
  {journal} {Journal of Physics: Condensed Matter}\ }\textbf {\bibinfo {volume}
  {35}},\ \bibinfo {pages} {035601} (\bibinfo {year} {2022})}\BibitemShut
  {NoStop}%
\bibitem [{\citenamefont {Wang}\ \emph {et~al.}(2016)\citenamefont {Wang},
  \citenamefont {Sun}, \citenamefont {Zhang}, \citenamefont {Pang},\ and\
  \citenamefont {Lei}}]{SupPhysRevB.94.075135}%
  \BibitemOpen
  \bibfield  {author} {\bibinfo {author} {\bibfnamefont {Q.}~\bibnamefont
  {Wang}}, \bibinfo {author} {\bibfnamefont {S.}~\bibnamefont {Sun}}, \bibinfo
  {author} {\bibfnamefont {X.}~\bibnamefont {Zhang}}, \bibinfo {author}
  {\bibfnamefont {F.}~\bibnamefont {Pang}},\ and\ \bibinfo {author}
  {\bibfnamefont {H.}~\bibnamefont {Lei}},\ }\href
  {https://doi.org/10.1103/PhysRevB.94.075135} {\bibfield  {journal} {\bibinfo
  {journal} {Phys. Rev. B}\ }\textbf {\bibinfo {volume} {94}},\ \bibinfo
  {pages} {075135} (\bibinfo {year} {2016})}\BibitemShut {NoStop}%
\bibitem [{\citenamefont {Roy}\ \emph {et~al.}(2020{\natexlab{b}})\citenamefont
  {Roy}, \citenamefont {Singha}, \citenamefont {Ghosh}, \citenamefont
  {Pariari},\ and\ \citenamefont {Mandal}}]{PhysRevB.102.085147}%
  \BibitemOpen
  \bibfield  {author} {\bibinfo {author} {\bibfnamefont {S.}~\bibnamefont
  {Roy}}, \bibinfo {author} {\bibfnamefont {R.}~\bibnamefont {Singha}},
  \bibinfo {author} {\bibfnamefont {A.}~\bibnamefont {Ghosh}}, \bibinfo
  {author} {\bibfnamefont {A.}~\bibnamefont {Pariari}},\ and\ \bibinfo {author}
  {\bibfnamefont {P.}~\bibnamefont {Mandal}},\ }\href
  {https://doi.org/10.1103/PhysRevB.102.085147} {\bibfield  {journal} {\bibinfo
   {journal} {Phys. Rev. B}\ }\textbf {\bibinfo {volume} {102}},\ \bibinfo
  {pages} {085147} (\bibinfo {year} {2020}{\natexlab{b}})}\BibitemShut
  {NoStop}%
\bibitem [{\citenamefont {Wang}\ \emph {et~al.}(2018)\citenamefont {Wang},
  \citenamefont {Xu}, \citenamefont {Lou}, \citenamefont {Liu}, \citenamefont
  {Li}, \citenamefont {Huang}, \citenamefont {Shen}, \citenamefont {Weng},
  \citenamefont {Wang},\ and\ \citenamefont {Lei}}]{SupWang2018}%
  \BibitemOpen
  \bibfield  {author} {\bibinfo {author} {\bibfnamefont {Q.}~\bibnamefont
  {Wang}}, \bibinfo {author} {\bibfnamefont {Y.}~\bibnamefont {Xu}}, \bibinfo
  {author} {\bibfnamefont {R.}~\bibnamefont {Lou}}, \bibinfo {author}
  {\bibfnamefont {Z.}~\bibnamefont {Liu}}, \bibinfo {author} {\bibfnamefont
  {M.}~\bibnamefont {Li}}, \bibinfo {author} {\bibfnamefont {Y.}~\bibnamefont
  {Huang}}, \bibinfo {author} {\bibfnamefont {D.}~\bibnamefont {Shen}},
  \bibinfo {author} {\bibfnamefont {H.}~\bibnamefont {Weng}}, \bibinfo {author}
  {\bibfnamefont {S.}~\bibnamefont {Wang}},\ and\ \bibinfo {author}
  {\bibfnamefont {H.}~\bibnamefont {Lei}},\ }\href
  {https://doi.org/10.1038/s41467-018-06088-2} {\bibfield  {journal} {\bibinfo
  {journal} {Nature Communications}\ }\textbf {\bibinfo {volume} {9}},\
  \bibinfo {pages} {3681} (\bibinfo {year} {2018})}\BibitemShut {NoStop}%
\bibitem [{\citenamefont {Liu}\ \emph {et~al.}(2019)\citenamefont {Liu},
  \citenamefont {Liang}, \citenamefont {Liu}, \citenamefont {Xu}, \citenamefont
  {Li}, \citenamefont {Chen}, \citenamefont {Pei}, \citenamefont {Shi},
  \citenamefont {Mo}, \citenamefont {Dudin}, \citenamefont {Kim}, \citenamefont
  {Cacho}, \citenamefont {Li}, \citenamefont {Sun}, \citenamefont {Yang},
  \citenamefont {Liu}, \citenamefont {Parkin}, \citenamefont {Felser},\ and\
  \citenamefont {Chen}}]{Liu2019}%
  \BibitemOpen
  \bibfield  {author} {\bibinfo {author} {\bibfnamefont {D.~F.}\ \bibnamefont
  {Liu}}, \bibinfo {author} {\bibfnamefont {A.~J.}\ \bibnamefont {Liang}},
  \bibinfo {author} {\bibfnamefont {E.~K.}\ \bibnamefont {Liu}}, \bibinfo
  {author} {\bibfnamefont {Q.~N.}\ \bibnamefont {Xu}}, \bibinfo {author}
  {\bibfnamefont {Y.~W.}\ \bibnamefont {Li}}, \bibinfo {author} {\bibfnamefont
  {C.}~\bibnamefont {Chen}}, \bibinfo {author} {\bibfnamefont {D.}~\bibnamefont
  {Pei}}, \bibinfo {author} {\bibfnamefont {W.~J.}\ \bibnamefont {Shi}},
  \bibinfo {author} {\bibfnamefont {S.~K.}\ \bibnamefont {Mo}}, \bibinfo
  {author} {\bibfnamefont {P.}~\bibnamefont {Dudin}}, \bibinfo {author}
  {\bibfnamefont {T.}~\bibnamefont {Kim}}, \bibinfo {author} {\bibfnamefont
  {C.}~\bibnamefont {Cacho}}, \bibinfo {author} {\bibfnamefont
  {G.}~\bibnamefont {Li}}, \bibinfo {author} {\bibfnamefont {Y.}~\bibnamefont
  {Sun}}, \bibinfo {author} {\bibfnamefont {L.~X.}\ \bibnamefont {Yang}},
  \bibinfo {author} {\bibfnamefont {Z.~K.}\ \bibnamefont {Liu}}, \bibinfo
  {author} {\bibfnamefont {S.~S.~P.}\ \bibnamefont {Parkin}}, \bibinfo {author}
  {\bibfnamefont {C.}~\bibnamefont {Felser}},\ and\ \bibinfo {author}
  {\bibfnamefont {Y.~L.}\ \bibnamefont {Chen}},\ }\href
  {https://doi.org/10.1126/science.aav2873} {\bibfield  {journal} {\bibinfo
  {journal} {Science}\ }\textbf {\bibinfo {volume} {365}},\ \bibinfo {pages}
  {1282} (\bibinfo {year} {2019})}\BibitemShut {NoStop}%
\bibitem [{fle()}]{Supfleur}%
  \BibitemOpen
  \href {https://www.flapw.de/MaX-6.0/} {\bibinfo {title}
  {{www.flapw.de}}}\BibitemShut {NoStop}%
\bibitem [{\citenamefont {Kresse}\ and\ \citenamefont {Joubert}(1999)}]{Supvasp}%
  \BibitemOpen
  \bibfield  {author} {\bibinfo {author} {\bibfnamefont {G.}~\bibnamefont
  {Kresse}}\ and\ \bibinfo {author} {\bibfnamefont {D.}~\bibnamefont
  {Joubert}},\ }\href {https://doi.org/10.1103/PhysRevB.59.1758} {\bibfield
  {journal} {\bibinfo  {journal} {Phys. Rev. B}\ }\textbf {\bibinfo {volume}
  {59}},\ \bibinfo {pages} {1758} (\bibinfo {year} {1999})}\BibitemShut
  {NoStop}%
\bibitem [{\citenamefont {Perdew}\ \emph {et~al.}(1996)\citenamefont {Perdew},
  \citenamefont {Burke},\ and\ \citenamefont {Ernzerhof}}]{Supburke}%
  \BibitemOpen
  \bibfield  {author} {\bibinfo {author} {\bibfnamefont {J.~P.}\ \bibnamefont
  {Perdew}}, \bibinfo {author} {\bibfnamefont {K.}~\bibnamefont {Burke}},\ and\
  \bibinfo {author} {\bibfnamefont {M.}~\bibnamefont {Ernzerhof}},\ }\href
  {https://doi.org/10.1103/PhysRevLett.77.3865} {\bibfield  {journal} {\bibinfo
   {journal} {Phys. Rev. Lett.}\ }\textbf {\bibinfo {volume} {77}},\ \bibinfo
  {pages} {3865} (\bibinfo {year} {1996})}\BibitemShut {NoStop}%
\bibitem [{\citenamefont {Perdew}\ and\ \citenamefont {Wang}(1992)}]{Supperdew}%
  \BibitemOpen
  \bibfield  {author} {\bibinfo {author} {\bibfnamefont {J.~P.}\ \bibnamefont
  {Perdew}}\ and\ \bibinfo {author} {\bibfnamefont {Y.}~\bibnamefont {Wang}},\
  }\href {https://doi.org/10.1103/PhysRevB.45.13244} {\bibfield  {journal}
  {\bibinfo  {journal} {Phys. Rev. B}\ }\textbf {\bibinfo {volume} {45}},\
  \bibinfo {pages} {13244} (\bibinfo {year} {1992})}\BibitemShut {NoStop}%
\bibitem [{\citenamefont {Monkhorst}\ and\ \citenamefont {Pack}(1976)}]{Supmonk}%
  \BibitemOpen
  \bibfield  {author} {\bibinfo {author} {\bibfnamefont {H.~J.}\ \bibnamefont
  {Monkhorst}}\ and\ \bibinfo {author} {\bibfnamefont {J.~D.}\ \bibnamefont
  {Pack}},\ }\href {https://doi.org/10.1103/PhysRevB.13.5188} {\bibfield
  {journal} {\bibinfo  {journal} {Phys. Rev. B}\ }\textbf {\bibinfo {volume}
  {13}},\ \bibinfo {pages} {5188} (\bibinfo {year} {1976})}\BibitemShut
  {NoStop}%
\bibitem [{\citenamefont {Pizzi}\ \emph {et~al.}(2020)\citenamefont {Pizzi},
  \citenamefont {Vitale}, \citenamefont {Arita}, \citenamefont {Bl{\"u}gel},
  \citenamefont {Freimuth}, \citenamefont {G{\'e}ranton}, \citenamefont
  {Gibertini}, \citenamefont {Gresch}, \citenamefont {Johnson}, \citenamefont
  {Koretsune} \emph {et~al.}}]{Suppizzi2020wannier90}%
  \BibitemOpen
  \bibfield  {author} {\bibinfo {author} {\bibfnamefont {G.}~\bibnamefont
  {Pizzi}}, \bibinfo {author} {\bibfnamefont {V.}~\bibnamefont {Vitale}},
  \bibinfo {author} {\bibfnamefont {R.}~\bibnamefont {Arita}}, \bibinfo
  {author} {\bibfnamefont {S.}~\bibnamefont {Bl{\"u}gel}}, \bibinfo {author}
  {\bibfnamefont {F.}~\bibnamefont {Freimuth}}, \bibinfo {author}
  {\bibfnamefont {G.}~\bibnamefont {G{\'e}ranton}}, \bibinfo {author}
  {\bibfnamefont {M.}~\bibnamefont {Gibertini}}, \bibinfo {author}
  {\bibfnamefont {D.}~\bibnamefont {Gresch}}, \bibinfo {author} {\bibfnamefont
  {C.}~\bibnamefont {Johnson}}, \bibinfo {author} {\bibfnamefont
  {T.}~\bibnamefont {Koretsune}}, \emph {et~al.},\ }\href
  {https://doi.org/10.1088/1361-648X/ab51ff} {\bibfield  {journal} {\bibinfo
  {journal} {Journal of Physics: Condensed Matter}\ }\textbf {\bibinfo {volume}
  {32}},\ \bibinfo {pages} {165902} (\bibinfo {year} {2020})}\BibitemShut
  {NoStop}%
\bibitem [{\citenamefont {Marzari}\ and\ \citenamefont
  {Vanderbilt}(1997)}]{SupmarzariPhysRevB.56.12847}%
  \BibitemOpen
  \bibfield  {author} {\bibinfo {author} {\bibfnamefont {N.}~\bibnamefont
  {Marzari}}\ and\ \bibinfo {author} {\bibfnamefont {D.}~\bibnamefont
  {Vanderbilt}},\ }\href {https://doi.org/10.1103/PhysRevB.56.12847} {\bibfield
   {journal} {\bibinfo  {journal} {Phys. Rev. B}\ }\textbf {\bibinfo {volume}
  {56}},\ \bibinfo {pages} {12847} (\bibinfo {year} {1997})}\BibitemShut
  {NoStop}%
\bibitem [{\citenamefont {Yao}\ \emph {et~al.}(2004)\citenamefont {Yao},
  \citenamefont {Kleinman}, \citenamefont {MacDonald}, \citenamefont {Sinova},
  \citenamefont {Jungwirth}, \citenamefont {Wang}, \citenamefont {Wang},\ and\
  \citenamefont {Niu}}]{Supkubo}%
  \BibitemOpen
  \bibfield  {author} {\bibinfo {author} {\bibfnamefont {Y.}~\bibnamefont
  {Yao}}, \bibinfo {author} {\bibfnamefont {L.}~\bibnamefont {Kleinman}},
  \bibinfo {author} {\bibfnamefont {A.~H.}\ \bibnamefont {MacDonald}}, \bibinfo
  {author} {\bibfnamefont {J.}~\bibnamefont {Sinova}}, \bibinfo {author}
  {\bibfnamefont {T.}~\bibnamefont {Jungwirth}}, \bibinfo {author}
  {\bibfnamefont {D.-s.}\ \bibnamefont {Wang}}, \bibinfo {author}
  {\bibfnamefont {E.}~\bibnamefont {Wang}},\ and\ \bibinfo {author}
  {\bibfnamefont {Q.}~\bibnamefont {Niu}},\ }\href
  {https://doi.org/10.1103/PhysRevLett.92.037204} {\bibfield  {journal}
  {\bibinfo  {journal} {Phys. Rev. Lett.}\ }\textbf {\bibinfo {volume} {92}},\
  \bibinfo {pages} {037204} (\bibinfo {year} {2004})}\BibitemShut {NoStop}%
\bibitem [{\citenamefont {Wu}\ \emph {et~al.}(2018)\citenamefont {Wu},
  \citenamefont {Zhang}, \citenamefont {Song}, \citenamefont {Troyer},\ and\
  \citenamefont {Soluyanov}}]{Wu_2018}%
  \BibitemOpen
  \bibfield  {author} {\bibinfo {author} {\bibfnamefont {Q.}~\bibnamefont
  {Wu}}, \bibinfo {author} {\bibfnamefont {S.}~\bibnamefont {Zhang}}, \bibinfo
  {author} {\bibfnamefont {H.-F.}\ \bibnamefont {Song}}, \bibinfo {author}
  {\bibfnamefont {M.}~\bibnamefont {Troyer}},\ and\ \bibinfo {author}
  {\bibfnamefont {A.~A.}\ \bibnamefont {Soluyanov}},\ }\href
  {https://doi.org/10.1016/j.cpc.2017.09.033} {\bibfield  {journal} {\bibinfo
  {journal} {Computer Physics Communications}\ }\textbf {\bibinfo {volume}
  {224}},\ \bibinfo {pages} {405} (\bibinfo {year} {2018})}\BibitemShut
  {NoStop}%
\end{thebibliography}
\end{document}